\documentclass[final,3p,times,sort&compress]{elsarticle}
\usepackage{multirow,setspace,times,amssymb,amsmath,graphicx,color,rotating,subfigure,url}
\usepackage{lineno}
\usepackage{natbib}
\usepackage{booktabs}%
\usepackage{longtable}%
\usepackage{soul}
\usepackage{subfigure}
\usepackage[table]{xcolor}%
\usepackage{tabularx}
\usepackage{graphicx} 
\usepackage{epstopdf}
\usepackage{mathrsfs}
\usepackage{makecell}
\usepackage{pifont}
\usepackage[bookmarks=true,colorlinks,linkcolor=blue,anchorcolor=blue,citecolor=blue,unicode]{hyperref}
\usepackage{bookmark}
\graphicspath{{Figures/}}

\journal{Fractals}

\begin{document}

\begin{frontmatter}
\title{Testing for intrinsic multifractality in the global grain spot market indices:\\A multifractal detrended fluctuation analysis}

\author[SB,RCE,SPhys]{Li Wang}
\author[SB]{Xing-Lu Gao}
\author[SB,RCE,Math]{Wei-Xing Zhou\corref{cor1}}
\ead{wxzhou@ecust.edu.cn}
\cortext[cor1]{Corresponding author.}
\address[SB]{School of Business, East China University of Science and Technology, Shanghai 200237, China}
\address[RCE]{Research Center for Econophysics, East China University of Science and Technology, Shanghai 200237, China}
\address[SPhys]{International Elite Engineering School, East China University of Science and Technology, Shanghai 200237, China}
\address[Math]{School of Mathematics, East China University of Science and Technology, Shanghai 200237, China}


\begin{abstract}
   Grains account for more than 50\% of the calories consumed by people worldwide, and military conflicts, pandemics, climate change, and soaring grain prices all have vital impacts on food security. However, the complex price behavior of the global grain spot markets has not been well understood. A recent study performed multifractal moving average analysis (MF-DMA) of the Grains \& Oilseeds Index (GOI) and its sub-indices of wheat, maize, soyabeans, rice, and barley and found that only the maize and barley sub-indices exhibit an intrinsic multifractal nature with convincing evidence. Here, we utilize multifractal fluctuation analysis (MF-DFA) to investigate the same problem. Extensive statistical tests confirm the presence of intrinsic multifractality in the maize and barley sub-indices and the absence of intrinsic multifractality in the wheat and rice sub-indices. Different from the MF-DMA results, the MF-DFA results suggest that there is also intrinsic multifractality in the GOI and soyabeans sub-indices. Our comparative analysis does not provide conclusive information about the GOI and soyabeans and highlights the high complexity of the global grain spot markets.
\end{abstract}


\end{frontmatter}


\section{Introduction}
\label{S1:Introduction}

Food is a basic human need. However, food crises happened from time to time in history. The past few years have witnessed the accumulation of the ongoing food crisis, mainly driven by wars/conflicts/insecurity, the COVID-19 pandemic, weather extremes, and domestic food price inflation. Ant{\'o}nio Guterres, Secretary-General of the United Nations, remarked that\footnote{Forward of the 2022 {\it{Global Report on Food Crises}} (GRFC 2022), available at {\url{https://www.fao.org/}}.}, ``We are facing hunger on an unprecedented scale, food prices have never been higher, and millions of lives and livelihoods are hanging in the balance.'' Grains account for more than 50\% of the calories consumed by people worldwide. However, the complex behavior of the price evolution of the global grain spot markets is less researched from the perspective of complexity science, such as multifractal analysis of the grain markets.

Multifractal nature is ubiquitously present in financial time series for different financial variables \cite{Jiang-Xie-Zhou-Sornette-2019-RepProgPhys}. Concerning agricultural assets, researchers have carried out multifractal analyses of various agricultural commodity futures in different markets \cite{Jiang-Xie-Zhou-Sornette-2019-RepProgPhys}, including returns or prices \cite{He-Chen-2010-PhysicaA,FJ-He-Chen-2011-ChaosSolitonsFractals,Wang-Hu-2015-PhysicaA,Wang-Feng-2020-JStatMech,Yin-Wang-2021-AgricEcon}, volatility \cite{Chen-He-2010-PhysicaA,Li-Lu-2012-PhysicaA}, and trading volumes \cite{He-Chen-2011-PhysicaA}.
Multifractal analysis has also been performed of agricultural commodity spot markets, including returns or prices \cite{Kim-Oh-Kim-2011-PhysicaA,Liu-2014-PhysicaA,Delbianco-Tohme-Stosic-Stosic-2016-PhysicaA,Wang-Feng-2020-JStatMech,Stosic-Nejad-Stosic-2020-Fractals} and volatility \cite{Liu-2014-PhysicaA}.
There are also studies that use joint multifractal analysis methods to investigate the relationship between two agricultural time series, including the price-volume relationships in agricultural commodity futures markets in the U.S.A. and China  \cite{He-Chen-2011-PhysicaA}, the interest rates and commodity futures prices in the U.S.A. \cite{Wang-Hu-2015-PhysicaA}, the futures prices of the same agricultural commodities in the U.S.A. and China \cite{Li-Lu-2012-PhysicaA,FJ-Feng-Cao-2022-FluctNoiseLett}, the prices of agricultural futures and spot markets in the U.S.A. \cite{Wang-Feng-2020-JStatMech}, and the crude oil and agricultural futures prices \cite{Wang-Shao-Kim-2020-ChaosSolitonsFractals}.
All these studies reported the presence of multifractal behavior in the considered agricultural futures and in the spot prices of the considered agricultural commodities.

The multifractal behavior of several agricultural commodity spot markets is also confirmed, including the daily averaging prices of four agricultural commodities (leek, radish, onion, and Korean cabbage) in Korea \cite{Kim-Oh-Kim-2011-PhysicaA}, the daily return and volatility of four agricultural commodities (corn, soybean, oat, and wheat) in the U.S.A.  \cite{Liu-2014-PhysicaA}, the monthly soybean index of the World Bank \cite{Delbianco-Tohme-Stosic-Stosic-2016-PhysicaA}, the daily closing prices of five agricultural commodities (soybean, corn, wheat, soybean meal, and soybean oil) in the U.S.A. spot markets  \cite{Wang-Feng-2020-JStatMech},  and 12 agricultural commodities (ethanol, sugar, coffee, corn, cotton, rice, soybean, wheat, cattle, calf, and pork) in Brazil \cite{Stosic-Nejad-Stosic-2020-Fractals}. All these agricultural commodities are from local markets.
Recently, Gao et al. performed multifractal moving average analysis (MF-DMA) of the Grains \& Oilseeds Index (GOI) and its sub-indices of wheat, maize, soyabeans, rice, and barley \cite{Gao-Shao-Yang-Zhou-2022-ChaosSolitonsFractals}. Intriguingly, through statistical tests based on surrogates generated by the iterative amplitude adjusted Fourier transform (IAAFT) algorithm, they found that the maize and barley sub-indices exhibit intrinsic multifractality, the GOI, wheat, and rice price indices do not possess multifractality, while the soyabeans sub-index might have multifractal behavior \cite{Gao-Shao-Yang-Zhou-2022-ChaosSolitonsFractals}. The study highlights the importance of statistical tests in confirming or denying the presence of multifractal nature in real-world time series.

Indeed, performing multifractal analysis on a time series will always result in the characteristic multifractal functions $H(q)$, $\tau(q)$ and $f(\alpha)$, and they usually deviate from the theoretical results of monofractal time series. However, it does not guarantee the presence of multifractality; rather, one needs to perform statistical tests to confirm if the extracted empirical multifractality is intrinsic or not \cite{Jiang-Xie-Zhou-Sornette-2019-RepProgPhys}. There are three sources of the empirical multifractality or apparent multifractality extracted from real time series \cite{Jiang-Xie-Zhou-Sornette-2019-RepProgPhys}, namely, the non-Gaussian distribution of the innovations (via shuffled time series, which retains the values of the original time series but destroys any linear and nonlinear correlations) \cite{Matia-Ashkenazy-Stanley-2003-EPL,Oswiecimka-Kwapien-Drozdz-Rak-2005-APPB,Lee-Lee-2005b-JKPS,Kwapien-Oswiecimka-Drozdz-2005-PhysicaA,Kumar-Deo-2009-PhysicaA,deSouza-Queiros-2009-ChaosSolitonsFractals}, the linear long-range correlations (via the phase randomization algorithm , which retains the linear correlation but destroys the distribution and nonlinear correlations)
\cite{Theiler-Eubank-Longtin-Galdrikian-Farmer-1992-PhysicaD,Ruan-Wang-Liu-Lv-2020-FluctNoiseLett,Feng-Wang-2021-FluctNoiseLett,Zhao-Dai-2021-FluctNoiseLett}, and the nonlinear long-range correlations \cite{Zhou-2009-EPL,Drozdz-Kwapien-Oswiecimka-Rak-2009-EPL,Zhou-2012-ChaosSolitonsFractals}.
Careful investigations based on the multifractal detrended fluctuation analysis (MF-DFA) \cite{Bogachev-Eichner-Bunde-2007-PhysRevLett,Drozdz-Kwapien-Oswiecimka-Rak-2009-EPL,Zhou-2009-EPL,Zhou-2012-ChaosSolitonsFractals}, the wavelet transform modulus maxima (WTMM) approach \cite{Drozdz-Kwapien-Oswiecimka-Rak-2009-EPL}, and the partition function approach \cite{Zhou-2012-ChaosSolitonsFractals} have unveiled that the key source of intrinsic multifractality in time series is the nonlinear long-range correlations, while the fat tails and/or linear correlations play a part only when nonlinear correlations are present \cite{Kwapien-Blasiak-Drozdz-Oswiecimka-2023-PhysRevE}.

Therefore, to statistically test the presence of intrinsic multifractality, the null model should be able to generate surrogates that are randomized copies of the original time series with the same power spectrum, which do not possess any nonlinear correlations \cite{Jiang-Xie-Zhou-Sornette-2019-RepProgPhys}. In other words, the surrogates should preserve the values and linear long-term correlations of the original time series, but not the nonlinear correlations. In this vein, the iterated amplitude adjusted Fourier transform algorithm of Schreiber and Schmitz \cite{Schreiber-Schmitz-1996-PhysRevLett,Schreiber-Schmitz-2000-PhysicaD} perfectly meets the requirements to generate the surrogates \cite{Jiang-Xie-Zhou-Sornette-2019-RepProgPhys}. Indeed, this idea has been adopted and implemented in the multifractal analysis of real-world time series using the multifractal detrended fluctuation analysis \cite{Zhou-2009-EPL,Oswiecimka-Drozdz-Frasca-Gebarowski-Yoshimura-Zunino-Minati-2020-NonlinearDyn} and the multifractal detrending moving average analysis \cite{Gao-Shao-Yang-Zhou-2022-ChaosSolitonsFractals}. With the IAAFT surrogates, we can design statistics for $H(q)$, $\tau(q)$, $\alpha(q)$, and $f(\alpha)$ and perform statistical tests.


Different multifractal analysis methods for univariate and multivariate time series have been developed \cite{Jiang-Xie-Zhou-Sornette-2019-RepProgPhys}, such as the multifractal detrended fluctuation analysis \cite{Kantelhardt-Zschiegner-KoscielnyBunde-Havlin-Bunde-Stanley-2002-PhysicaA}, the multifractal detrending moving average analysis \cite{Gu-Zhou-2010-PhysRevE} as a generalization of the detrending moving average analysis \cite{Alessio-Carbone-Castelli-Frappietro-2002-EurPhysJB,Carbone-Castelli-Stanley-2004-PhysicaA}, the partition function approach \cite{Grassberger-1983-PLA,Grassberger-Procaccia-1983-PhysicaD,Grassberger-1985-PLA}, the structure function approach \cite{Kolmogorov-1962-JFM,VanAtta-Chen-1970-JFM,DiMatteo-2007-QuantFinanc}, the wavelet analysis \cite{Arneodo-Grasseau-Holschneider-1988-PhysRevLett},
the multifractal Higuchi's dimension analysis (MF-HDA) \cite{CarrizalesVelazquez-Donner-GuzmanVargas-2022-NonlinearDyn}, and their variants such as the multifractal detrended fluctuation analysis based on (optimized) empirical mode decomposition  \cite{Qian-Gu-Zhou-2011-PhysicaA,Lin-Dou-Liu-2021-NonlinearDyn}, the time-singularity multifractal spectrum distribution \cite{UrdaBenitez-CastroOspina-OrozcoDuque-2021-CommunNonlinearSciNumerSimul}, and the extended self-similarity based multifractal detrended fluctuation analysis \cite{Nian-Fu-2019-CommunNonlinearSciNumerSimul}. In this work, we will adopt the most popular method, MF-DFA, to investigate the Grains \& Oilseeds Index (GOI) and its sub-indices of wheat, maize, soyabeans, rice, and barley and compare with the MF-DMA results \cite{Gao-Shao-Yang-Zhou-2022-ChaosSolitonsFractals}.
MF-DFA is the multifractal extension of the detrended fluctuation analysis (DFA), which was introduced originally to study the long-range correlations in coding and noncoding DNA nucleotide sequences \cite{Peng-Buldyrev-Havlin-Simons-Stanley-Goldberger-1994-PhysRevE}. We note that, although MF-DFA became the most popular method for multifractal analysis of time series in diverse fields after the work of Kantelhardt et al. \cite{Kantelhardt-Zschiegner-KoscielnyBunde-Havlin-Bunde-Stanley-2002-PhysicaA}, it was developed earlier independently by Castro e Silva and Moreira in 1997 \cite{CastroESilva-Moreira-1997-PhysicaA} and by Weber and Talkner in 2001 \cite{Weber-Talkner-2001-JGR}.



In this work, we will utilize the multifractal fluctuation analysis to investigate the GOI and its sub-indices of wheat, maize, soyabeans, rice, and barley by extensive statistical tests based on their IAAFT surrogates and compare with the MF-DMA results \cite{Gao-Shao-Yang-Zhou-2022-ChaosSolitonsFractals}. We confirm the presence of intrinsic multifractality in the maize and barley indices and the absence of intrinsic multifractality in the wheat and rice indices. Different from the MF-DMA results, the MF-DFA results suggest that there is also intrinsic multifractality in the GOI and soyabeans indices. The partial discrepancies between the MF-DFA results and the MF-DMA results highlight the high complexity of the global grain spot markets and call for further studies on the underlying mechanisms.

The remainder of this work is organized as follows. Section~\ref{S1:Data} describes the data sets we investigate and the MF-DFA method we use. Section~\ref{S1:GOI:return:MFDFA:Empirics} presents the empirical results and extensive statistical tests on several characteristic multifractal variables using the IAAFT surrogate time series of the six indices. We summarize, discuss, and conclude in Section~\ref{S1:Summary}.

\section{Data and methods}
\label{S1:Data}

\subsection{Data description}

\begin{figure}[!ht]
    \centering
    \includegraphics[width=0.325\linewidth]{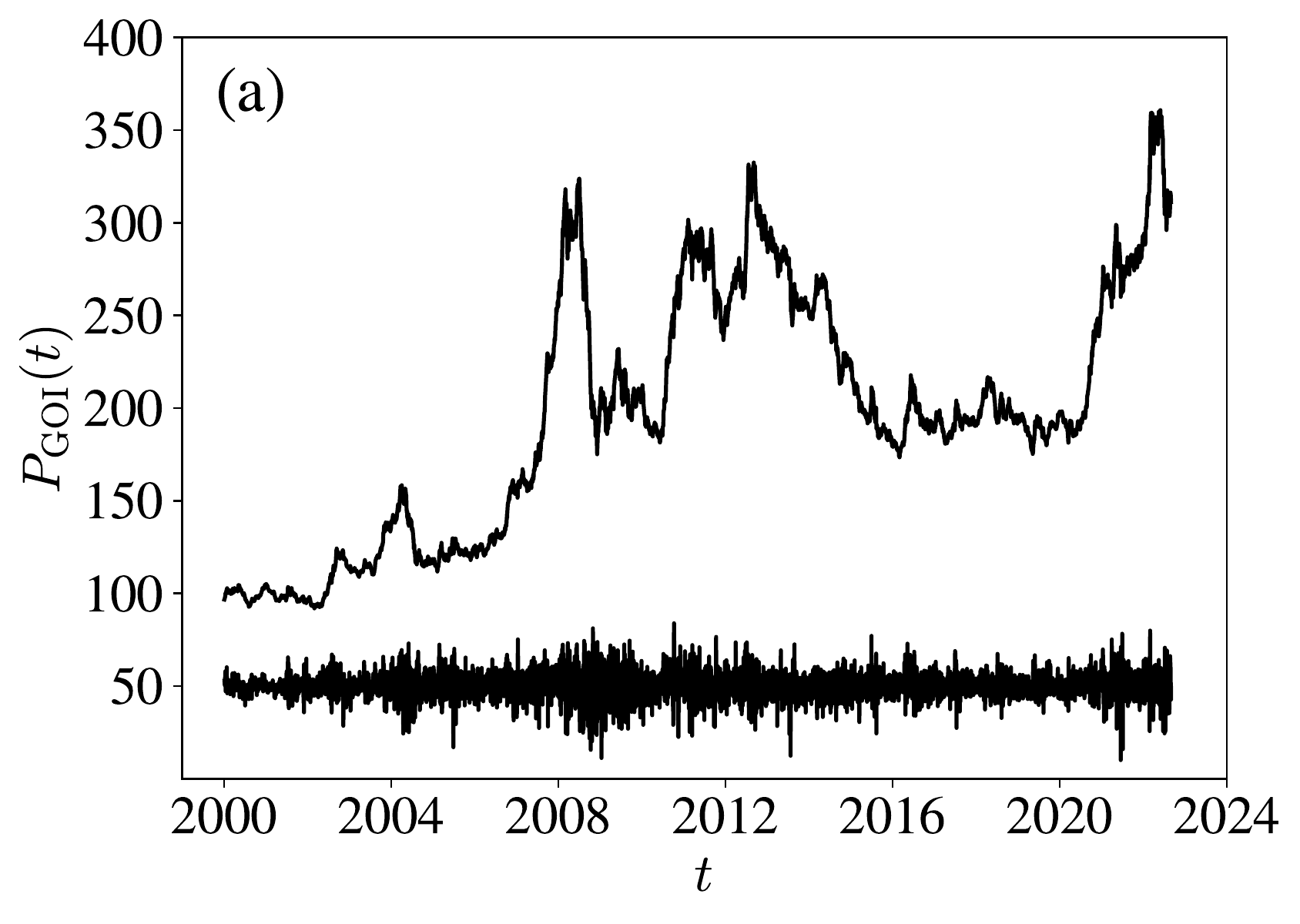}
    \includegraphics[width=0.325\linewidth]{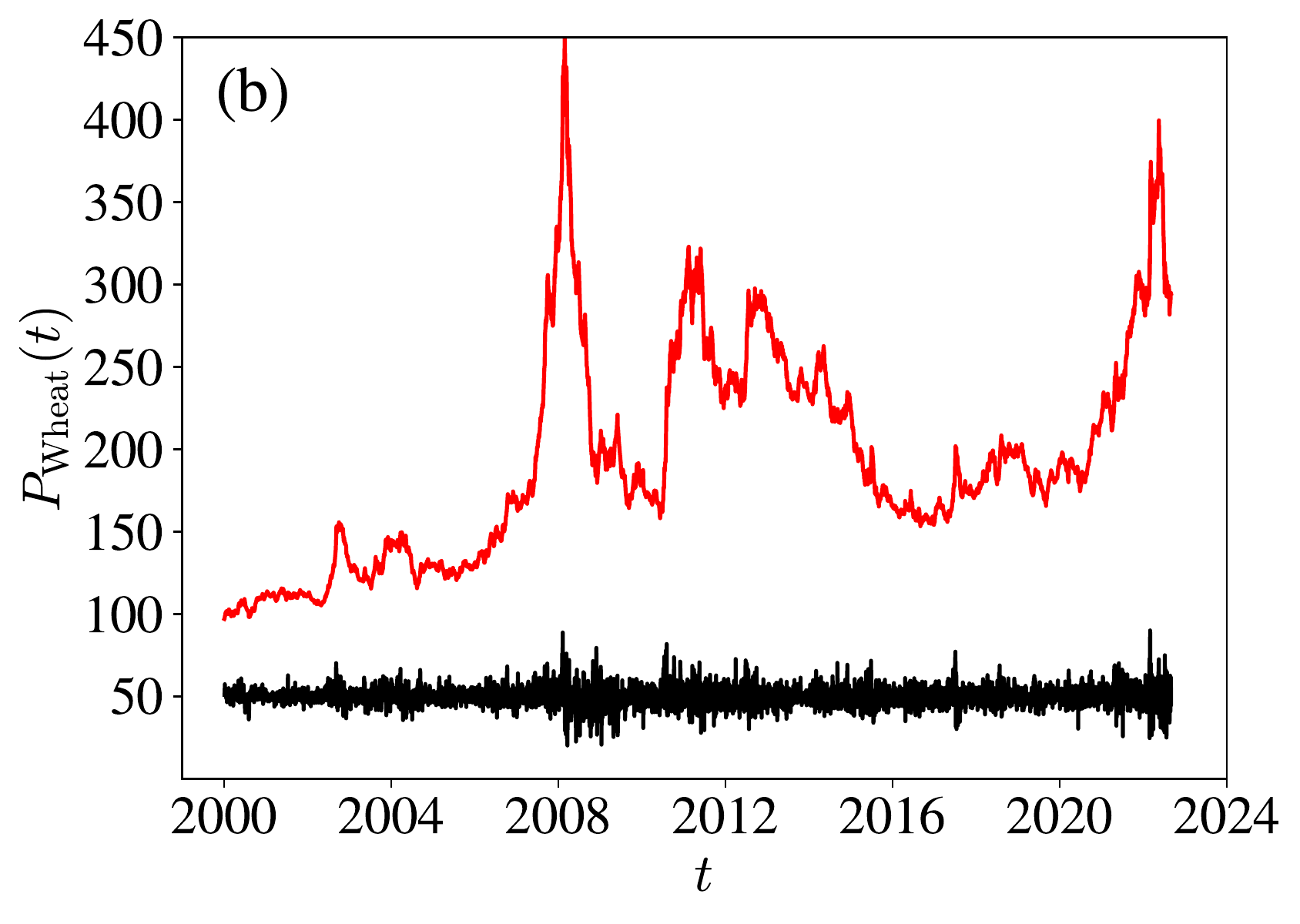}
    \includegraphics[width=0.325\linewidth]{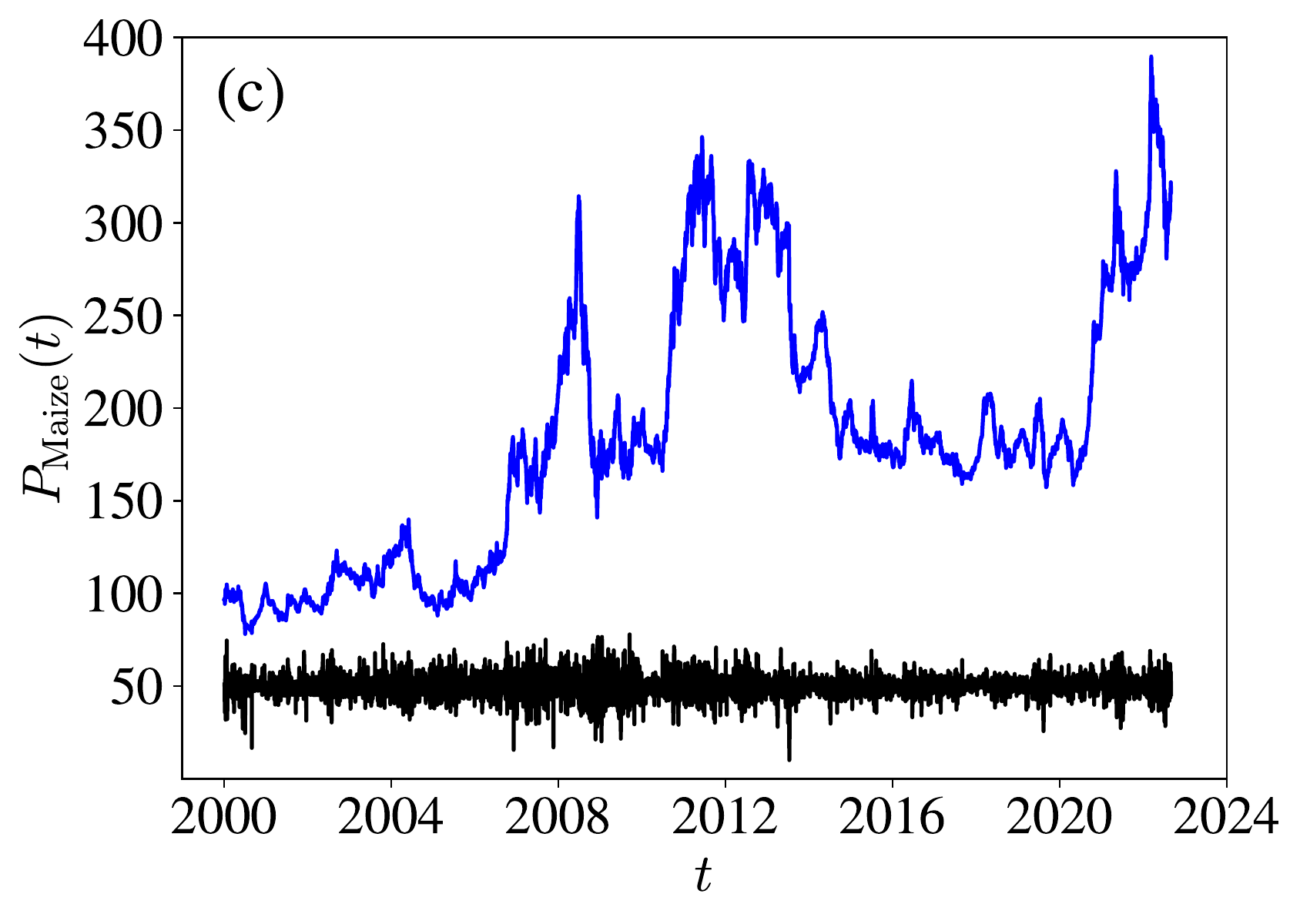}
    \includegraphics[width=0.325\linewidth]{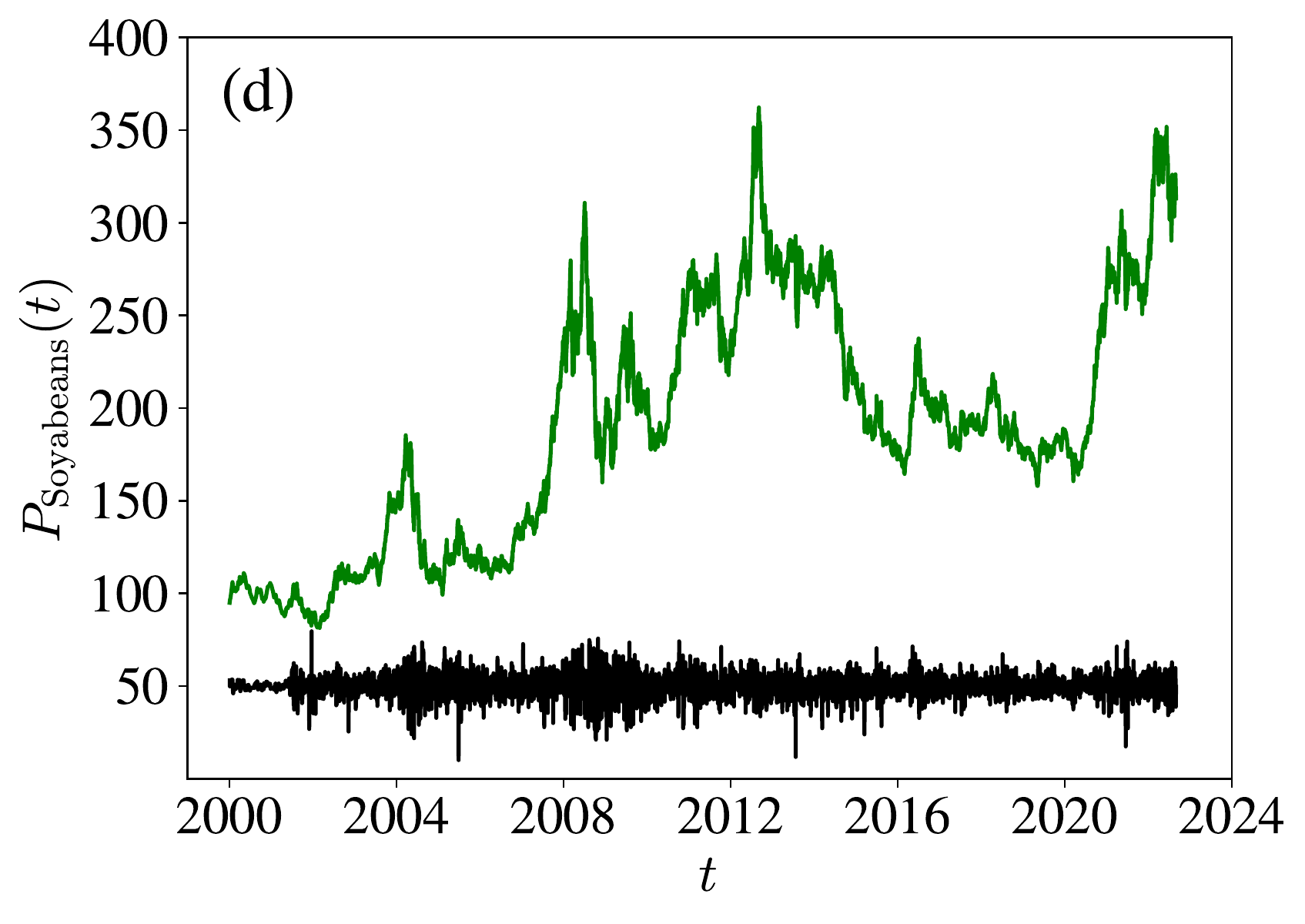}
    \includegraphics[width=0.325\linewidth]{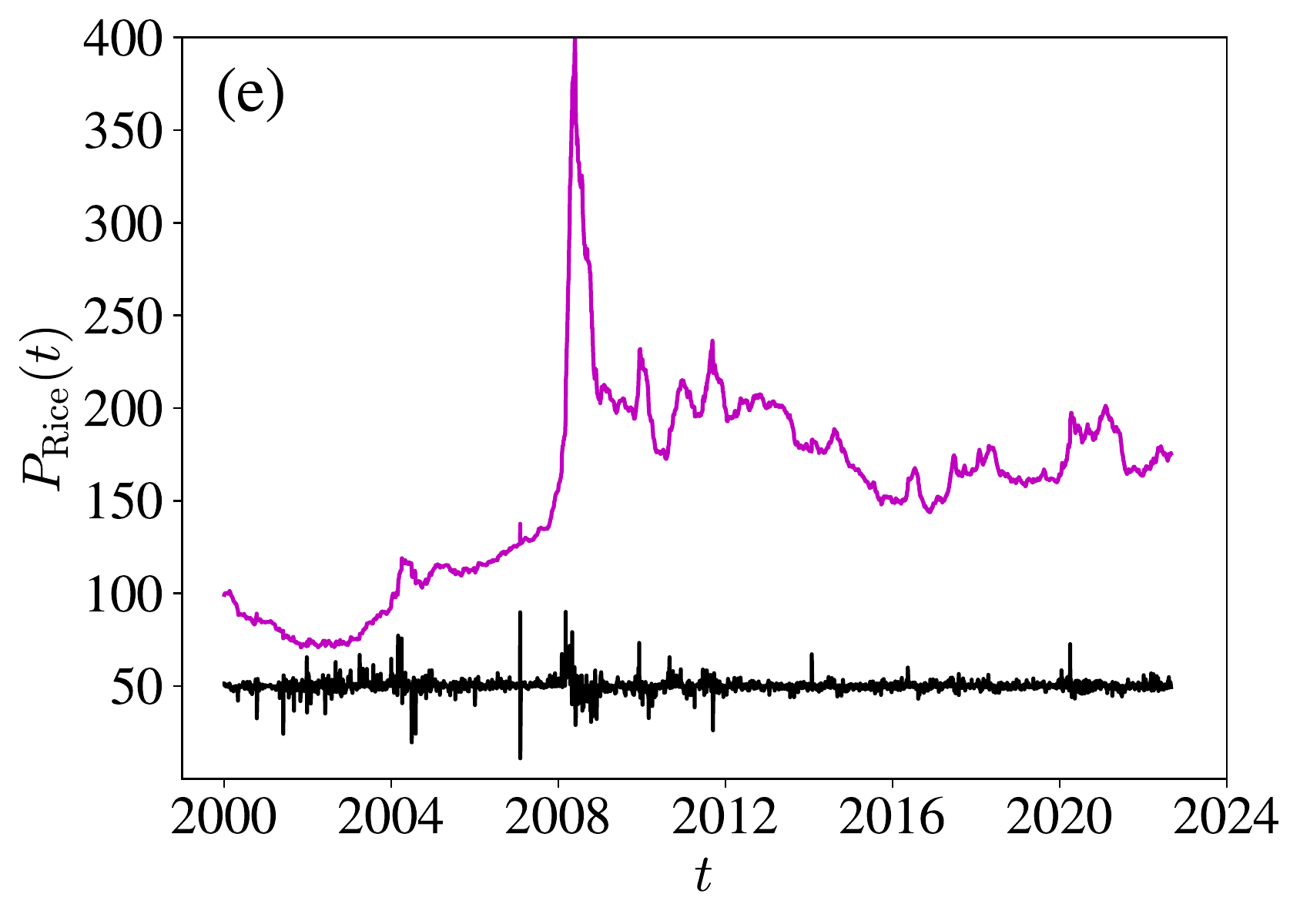}
    \includegraphics[width=0.325\linewidth]{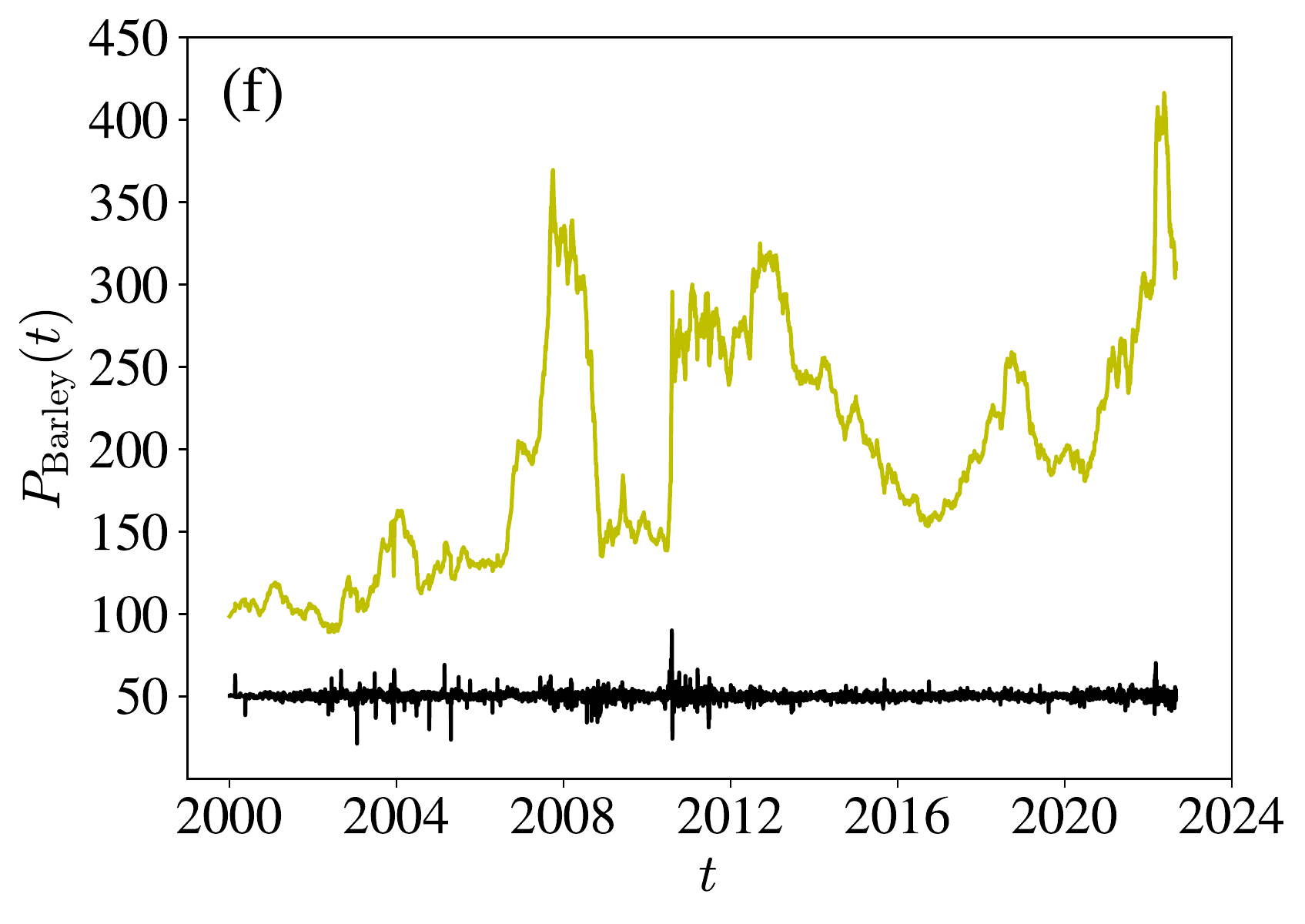}
    \caption{Daily price time series $P(t)$ and return time series $r(t)$ of the Grains \& Oilseeds Index GOI (a), the wheat sub-index (b), the maize sub-index (c), the soyabeans sub-index (d), the rice sub-index (e), and the barley sub-index (f) from 3 January 2000 to 31 August 2022. Source: The official website of the International Grains Council is available at http://www.igc.int.}
    \label{Fig:GOI:Pt:Rt}
\end{figure}

We retrieved the daily price time series of the Grains \& Oilseeds Index and its five sub-indices for wheat, maize, soyabeans, rice, and barley from the International Grains Council (publicly available at http://www.igc.int). All six time series cover the time period from January 3, 2000 to August 31, 2022.
In Fig.~\ref{Fig:GOI:Pt:Rt}, we show the daily price time series $P(t)$ of the GOI index $P_{\mathrm{GOI}}(t)$, the wheat sub-index $P_{\mathrm{Wheat}}(t)$, the maize sub-index $P_{\mathrm{Maize}}(t)$, the soyabeans sub-index $P_{\mathrm{Soyabeans}}(t)$, the rice sub-index $P_{\mathrm{Rice}}(t)$, and the barley sub-index $P_{\mathrm{Barley}}(t)$. All six indices suffered from the global food crisis of 2007-2008 with sharp price rises \cite{Headey-2011-FoodPolicy,Abbott-2012-AmJAgrEcon,Bouet-Debucquet-2012-RevWorldEcon}. The prices of wheat, maize, soyabeans, and barley also rose remarkably during 2010-2012 and 2020-2021. The recent spike in food prices is mainly caused by the ongoing COVID-19 pandemic \cite{Barrett-2020-NatFood,Akter-2020-FoodSecur,Falkendal-Otto-Schewe-Jagermeyr-Konar-Kummu-Watkins-Puma-2021-NatFood}. In addition, Russia started the so-called ``special military operation'' in Ukraine since February 24, 2022 and the Russia-Ukraine crisis sparked the further rise of food prices since the two countries are the third and fourth largest exporters of agricultural products and the main exporters of fertilizers. Brokered by the United Nations and Turkey on July 22, 2022, Ukraine and Russia signed an agreement in Istanbul that would allow the resumption of vital grain exports from Ukrainian Black Sea ports, which eased the global food crisis and reversed the rising trend of food prices.

The logarithmic return $r(t)$ of each index on date $t$ is defined as follows
\begin{equation}
    r(t) = \ln P(t) - \ln P(t-1).
\end{equation}
Figure~\ref{Fig:GOI:Pt:Rt} also shows the daily return time series $r(t)$ of the six indices. For better visibility, we manipulate $r(t)$ by stretching and vertical translation and plot the following quantity:
\begin{equation}
    \frac{40}{\max_t(|r(t)|)}r(t) + 50.
\end{equation}
The volatility clustering phenomenon can be observed in all return time series.

\subsection{Multifractal detrended fluctuation analysis}

The multifractal detrended fluctuation analysis is described briefly below.

For an index time series $P(t)$ ($t=1,2,\ldots,N$), we first determine the polynomial trend functions $\widetilde{P}_s(t)$ with respect to preset covering box sizes $s$  \cite{Kantelhardt-Zschiegner-KoscielnyBunde-Havlin-Bunde-Stanley-2002-PhysicaA}. For each point $t$, its trend value $\widetilde{P}(t)$ is determined together with other points belonging to the same segment $S_v$ based on a polynomial fitting of order $\ell$ to the $s$ data points in segment $S_v$:
\begin{equation}
  \widetilde{P}(t) = \sum_{k=0}^\ell a_k t^k, ~~~\ell=1,2, \cdots.
  \label{Eq:MF:Trend:Polynomial}
\end{equation}
We then detrend the index time series $P(t)$ by removing the polynomial trend function $\widetilde{P}_s(i)$ from $P(t)$ as follows
\begin{equation}
  \epsilon_s(t)=P(t)-\widetilde{P}_s(t).
  \label{Eq:1ddfa:epsilon}
\end{equation}
which results in the residual series $\epsilon_s(t)$.

We divide the residual time series $\{\epsilon_s(t)\}$ into $N_s=\mathrm{ceil}(N/s)$ non-overlapping segments with boxes of size $s$, where $\mathrm{ceil}(y)$ is the largest integer that is no larger than $y$. We denote the $v$th segment as $S_v=\{\epsilon_s((v-1)s+j)| j=1,2, \cdots, s\}$.
The local detrended fluctuation function $F_v(s)$ in the $v$th box is defined as the root mean square of the residuals:
\begin{equation}
  \left[F_v(s)\right]^2 = \frac{1}{s}\sum_{i=1}^{s} \left[\epsilon_s((v-1)s+j)\right]^2~.
  \label{Eq:MF:Detrend:Fv:s}
\end{equation}
The $q$th-order overall detrended fluctuation is
\begin{equation}
  F_q(s) = \left\{\frac{1}{N_s}\sum_{v=1}^{N_s} {F_v^q(s)}\right\}^{\frac{1}{q}},
  \label{Eq:MF:Detrend:Fqs}
\end{equation}
where $q$ can take any real value except $q=0$.
When $q=0$, we have
\begin{equation}
  \ln[F_0(s)] = \frac{1}{N_s}\sum_{v=1}^{N_s}{\ln[F_v(s)]},
  \label{Eq:MF:Detrend:Fq0}
\end{equation}
according to L'H\^{o}spital's rule. When the whole series $\{\epsilon_s(i)\}$ cannot be completely covered by $N_s$ boxes, one can use $2N_s$ boxes to cover the series from both ends of the series, where $N_s=\mathrm{ceil}[N/s]$ and a short part at each end of the residuals remains uncovered \cite{Kantelhardt-Zschiegner-KoscielnyBunde-Havlin-Bunde-Stanley-2002-PhysicaA}. We note that an alternative way is to cover the time series randomly, which is especially suitable for short time series \cite{Ji-Zhou-Liu-Gong-Wang-Yu-2009-PhysicaA,Jiang-Xie-Zhou-Sornette-2019-RepProgPhys}.


If the time series possesses fractal or multifractal nature, through varying the values of segment size $s$, we would obtain power-law relations between the function $F_q(s)$ and the size scale $s$,
\begin{equation}
  F_q(s) \sim {s}^{H(q)}.
  \label{Eq:MF:Detrend:hq}
\end{equation}
According to the standard multifractal formalism, the multifractal scaling exponents $\tau(q)$ can also be used to characterize the multifractal properties, which reads
\begin{equation}
  \tau(q)=qH(q)-D_f,
  \label{Eq:MF:Hq:2:tau}
\end{equation}
where $D_f$ is the fractal dimension of the geometric support of the multifractal measure \cite{Kantelhardt-Zschiegner-KoscielnyBunde-Havlin-Bunde-Stanley-2002-PhysicaA}. For time series, we have $D_f=1$. If the scaling exponent function $\tau(q)$ is a nonlinear concave function of $q$, the investigated time series is regarded to be multifractal. We can further calculate the singularity strength function $\alpha(q)$
\begin{equation}
   \alpha(q)={\rm{d}}\tau(q)/{\rm{d}}q,
\label{Eq:MF:tau:2:alpha}
\end{equation}
and the multifractal spectrum or singularity spectrum $f(\alpha)$
\begin{equation}
        f(q) = q\alpha(q) - \tau(q).
\label{Eq:MF:alpha:tau:2:f}
\end{equation}
All these functions characterize equivalently the multifractal nature of the time series.

In practice, it is not clear which polynomial order should be used for detrending, which may impact the results. Indeed, O{\'{s}}wi{\c{e}}cimka et al. investigated several mathematical models (fractional Brownian motion, L{\'e}vy process, and binomial measure) and real-world time series and reported that the calculated singularity spectra could be very sensitive to the order of the detrending polynomial \cite{Oswiecimka-Drozdz-Kwapien-Gorski-2013-APPA}. This sensitivity certainly depends on the time series under investigation. In this work, we investigated and compared the results based on the linear trend ($\ell=1$) and the quadratic polynomial trend ($\ell=2$). We did observe differences, which are however minor.

\subsection{Generation of IAAFT surrogates}

The IAAFT algorithm is implemented as follows \cite{Schreiber-Schmitz-1996-PhysRevLett,Schreiber-Schmitz-2000-PhysicaD}. First, we obtain an initial iterated time series as a random shuffle of the original time series. Second, the Fourier spectra of the original time series and the iterated time series are calculated. Third, the squared coefficients of the iterated time series are replaced by those of the original time series, while the phases remain unchanged. Fourth, the amplitudes of the generated time series are replaced by the values of the original sorted list with the same ranking. The last four steps iterate until the preset convergence condition is satisfied.

\section{Empirical analysis}
\label{S1:GOI:return:MFDFA:Empirics}

In this section, we perform multifractal detrended fluctuation analysis of each index time series. To investigate the possible impact of the choice of trend functions, we adopt the first-order ($\ell=1$) and second-order ($\ell=2$) polynomials to detrend the index price time series. We find that the results are almost the same qualitatively, with only slight differences. Hence, we show below the results based on the linear trend function in many places and all the comparative results when necessary. 
There are three issues we need to clarify:
\begin{enumerate}
    \item[(1)] Determine the range of $q$. Since higher-order moments in multifractal analysis diverge for large $q$'s \cite{Lvov-Podivilov-Pomyalove-Procaccia-Vandembroucq-1998-PhysRevE,Zhou-Sornette-Yuan-2006-PhysicaD}, we use  $q\in[-5,5]$ in our analysis, which also makes our MF-DFA results comparable to the MF-DMA results in Ref.~\cite{Gao-Shao-Yang-Zhou-2022-ChaosSolitonsFractals}.

    \item[(2)] Determine the scaling range. Kantelhardt et al. suggested to perform multifractal analysis in the scaling range $s\in[10, N/4]$ \cite{Kantelhardt-Zschiegner-KoscielnyBunde-Havlin-Bunde-Stanley-2002-PhysicaA}. In the MF-DMA analysis of GOI indices, Gao et al. used $s\in[10, 10^{2.5}]=[10, 316]$ \cite{Gao-Shao-Yang-Zhou-2022-ChaosSolitonsFractals}, which is consistent with Ref.~\cite{Kantelhardt-Zschiegner-KoscielnyBunde-Havlin-Bunde-Stanley-2002-PhysicaA} and slightly wider than the average width of scaling ranges in many practical studies \cite{Malcai-Lidar-Biham-Avnir-1997-PhysRevE,Avnir-Biham-Lidar-Malcai-1998-Science}. In this work, we use $s\in[20, 10^{2.5}]=[20, 316]$, which is a little bit narrower. The reason is that, for smaller scales $s$, there are very large residuals which result in very small negative-order moments such that $F_q(s)\to 0$. In this sense, MF-DMA performs better than MF-DFA.

    \item[(3)] Generate surrogates. To perform statistical tests, we generate 1000 IAAFT surrogates for each index time series. We also find that, in several situations, about 100 surrogates can already provide decisive results.
\end{enumerate}

\subsection{Scaling plots}

\begin{figure}[!ht]
    \centering
    \includegraphics[width=0.325\linewidth]{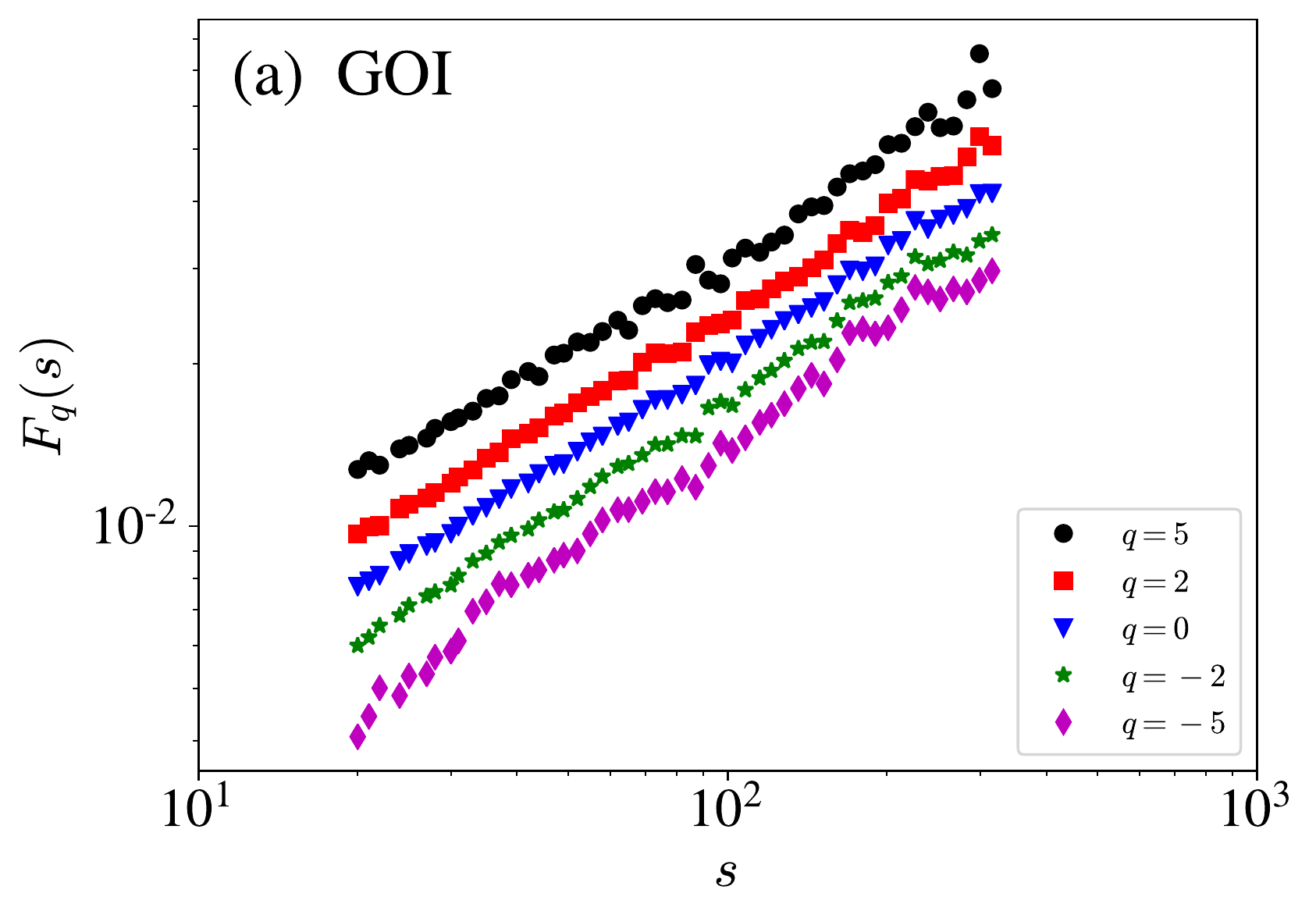}
    \includegraphics[width=0.325\linewidth]{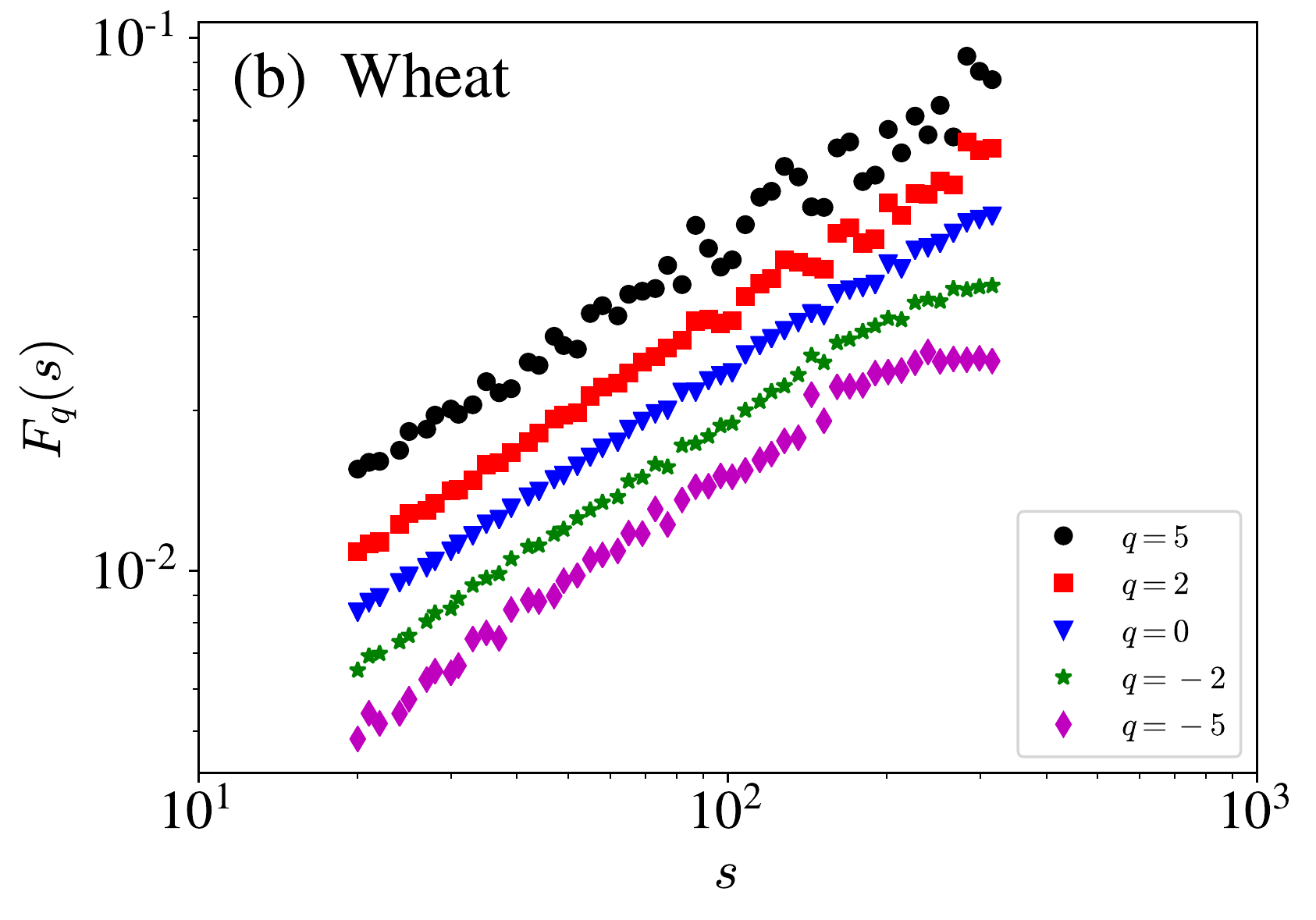}
    \includegraphics[width=0.325\linewidth]{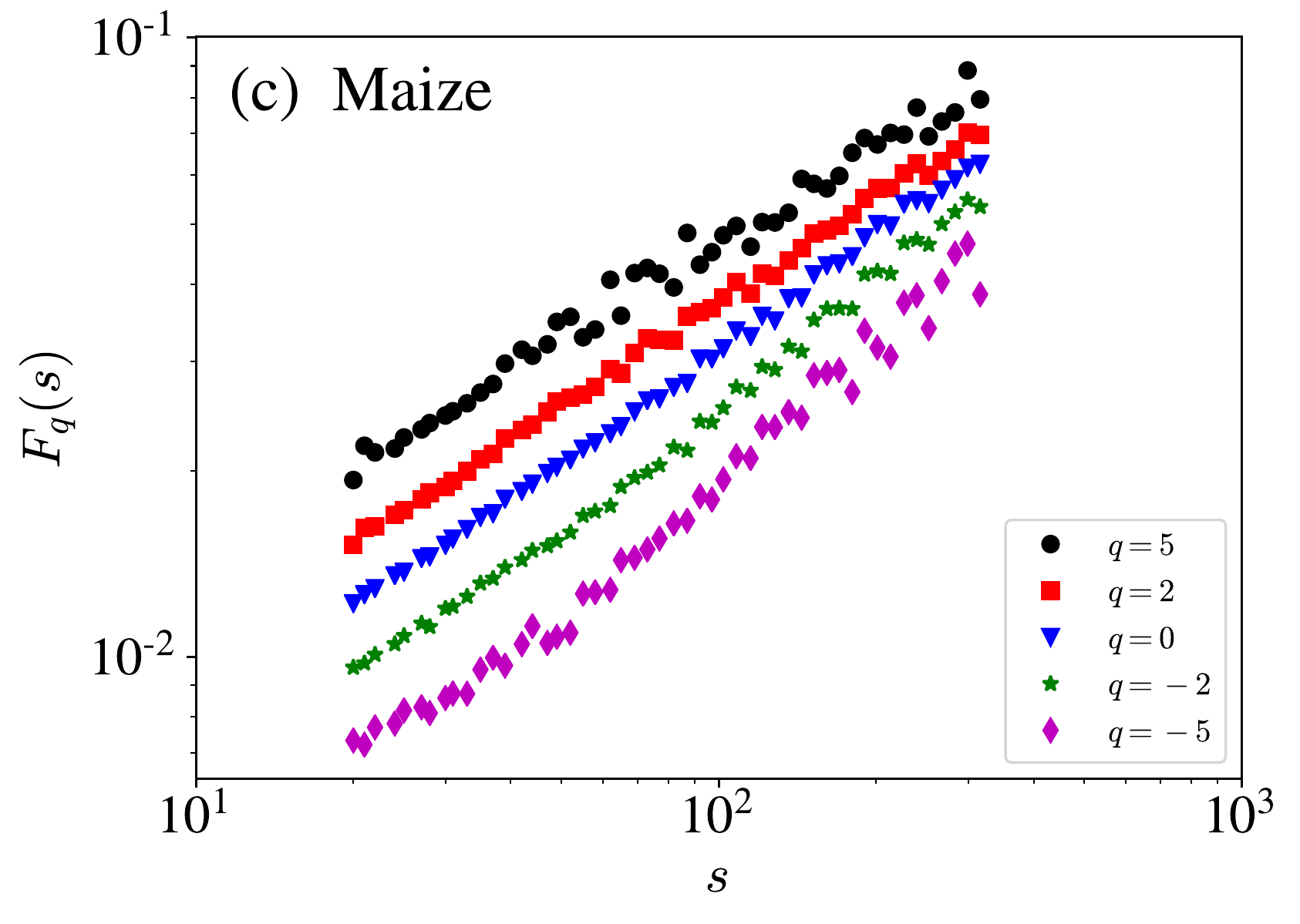}
    \includegraphics[width=0.325\linewidth]{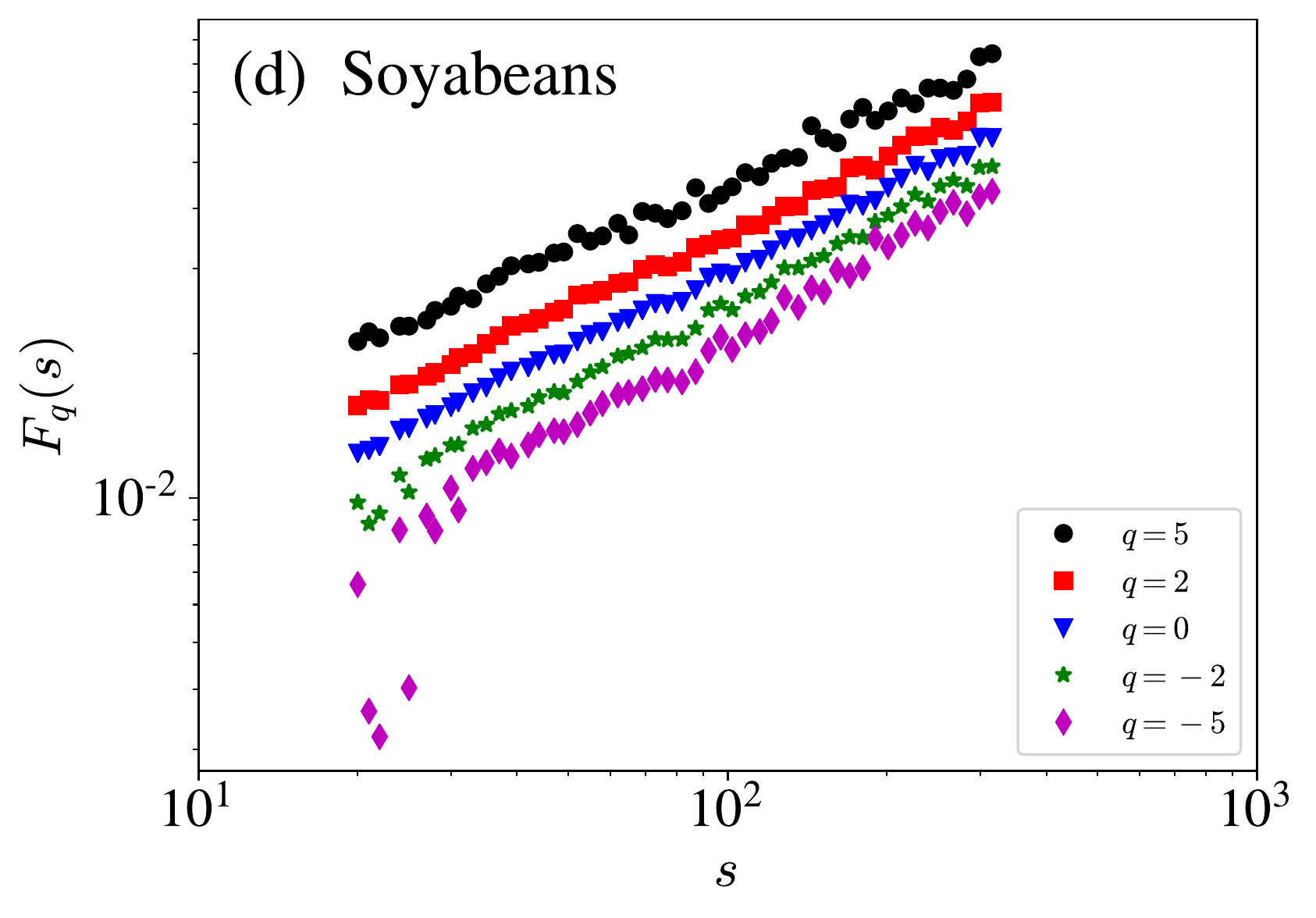}
    \includegraphics[width=0.325\linewidth]{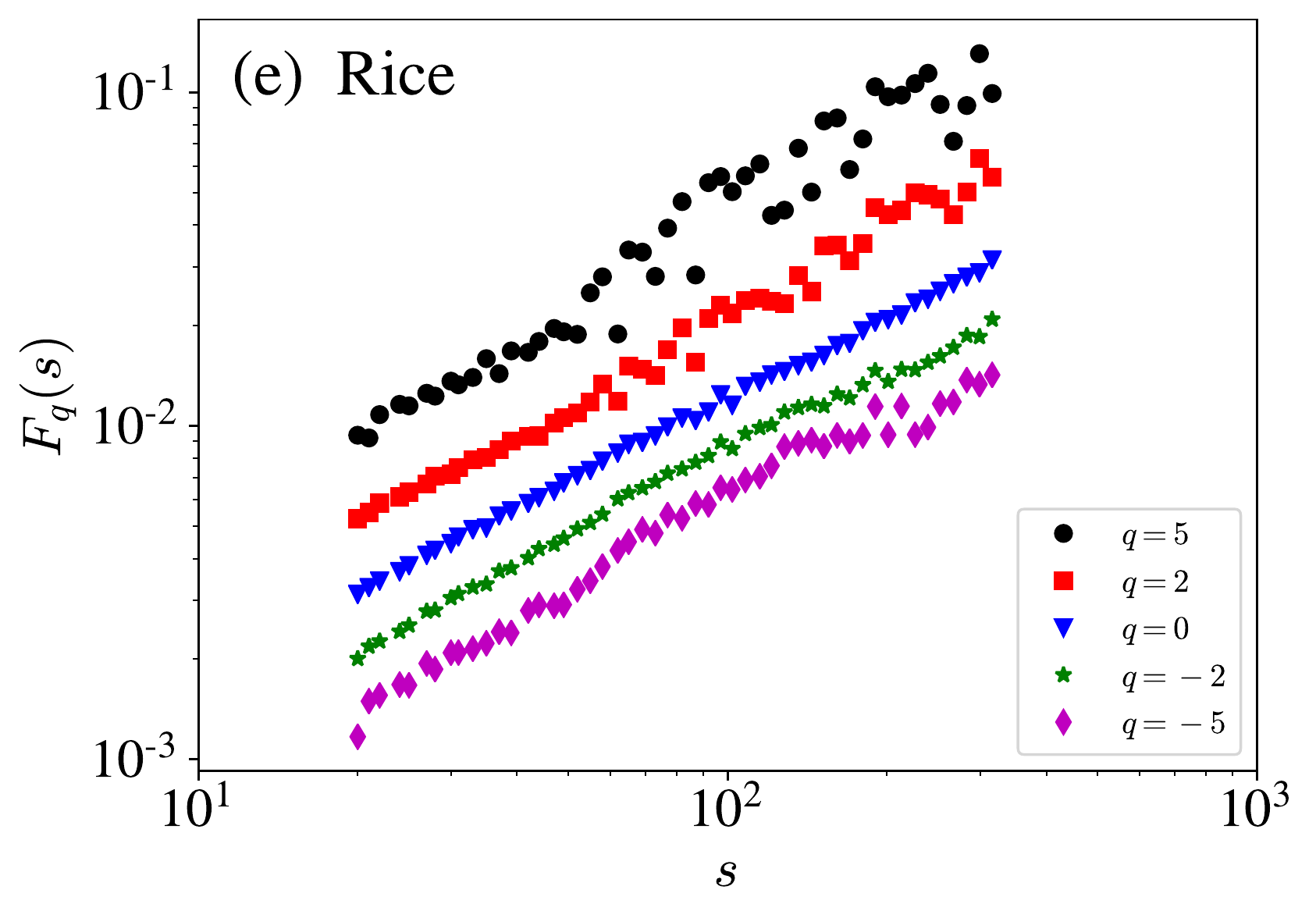}
    \includegraphics[width=0.325\linewidth]{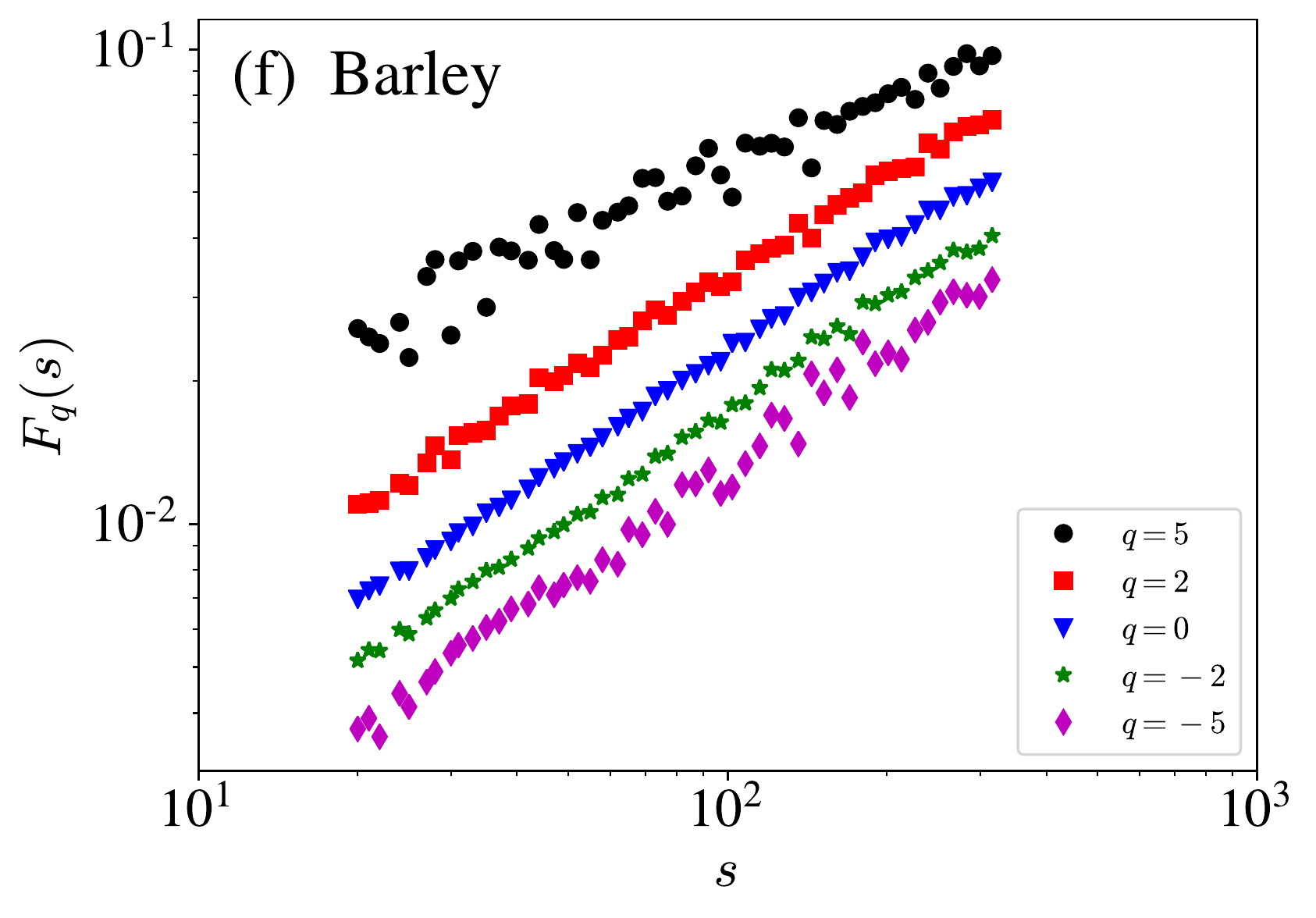}
    \caption{Scaling plots of the MF-DFA fluctuation function $F_q(s)$ with respect to the scale $s$ for the GOI index (a), the wheat sub-index (b), the maize sub-index (c), the soyabeans sub-index (d), the rice sub-index (e), and the barley sub-index (f). The polynomial used to detrend the indices is a linear function with $\ell=1$.}
    \label{Fig:GOI:Return:MFDFA:Fs}
\end{figure}

Fig.~\ref{Fig:GOI:Return:MFDFA:Fs} illustrates the scaling plots of the MF-DFA fluctuation functions $F_q(s)$ with respect to the scale $s$ for $q=- 5$, $-2$, 0, 2, and $5$. It can be observed that all the lines show evident power laws, except for soyabeans when $q$ is negative, where the scaling range should be narrower. Nevertheless, for comparability, we use the same scaling range for all $q$ values and all indices. In addition, we find that the scaling behavior is better for positive orders ($q>0$) than for negative orders ($q<0$). There may also be log-periodic oscillations around the power law \cite{Sornette-1998-PhysRep}, suggesting the possible presence of multiplicative cascades among different scales with a characteristic discrete scale ratio.

The scaling plots look similar when we detrend the indices with quadratic polynomials ($\ell=2$), as shown in Fig.~\ref{FigA:GOI:Return:MFDFA:Fs}. For instance, for negative $q$ the two $F_q(s)$ functions for $\ell=1$ and $\ell=2$ deviate remarkably when $s$ is small, and the deviation becomes larger when $|q|$ is larger.

\subsection{Generalized Hurst indexes}

\begin{figure}[!ht]
    \centering
    \includegraphics[width=0.325\linewidth]{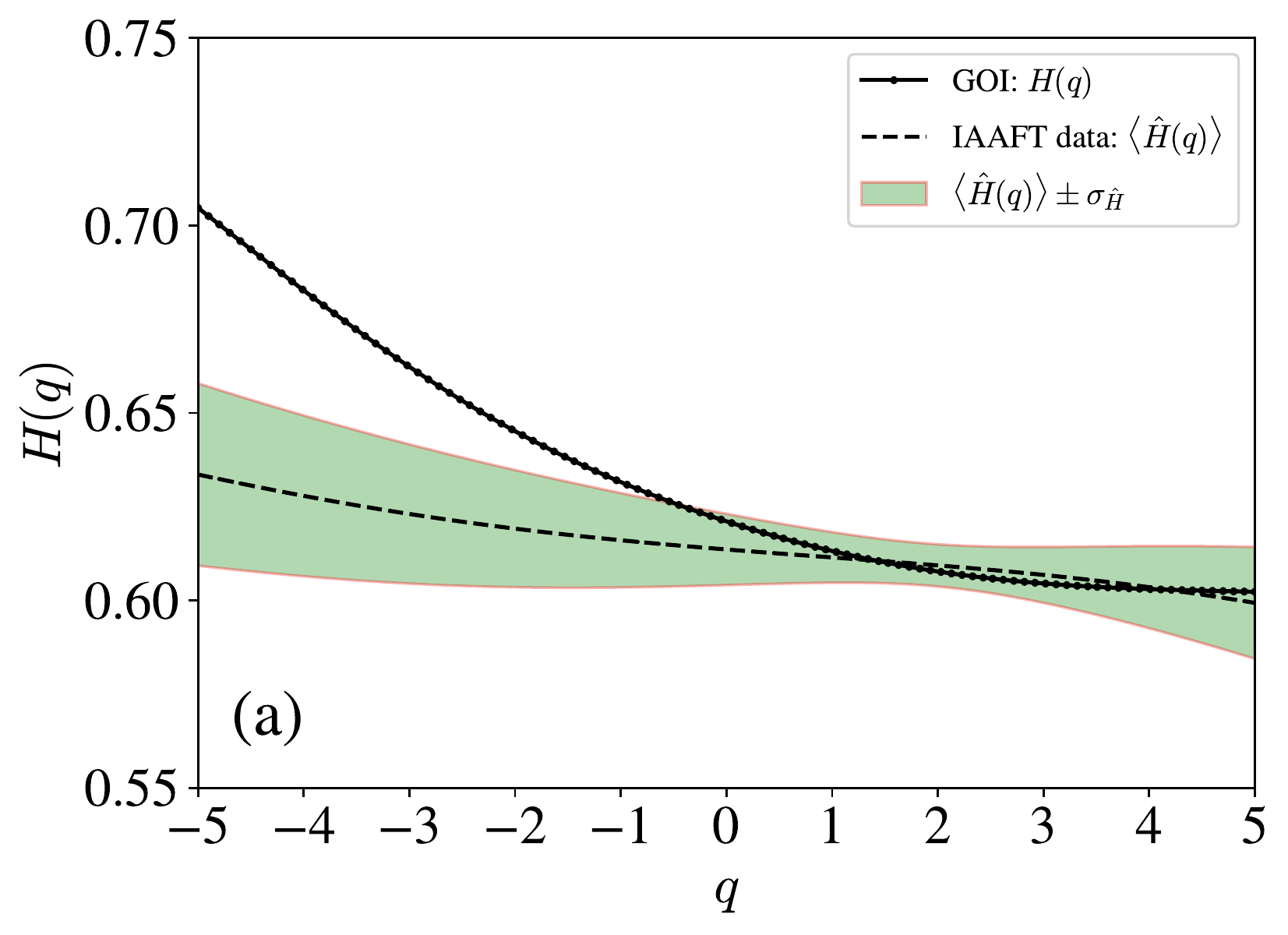}
    \includegraphics[width=0.325\linewidth]{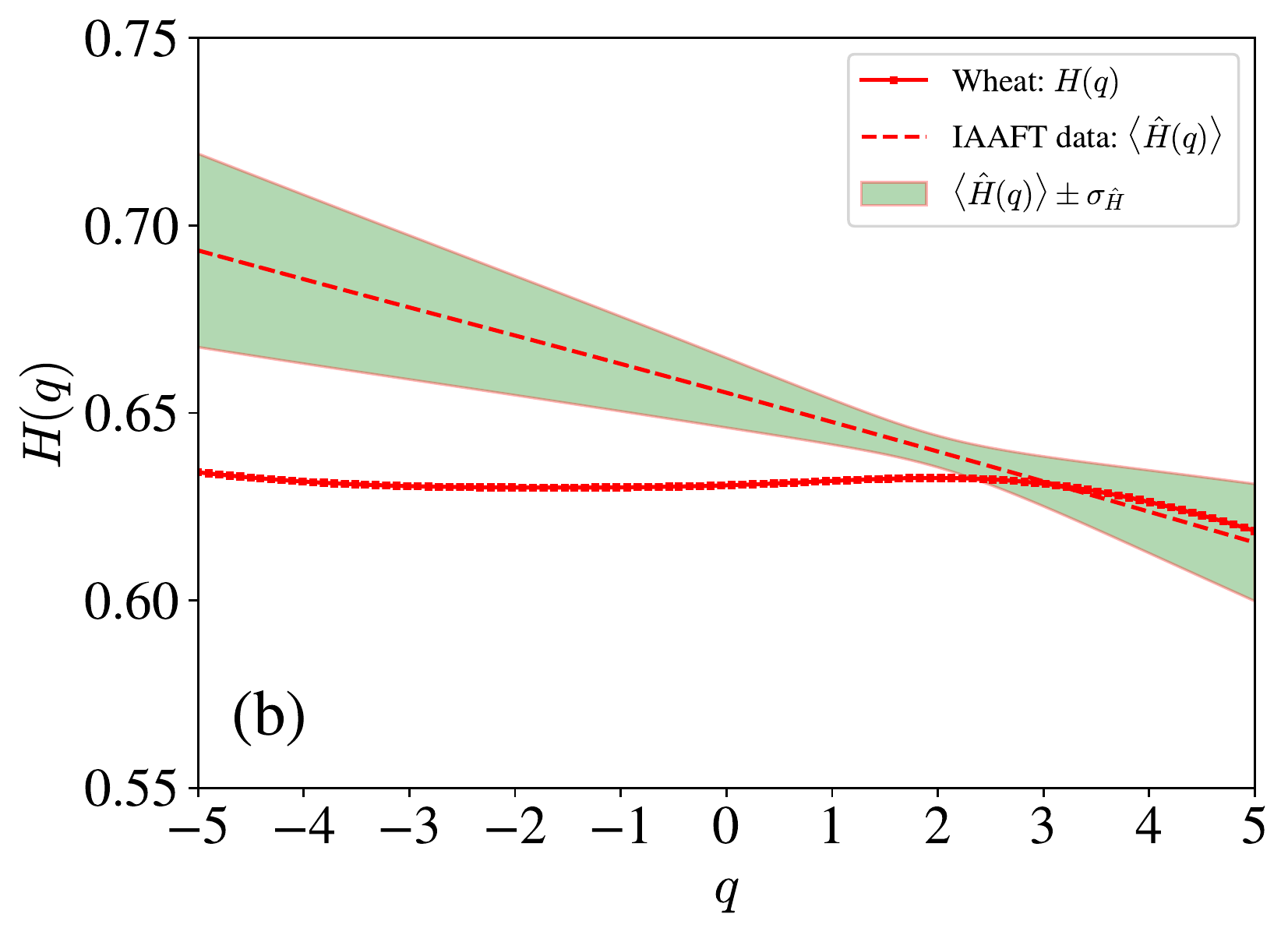}
    \includegraphics[width=0.325\linewidth]{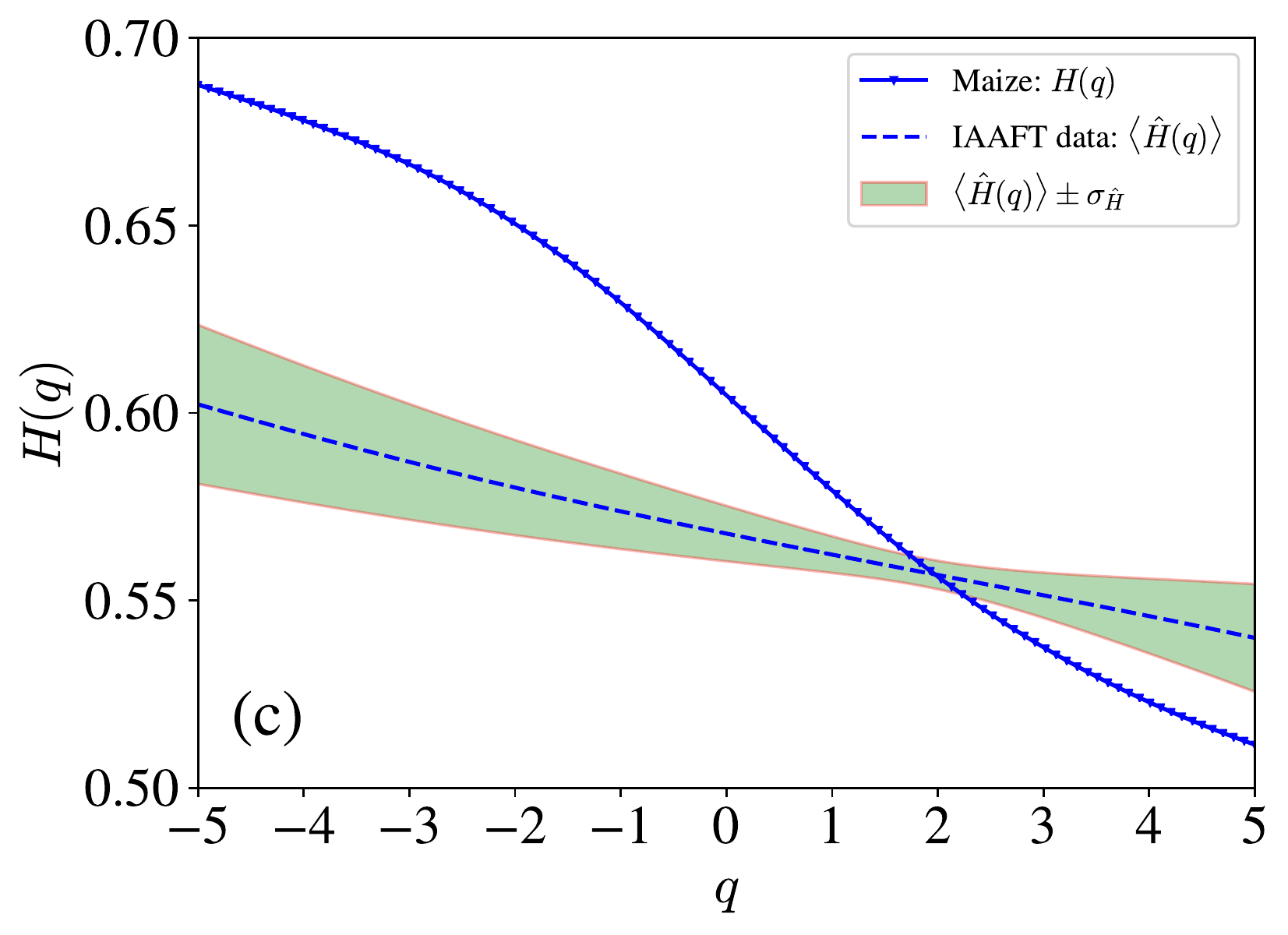}
    \includegraphics[width=0.325\linewidth]{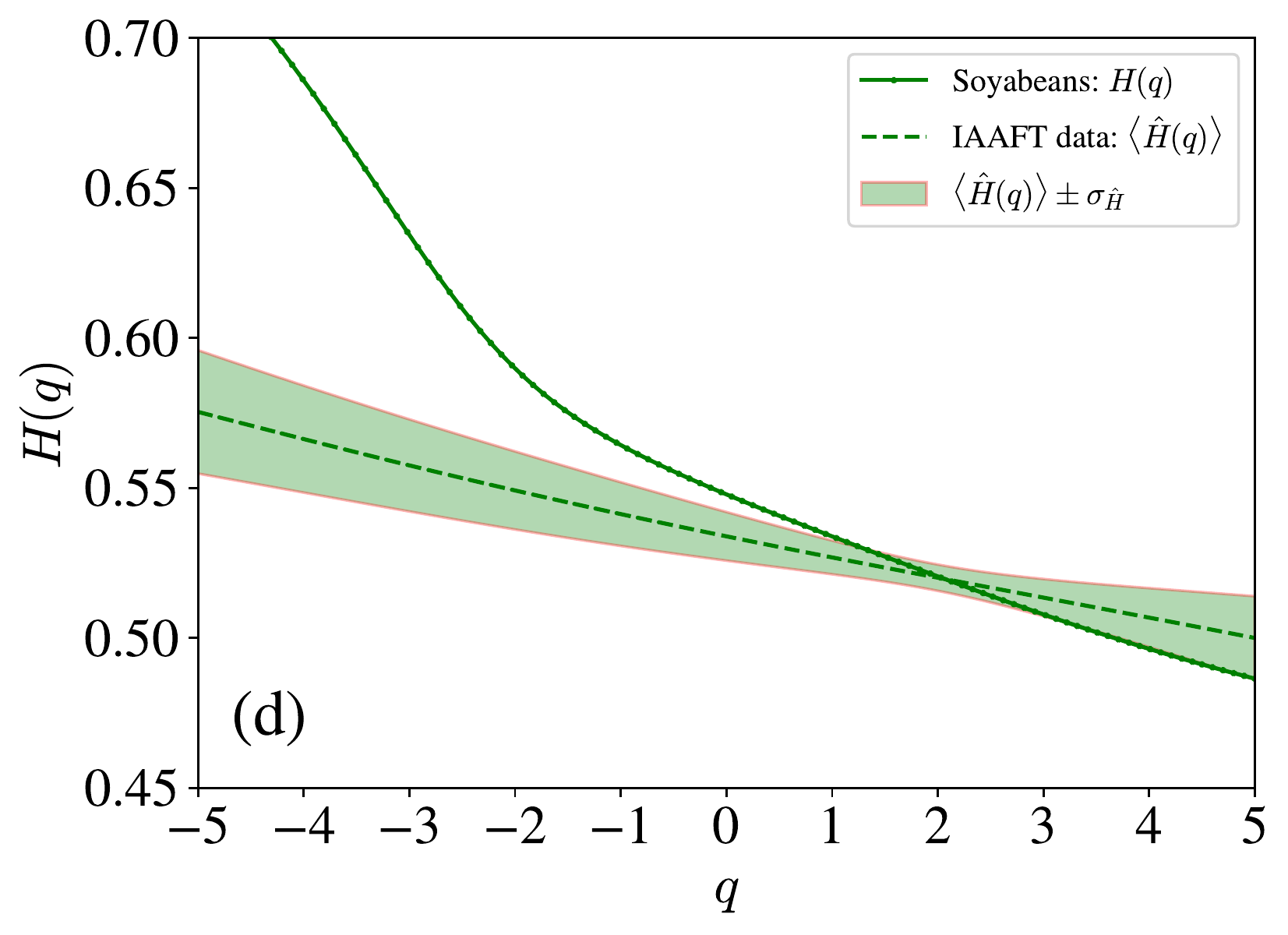}
    \includegraphics[width=0.325\linewidth]{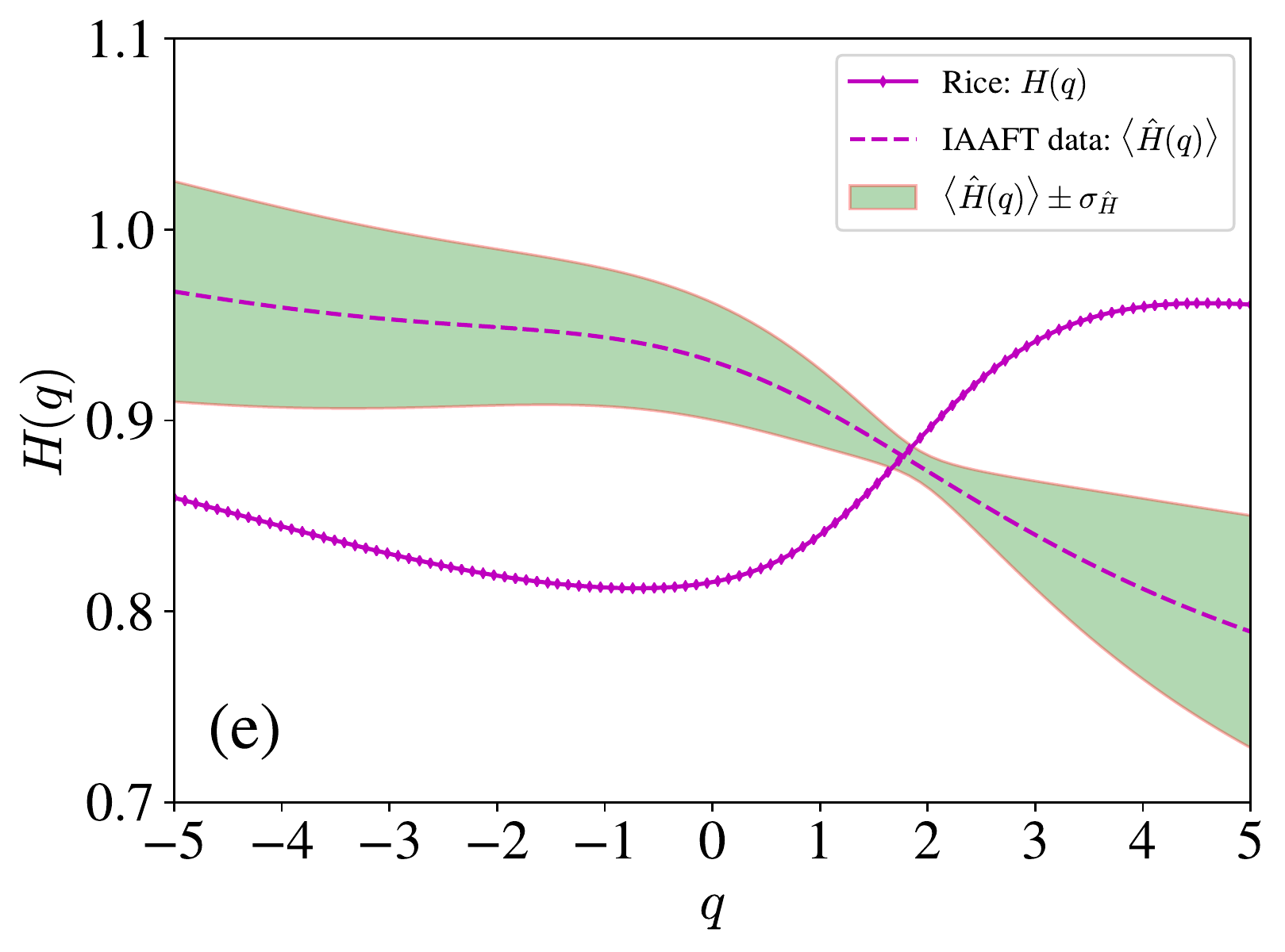}
    \includegraphics[width=0.325\linewidth]{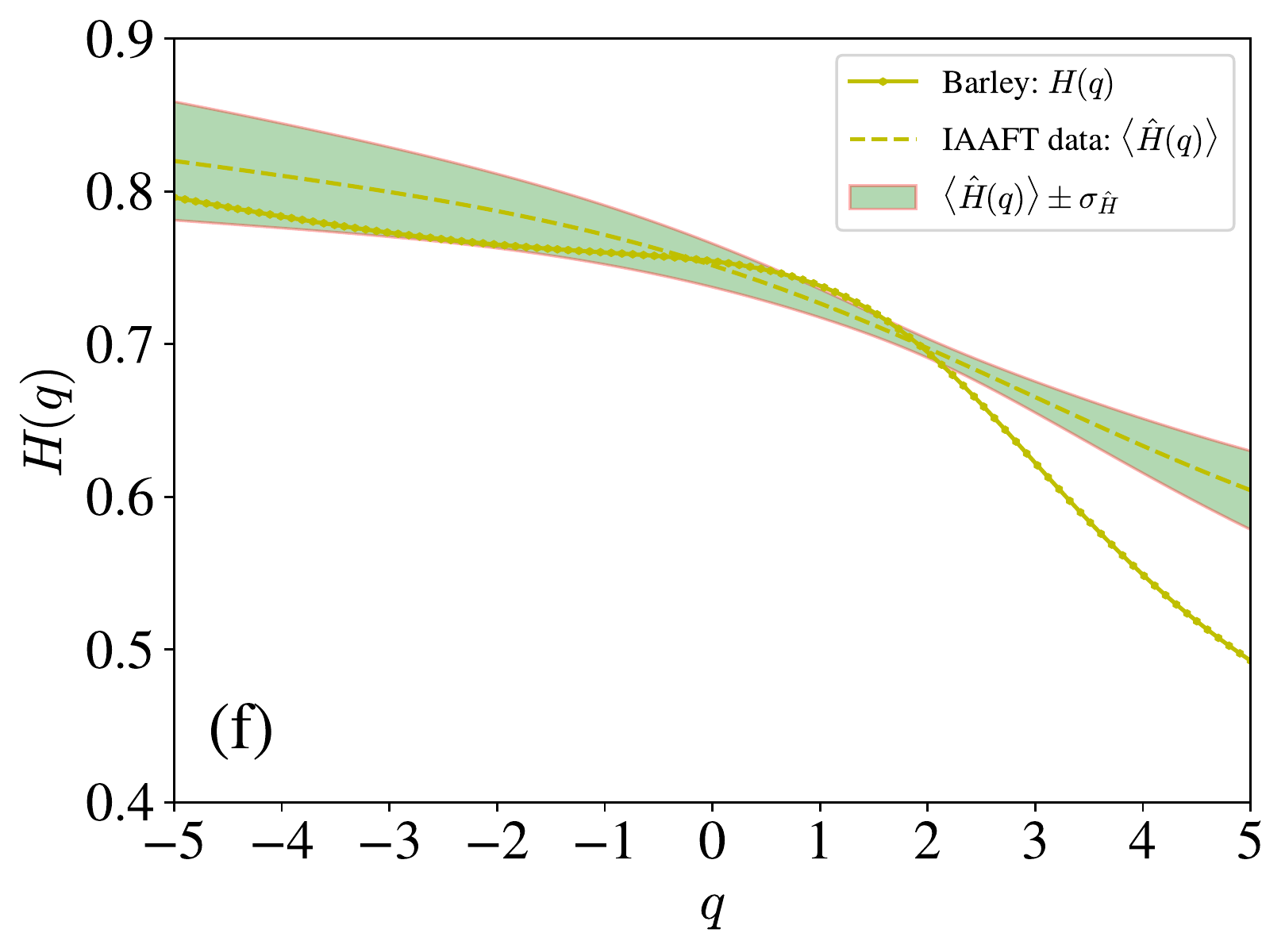}
    \caption{Generalized Hurst indexes $H(q)$ with respect to the order $q$ for the GOI index (a), the wheat sub-index (b), the maize sub-index (c), the soyabeans sub-index (d), the rice sub-index (e), and the barley sub-index (f). The polynomial used to detrend the indices is a linear function with $\ell=1$. For each GOI and sub-indices, we generate 1000 surrogate time series using the IAAFT algorithm and calculate the mean $\langle\hat{H}\rangle$ and standard deviation $\sigma_{\hat{H}}$.}
    \label{Fig:GOI:Return:MFDFA:Hq}
\end{figure}

We perform linear regression of $\ln F_q(s)$ against $\ln s$ for different values of $q$ to estimate the generalized Hurst indexes $H(q)$ of the six global grain spot price indices.  For each GOI and sub-indices, we generate 1000 IAAFT surrogate time series and calculate the  $\hat{H}(q)$ function for each surrogate to obtain the mean $\langle\hat{H}(q)\rangle$ and standard deviation $\sigma_{\hat{H}}$. The results for $\ell=1$ are illustrated in Fig.~\ref{Fig:GOI:Return:MFDFA:Hq}. The results for $\ell=2$ are similar, which are shown in Fig.~\ref{FigA:GOI:Return:MFDFA:Hq}.
For GOI in Fig.~\ref{Fig:GOI:Return:MFDFA:Hq}(a) and Soyabeans in Fig.~\ref{Fig:GOI:Return:MFDFA:Hq}(d), the $q>0$ part of $H(q)$ falls within the confidence interval such that $\langle\hat{H}(q)\rangle-\sigma_{\hat{H}}<H(q)<\langle\hat{H}(q)\rangle+\sigma_{\hat{H}}$, while the $q<0$ part of $H(q)$ is above the confidence interval such that $H(q)>\langle\hat{H}(q)\rangle+\sigma_{\hat{H}}$.
For wheat in Fig.~\ref{Fig:GOI:Return:MFDFA:Hq}(b), the $q>2$ part of $H(q)$ falls within the confidence interval such that $\langle\hat{H}(q)\rangle-\sigma_{\hat{H}}<H(q)<\langle\hat{H}(q)\rangle+\sigma_{\hat{H}}$, while the $q<2$ part of $H(q)$ is below the confidence interval such that $H(q)<\langle\hat{H}(q)\rangle-\sigma_{\hat{H}}$. In contrast, when $\ell=2$, $\langle\hat{H}(q)\rangle-\sigma_{\hat{H}}<H(q)<\langle\hat{H}(q)\rangle+\sigma_{\hat{H}}$ for all $q$ values.
For maize in Fig.~\ref{Fig:GOI:Return:MFDFA:Hq}(c), we find that $H(q)<\langle\hat{H}(q)\rangle-\sigma_{\hat{H}}$ when $q>2$ and $H(q)>\langle\hat{H}(q)\rangle+\sigma_{\hat{H}}$ when $q<2$.
For rice in Fig.~\ref{Fig:GOI:Return:MFDFA:Hq}(e), we find that $H(q)<\langle\hat{H}(q)\rangle-\sigma_{\hat{H}}$ when $q<2$ and $H(q)>\langle\hat{H}(q)\rangle+\sigma_{\hat{H}}$ when $q>2$.
For barley in Fig.~\ref{Fig:GOI:Return:MFDFA:Hq}(f), we find that $\langle\hat{H}(q)\rangle-\sigma_{\hat{H}}<H(q)<\langle\hat{H}(q)\rangle+\sigma_{\hat{H}}$ when $q<2$ and $H(q)<\langle\hat{H}(q)\rangle+\sigma_{\hat{H}}$ when $q>2$.
All the $H(q)$ curves deviate from the confidence intervals (shadow areas) $\langle\hat{H}(q)\rangle\pm\sigma_{\hat{H}}$, indicating that the complex behavior of the global grain spot price indices cannot be attributed only to the fat-tailed distributions and the linear correlations. However, the $H(q)$ curves of wheat and rice are not monotonically decreasing functions, showing that there is no multifractal nature or the multifractal nature has been ``contaminated''.

One common feature of the six plots in Fig.~\ref{Fig:GOI:Return:MFDFA:Hq} is that all the Hurst indexes $H(2)$ fall well within the confidence intervals:
\begin{equation}
    \langle\hat{H}(q)\rangle-\sigma_{\hat{H}}<H(2)<\langle\hat{H}(q)\rangle+\sigma_{\hat{H}}
\end{equation}
and $\sigma_{\hat{H}(2)}$ is the smallest among all $\sigma_{\hat{H}(q)}$, which
confirms the validation of the IAAFT method in generating required surrogates and the MF-DFA approach in determining the Hurst indexes since the surrogates has the power spectra as the original time series. Indeed, Fig.~\ref{Fig:GOI:Return:MFDFA:Fs} shows that the $\log{F_2(s)}\sim\log{s}$ curves are nice straight lines. In Table~\ref{Tab:MFDFA:H2}, we present the Hurst indexes of the six time series obtained using MF-DMA, MF-DFA ($\ell=1$) and MF-DFA ($\ell=2$). For each time series, the Hurst indexes estimated from three methods are close to each other. Specifically, only the Hurst indexes for maize and soyabeans are close to 0.5, while the Hurst indexes for GOI, wheat, and barley are greater than 0.6 and that for rice is greater than 0.7. It shows that the global rice spot price index has strong linear long-term correlations, and the global spot price indices of GOI, wheat, and barley also contain linear long-term correlations.

\begin{table}[!h]
    \centering
    \caption{Hurst indexes $H(2)$ of the six global grain spot price indices estimated using MF-DMA, MF-DFA ($\ell=1$) and MF-DFA ($\ell=2$).}
    \smallskip
    \renewcommand\tabcolsep{15pt}
    \begin{tabular}{ccccccccccccccccc}
    \toprule
        Method  & GOI & Wheat & Maize & Soyabeans & Rice & Barley  \\
    \midrule
        MF-DMA            & 0.6280 & 0.6239 & 0.5373 & 0.5355 & 0.7255 & 0.6448  \\
        MF-DFA ($\ell=1$) & 0.6080 & 0.6327 & 0.5579 & 0.5214 & 0.8908 & 0.6989  \\
        MF-DFA ($\ell=2$) & 0.5846 & 0.6437 & 0.5540 & 0.5138 & 0.8271 & 0.7113 \\
    \bottomrule
    \end{tabular}
    \label{Tab:MFDFA:H2}
\end{table}

\subsection{Mass exponents}

\begin{figure}[!ht]
    \centering
    \includegraphics[width=0.325\linewidth]{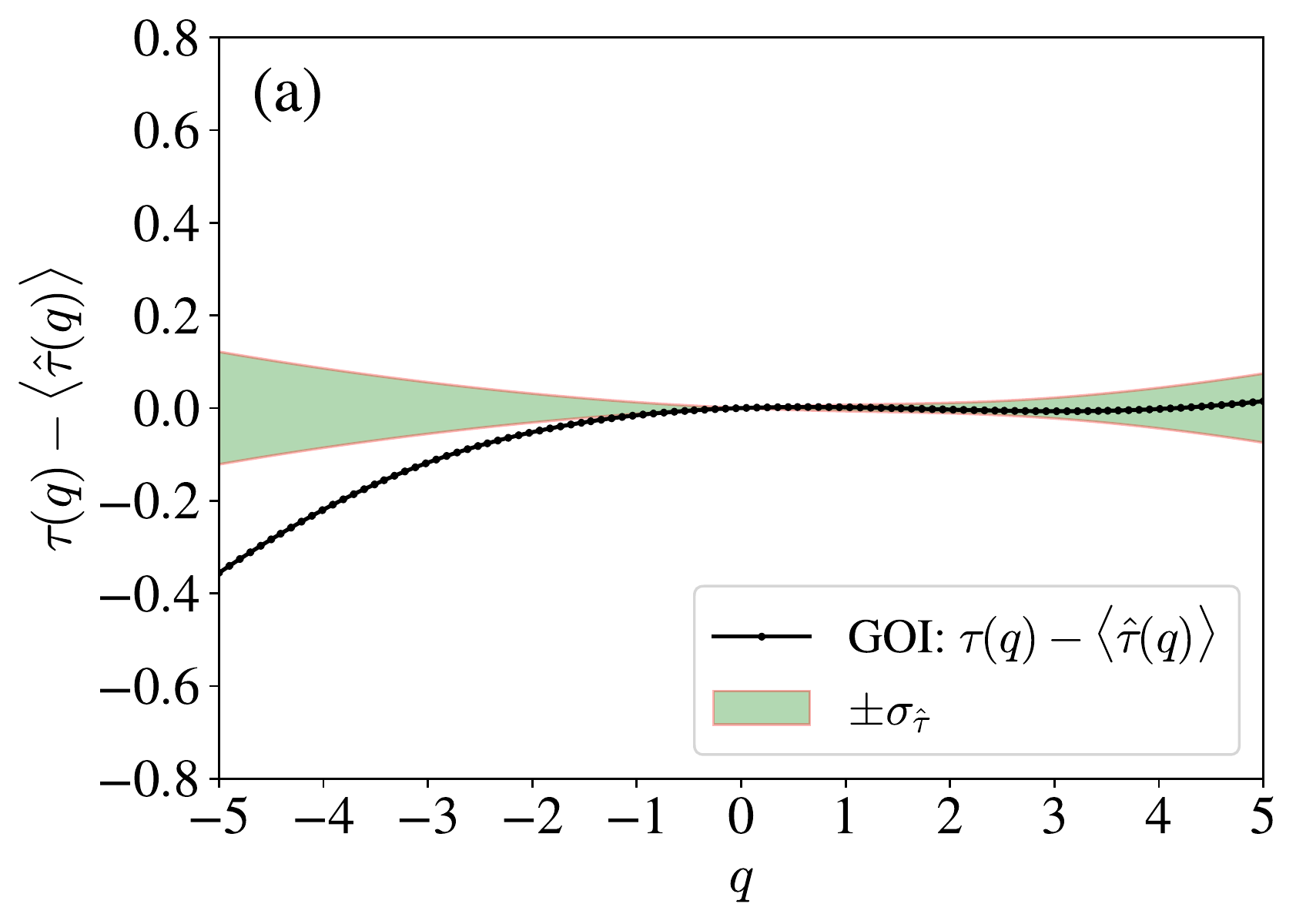}
    \includegraphics[width=0.325\linewidth]{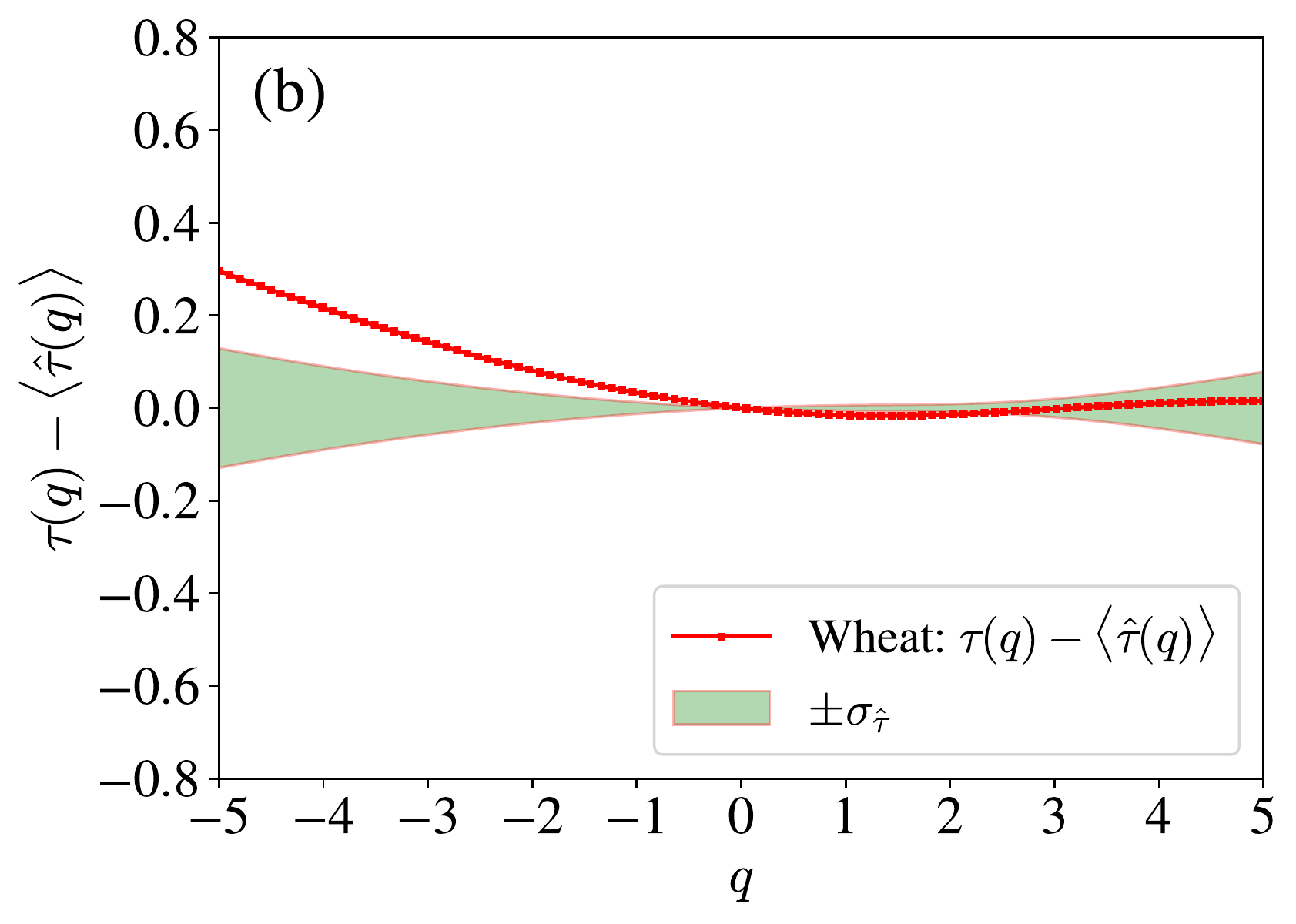}
    \includegraphics[width=0.325\linewidth]{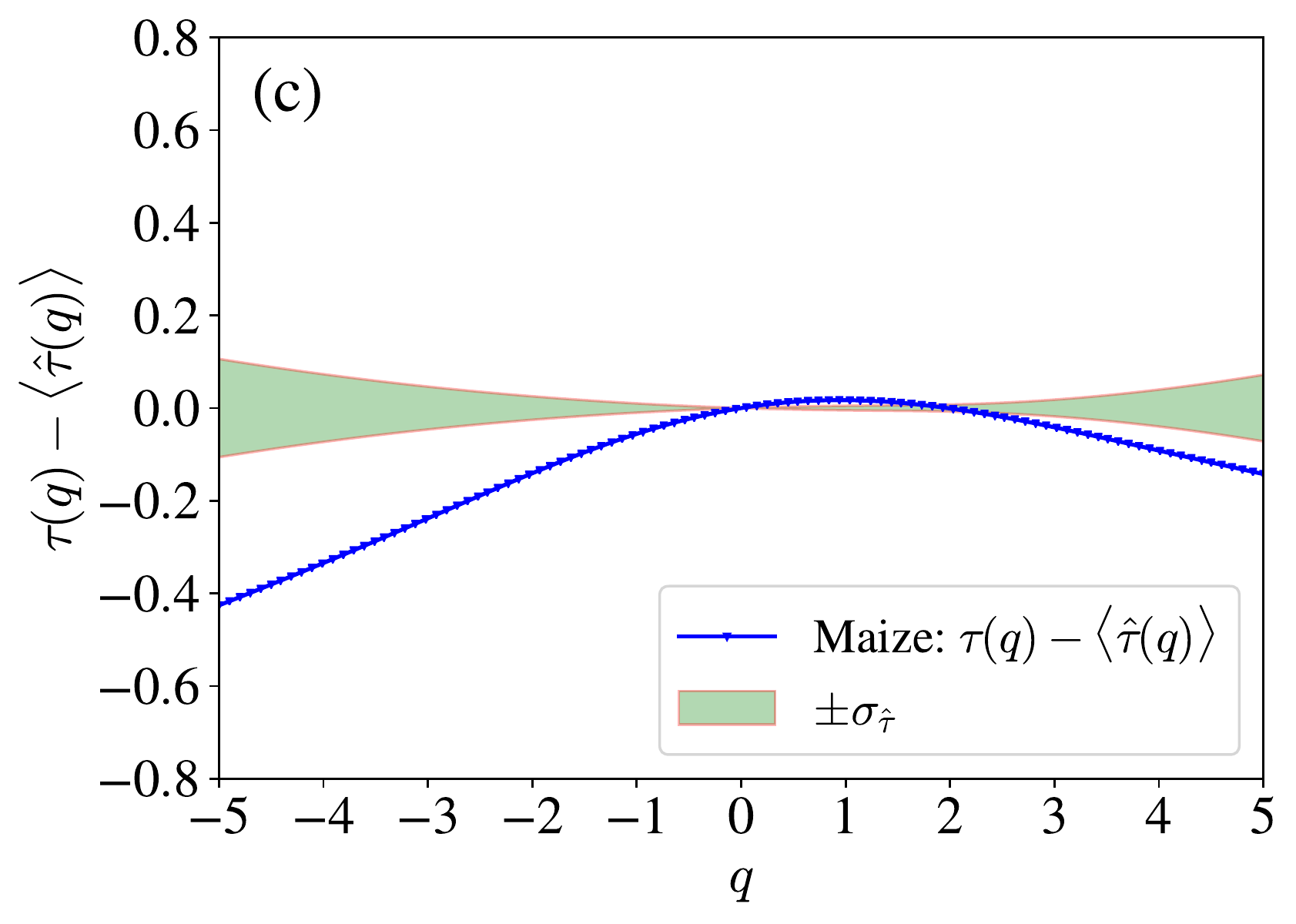}
    \includegraphics[width=0.325\linewidth]{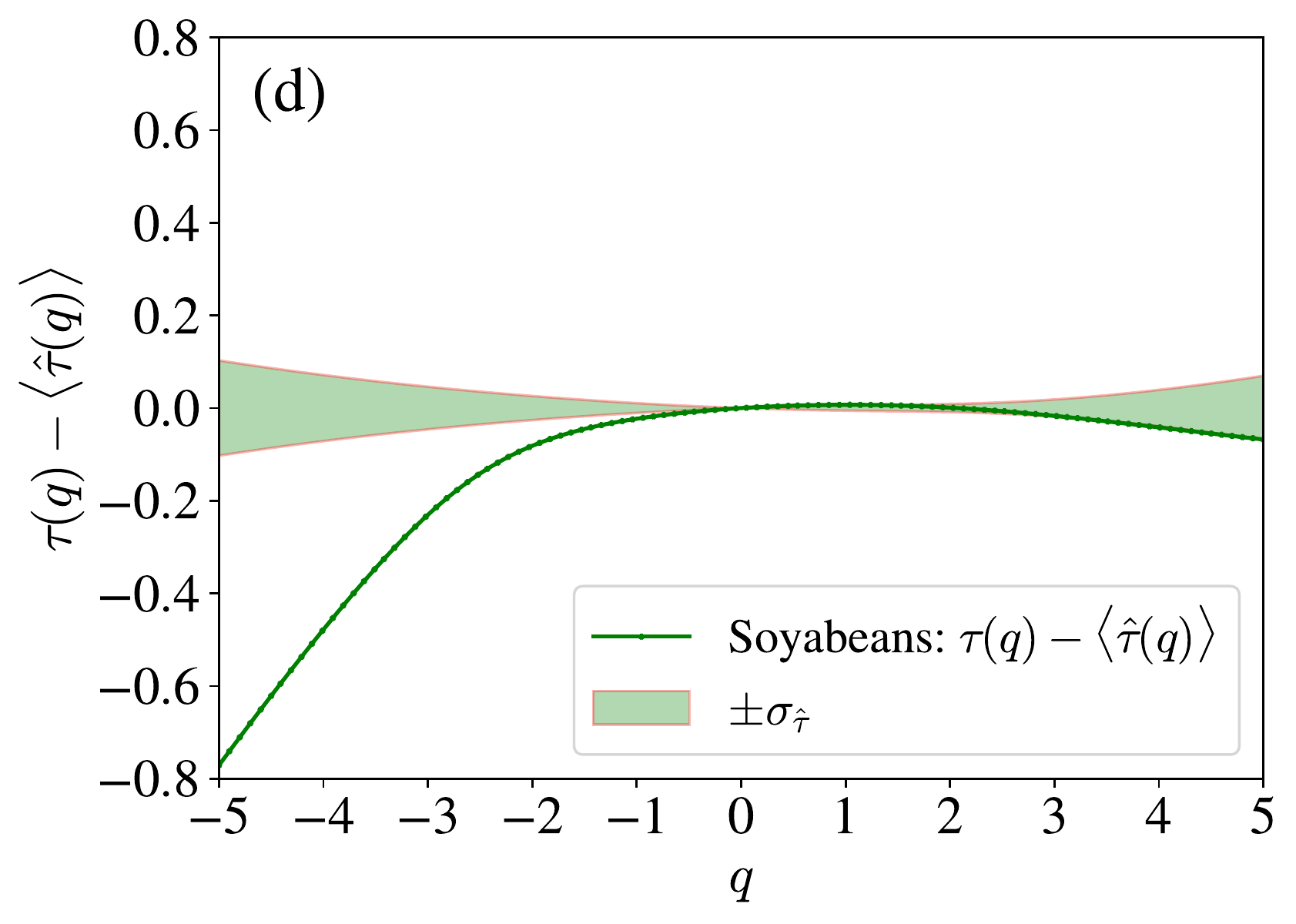}
    \includegraphics[width=0.325\linewidth]{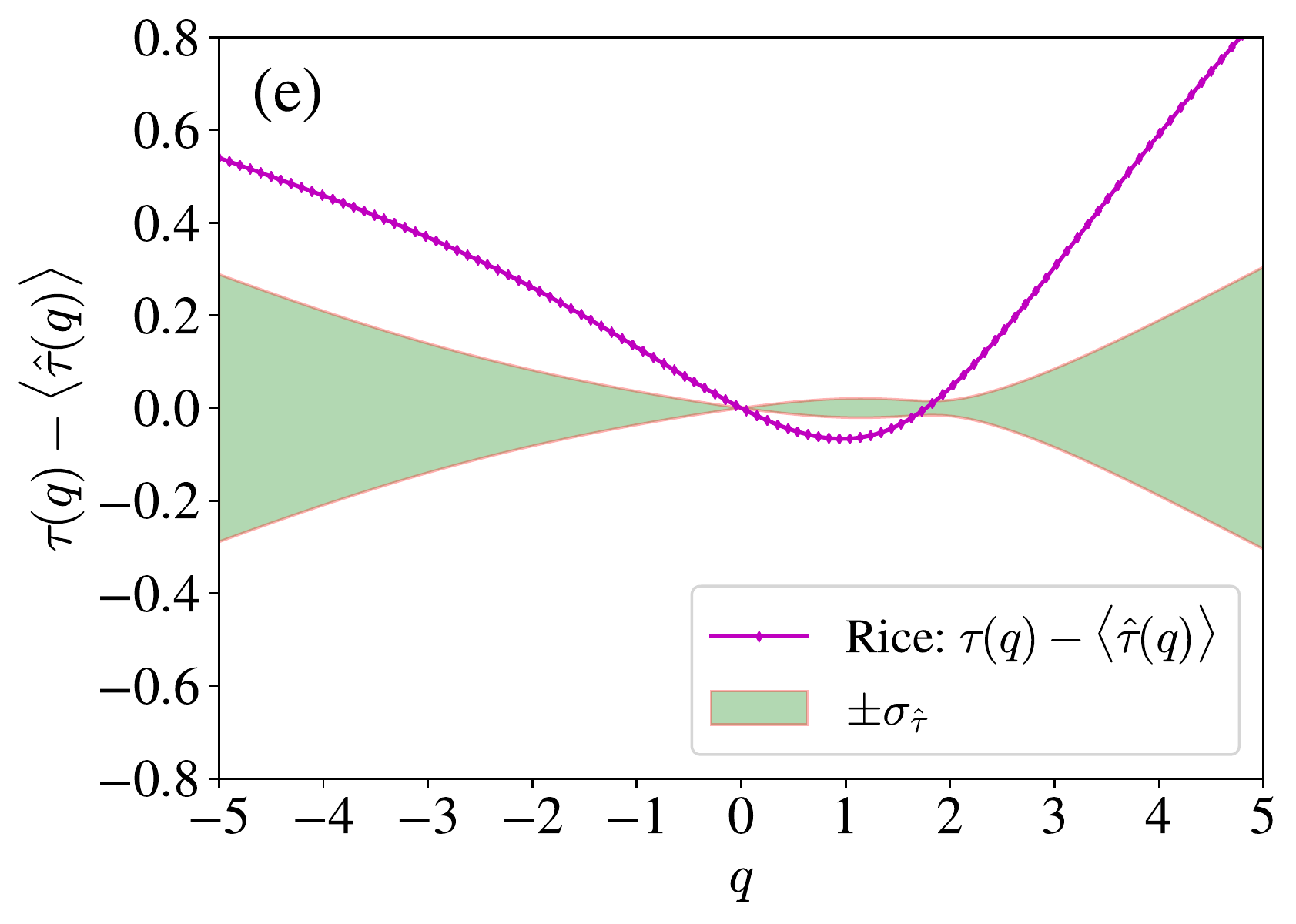}
    \includegraphics[width=0.325\linewidth]{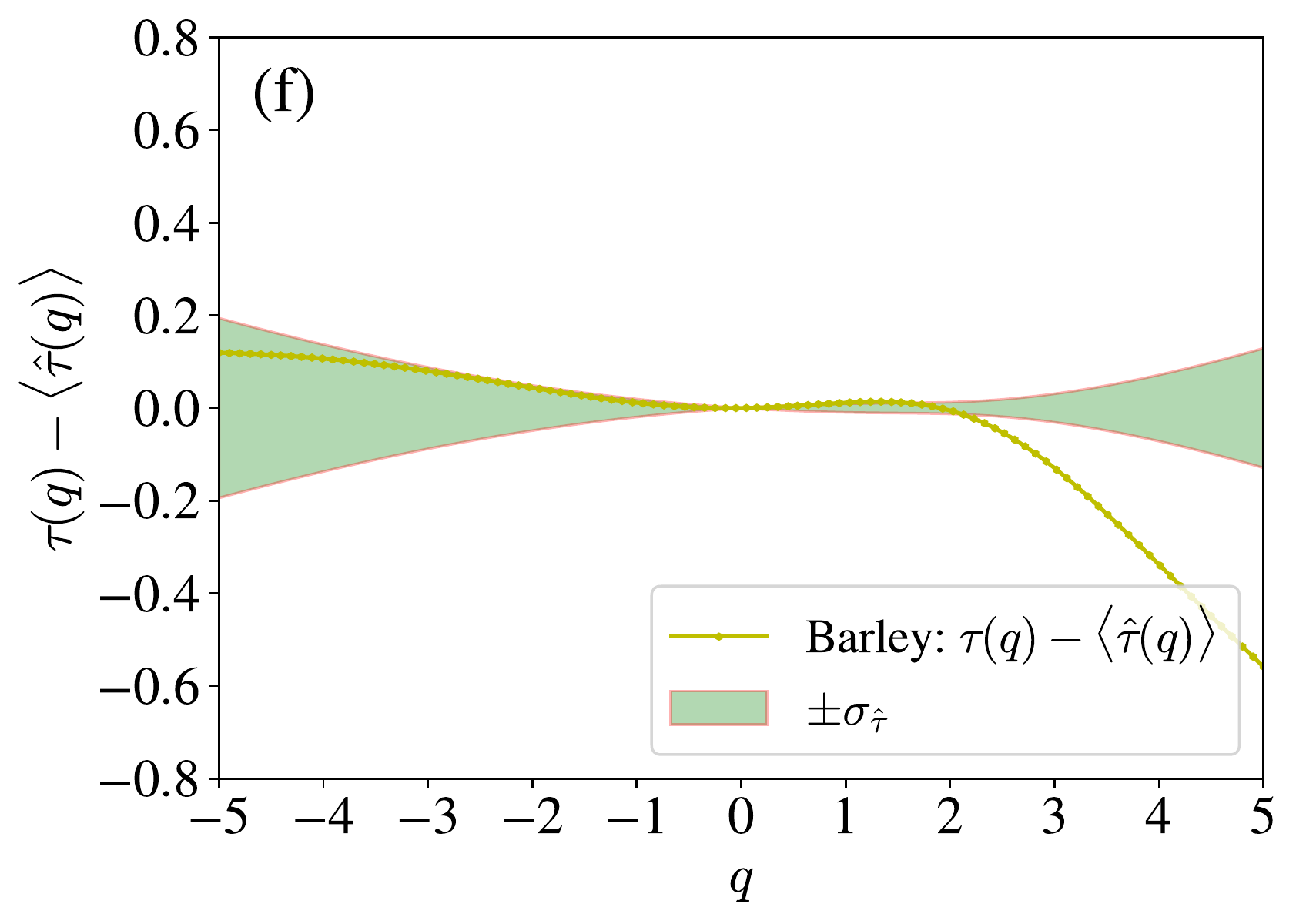}
    \caption{Deviations ($\tau(q)-\langle\hat{\tau}(q)\rangle$) of the mass exponents $\tau(q)$ of the original time series from the average mass exponents $\langle\hat{\tau}(q)\rangle$ of the IAAFT surrogates with respect to the order $q$ for the GOI index (a), the wheat sub-index (b), the maize sub-index (c), t)he soyabeans sub-index (d), the rice sub-index (e), and the barley sub-index (f). The polynomial used to detrend the indices is a linear function with $\ell=1$. For each GOI and sub-indices, we generate 1000 surrogate time series using the IAAFT algorithm and calculate the mean $\langle\hat{\tau}\rangle$ and standard deviation $\sigma_{\hat{\tau}}$.}
    \label{Fig:GOI:Return:MFDFA:dtau}
\end{figure}

According to Eq.~(\ref{Eq:MF:Hq:2:tau}), the mass exponents $\tau(q)$ can be calculated numerically from the generalized Hurst indexes $H(q)$. We calculate the deviation of the mass exponents $\tau(q)$ of the original time series from the average mass exponents $\langle\hat{\tau}(q)\rangle$ of the IAAFT surrogates, denoted as $\tau(q)-\langle\hat{\tau}(q)\rangle$. Fig.~\ref{Fig:GOI:Return:MFDFA:dtau} shows $\tau(q)-\langle\hat{\tau}(q)\rangle$ as a function of the order $q$ for the six indices. We find that all the $\tau(q)$ curves deviate from the confidence level.
For GOI in Fig.~\ref{Fig:GOI:Return:MFDFA:dtau}(a) and soyabeans in Fig.~\ref{Fig:GOI:Return:MFDFA:dtau}(d), $\tau(q) < \langle\hat{\tau}(q)\rangle-\sigma_{\hat{\tau}}$ when $q<-1$ and  $\langle\hat{\tau}(q)\rangle-\sigma_{\hat{\tau}} <\tau(q)< \langle\hat{\tau}(q)\rangle+\sigma_{\hat{\tau}}$ when $q>-1$.
For wheat in Fig.~\ref{Fig:GOI:Return:MFDFA:dtau}(b), $\tau(q) > \langle\hat{\tau}(q)\rangle+\sigma_{\hat{\tau}}$ when $q<0$ and  $\langle\hat{\tau}(q)\rangle-\sigma_{\hat{\tau}} <\tau(q)< \langle\hat{\tau}(q)\rangle+\sigma_{\hat{\tau}}$ when $q\geq 0$. However, when $\ell=2$, Fig.~\ref{FigA:GOI:Return:MFDFA:dtau}(d$^{\prime}$) shows that $\tau(q)$ falls within the confidence interval for all $q$ values.
For maize in Fig.~\ref{Fig:GOI:Return:MFDFA:dtau}(c), $\tau(q) < \langle\hat{\tau}(q)\rangle-\sigma_{\hat{\tau}}$ when $q<0$ and $q>2$.
For rice in Fig.~\ref{Fig:GOI:Return:MFDFA:dtau}(e), $\tau(q) > \langle\hat{\tau}(q)\rangle+\sigma_{\hat{\tau}}$ when $q<0$ and $q>2$ and $\tau(q) < \langle\hat{\tau}(q)\rangle-\sigma_{\hat{\tau}}$ when $0\leq{q}\leq2$.
For barley in Fig.~\ref{Fig:GOI:Return:MFDFA:dtau}(f), $\langle\hat{\tau}(q)\rangle-\sigma_{\hat{\tau}} <\tau(q)< \langle\hat{\tau}(q)\rangle+\sigma_{\hat{\tau}}$ when $q\leq 2$ and $\tau(q) < \langle\hat{\tau}(q)\rangle-\sigma_{\hat{\tau}}$ when $q>2$.
In all the six plots, we find that $\tau(0) = \langle\hat{\tau}(0)\rangle$ and $\tau(2) \approx \langle\hat{\tau}(2)\rangle$, and $\sigma_{\hat{\tau}(0)}=0$ and $\sigma_{\hat{\tau}(2)}\approx 0$.

\begin{table}[!h]
    \centering
    \caption{Testing the nonlinearities of the mass exponent function $\tau(q)$ of the six indices. The mass exponent function is assumed to have a quadratic form $\tau(q) = a_0 + a_1q + a_2q^2$. We test if $a_2=0$ or not. The upper panel is for $\ell=1$, while the lower panel is for $\ell=2$.}
    \smallskip
    \renewcommand\tabcolsep{7pt}
    \begin{tabular}{ccccccccccccccccc}
    \toprule
               & \multicolumn{3}{c}{Full model} &&  \multicolumn{3}{c}{Linear term} &&  \multicolumn{3}{c}{Quadratic term}  \\
               \cline{2-4} \cline{6-8} \cline{10-12}
        Index  & $F$-stat &  $p$-value &  $R^2$ &&  $a_1$ & $t$-stat & $p$-value  &&  $a_2$ & $t$-stat & $p$-value  \\
    \midrule
         GOI & 270026 & 0.0000 & 0.9998 &  &  0.6415 & 734 & 0.0000 &  &  -0.0102 & -30 & 0.0000   \\
         Wheat & 3347494 & 0.0000 & 1.0000 &  &  0.6290 & 2587 & 0.0000 &  &  -0.0013 & -13 & 0.0000   \\
         Maize & 990196 & 0.0000 & 1.0000 &  &  0.6009 & 1402 & 0.0000 &  &  -0.0180 & -109 & 0.0000   \\
         Soyabeans & 56755 & 0.0000 & 0.9991 &  &  0.5869 & 334 & 0.0000 &  &  -0.0247 & -36 & 0.0000   \\
         Rice & 92753 & 0.0000 & 0.9995 &  &  0.8948 & 430 & 0.0000 &  &  0.0118 & 14 & 0.0000   \\
         Barley & 28042 & 0.0000 & 0.9982 &  &  0.6749 & 235 & 0.0000 &  &  -0.0311 & -28 & 0.0000   \\\hline
         GOI & 206749 & 0.0000 & 0.9998 &  &  0.6401 & 641 & 0.0000 &  &  -0.0185 & -48 & 0.0000   \\
         Wheat & 3005762 & 0.0000 & 1.0000 &  &  0.6566 & 2450 & 0.0000 &  &  -0.0095 & -92 & 0.0000   \\
         Maize & 1466824 & 0.0000 & 1.0000 &  &  0.5874 & 1709 & 0.0000 &  &  -0.0144 & -108 & 0.0000   \\
         Soyabeans & 21590 & 0.0000 & 0.9977 &  &  0.6319 & 205 & 0.0000 &  &  -0.0391 & -33 & 0.0000   \\
         Rice & 143795 & 0.0000 & 0.9997 &  &  0.8717 & 536 & 0.0000 &  &  -0.0046 & -7 & 0.0000   \\
         Barley & 39694 & 0.0000 & 0.9988 &  &  0.6933 & 279 & 0.0000 &  &  -0.0355 & -37 & 0.0000   \\
    \bottomrule
    \end{tabular}
    \label{Tab:MFtest:tau:q:q^2}
\end{table}

For multifractal time series, the mass exponent function is nonlinear. To test the nonlinearity of the $\tau(q)$ function, we assume that that  \cite{Jiang-Xie-Zhou-Sornette-2019-RepProgPhys}
\begin{equation}
    \tau(q) = a_0 + a_1q + a_2q^2.
    \label{Eq:MF:tau:q:q^2}
\end{equation}
We test if $a_2\neq 0$ by regressing the the mass exponents $\tau(q)$ to Eq.~(\ref{Eq:MF:tau:q:q^2}). The regression results are shown in Table~\ref{Tab:MFtest:tau:q:q^2}. The upper panel is for $\ell=1$, while the lower panel is for $\ell=2$. For the full model, we have $p$-values close to 0 and the $R^2$ close to 1, indicating that the model fits the data very well. The coefficients $a_2$ of the quadratic term are significantly different from 0 with close-to-zero $p$-values for all the time series. However, we find that $a_2>0$ for rice when $\ell=1$. There are some differences when MF-DMA is adopted \cite{Gao-Shao-Yang-Zhou-2022-ChaosSolitonsFractals}, where $a_2$ for wheat is insignificantly different from 0 with the $p$-value being 0.2694 and $a_2>0$ for GOI and wheat.


\subsection{Singularity spectrum}

\subsubsection{Empirical singularity spectrum}

\begin{figure}[!ht]
    \centering
    \includegraphics[width=0.325\linewidth]{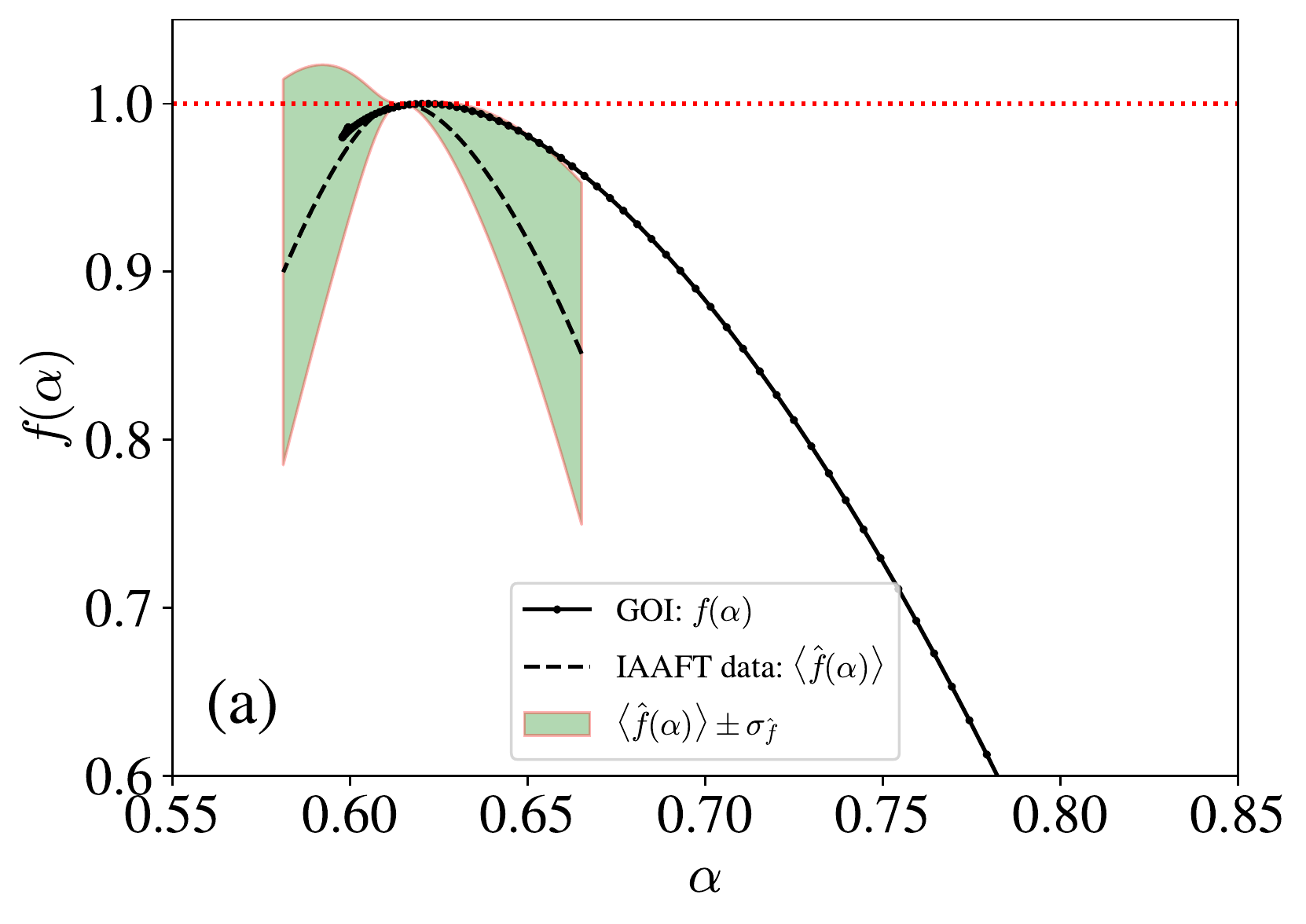}
    \includegraphics[width=0.325\linewidth]{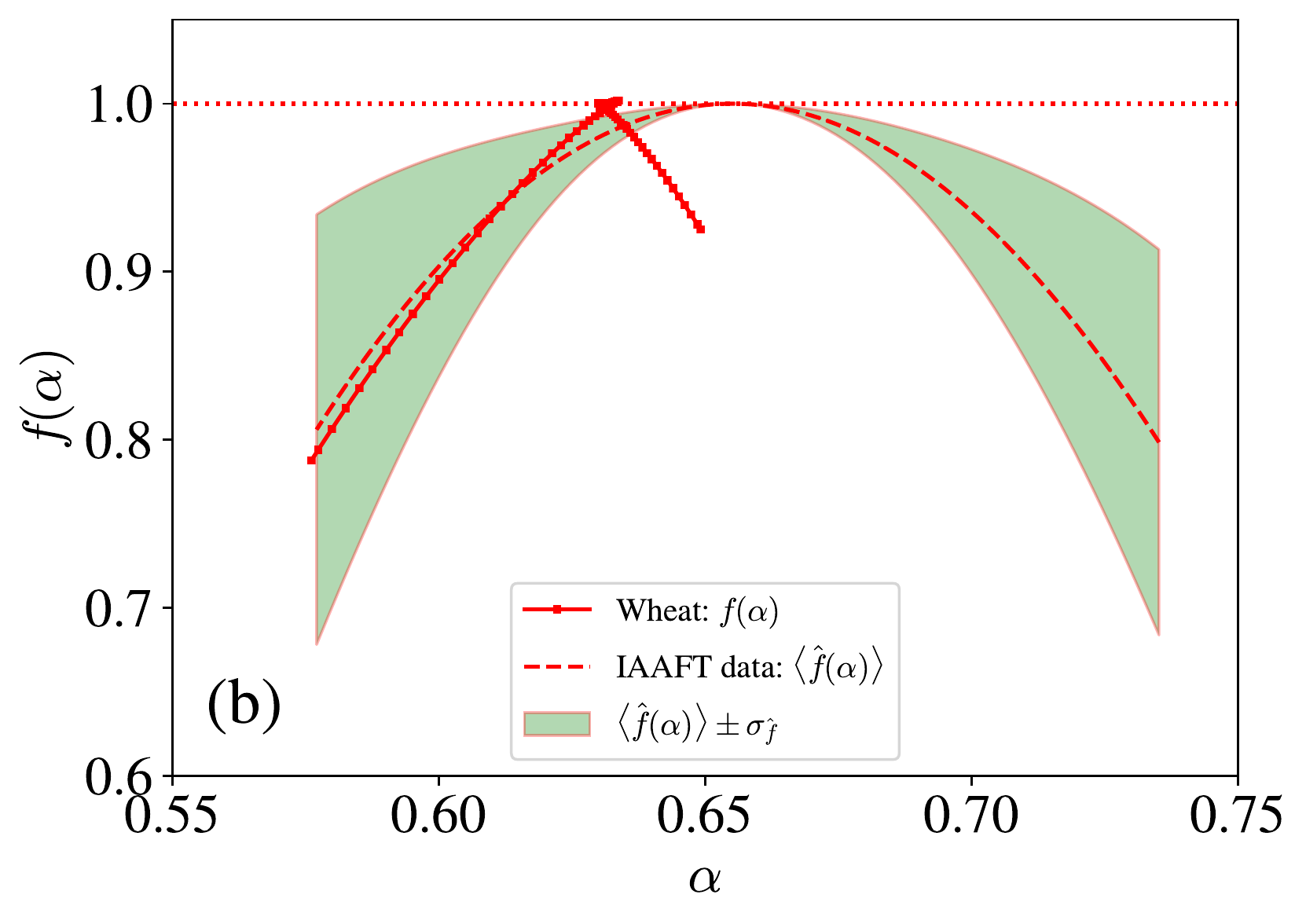}
    \includegraphics[width=0.325\linewidth]{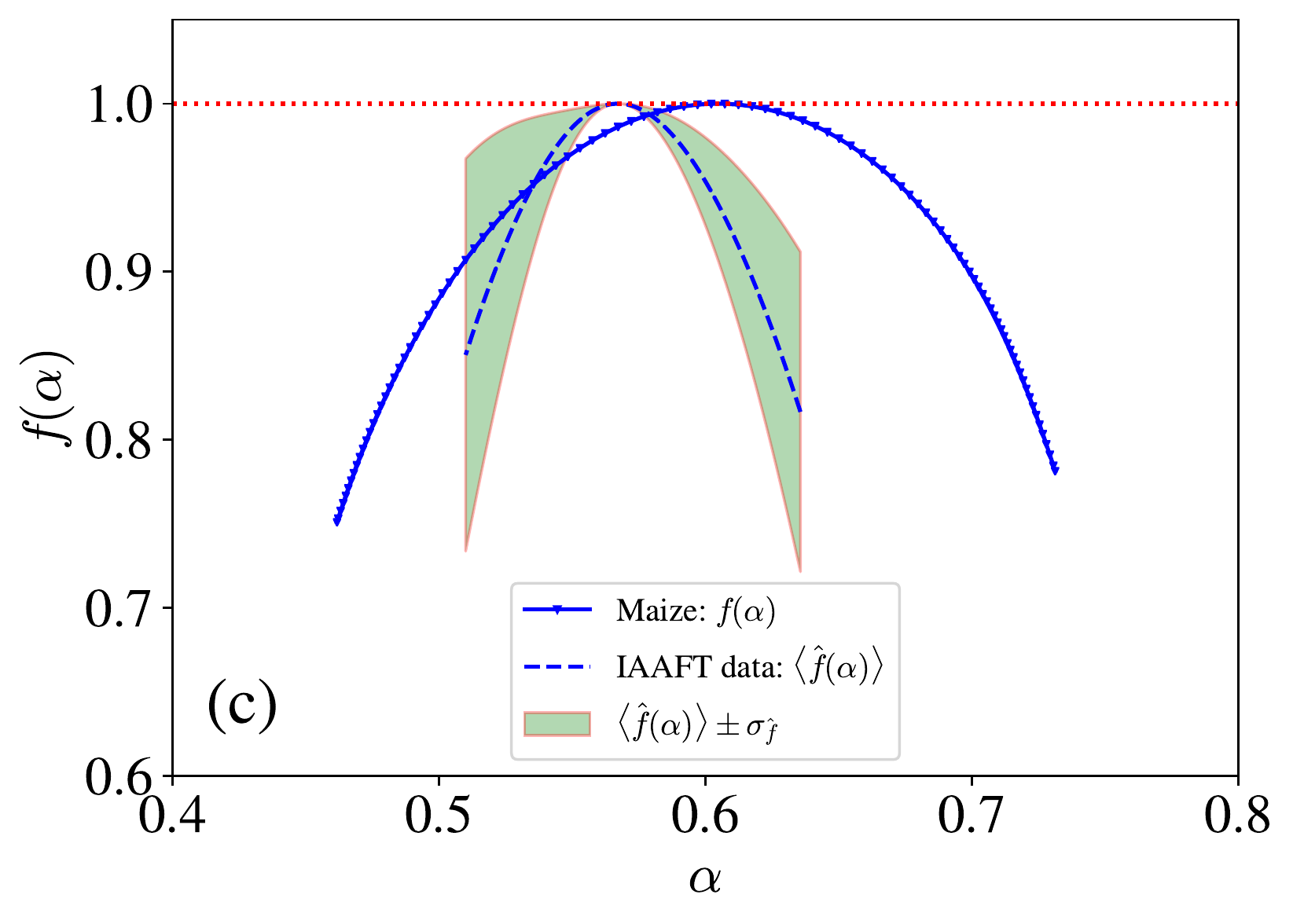}
    \includegraphics[width=0.325\linewidth]{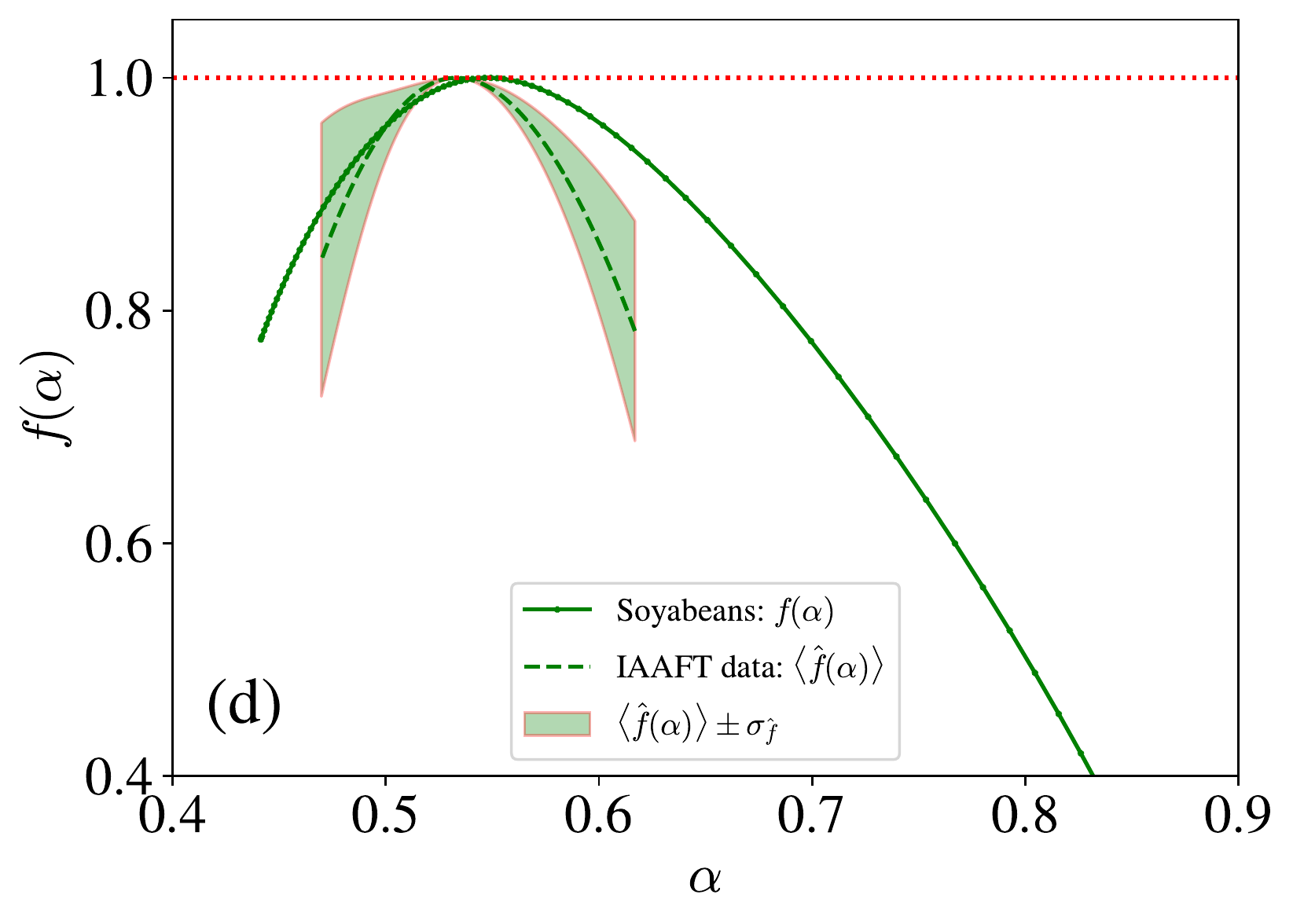}
    \includegraphics[width=0.325\linewidth]{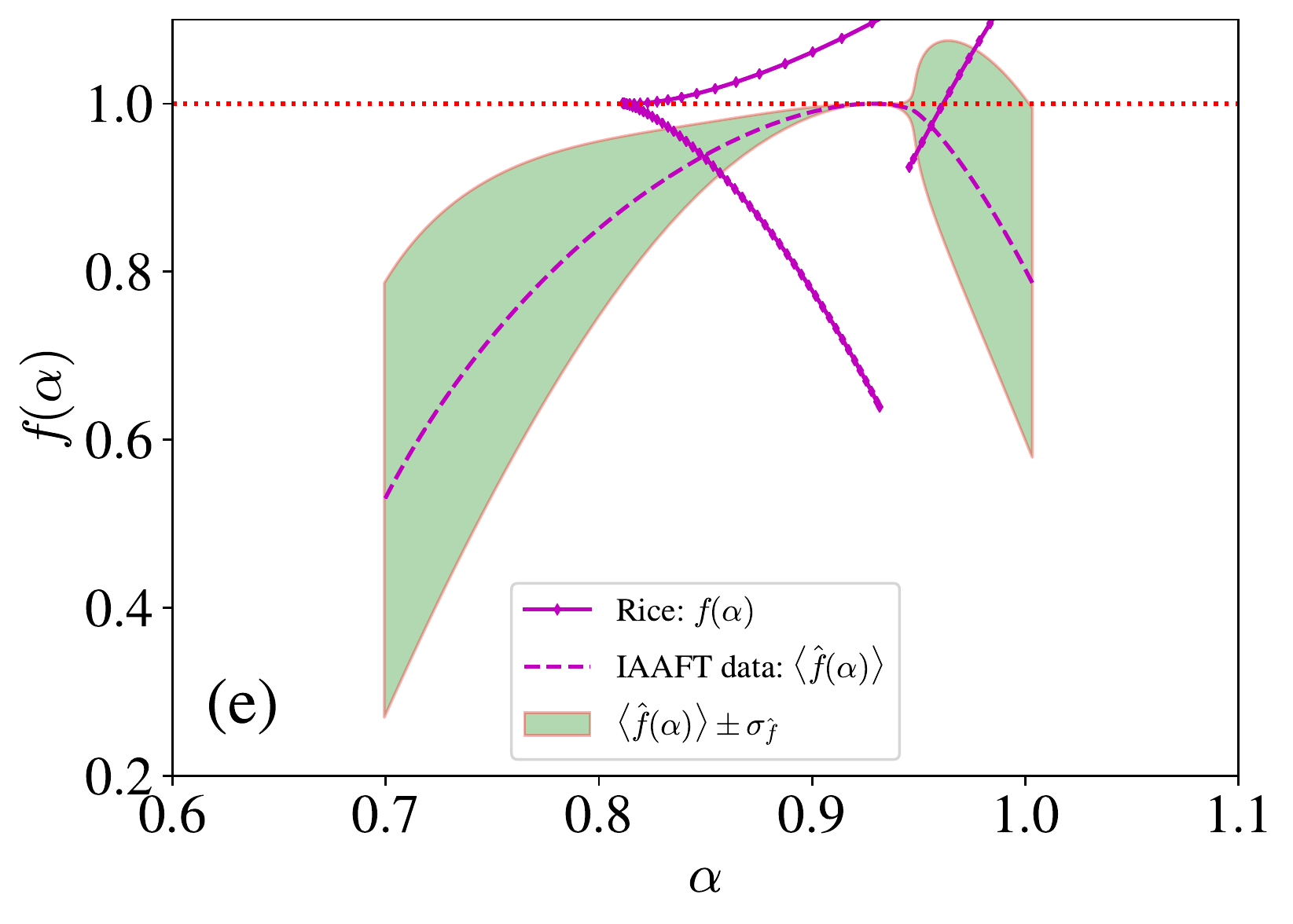}
    \includegraphics[width=0.325\linewidth]{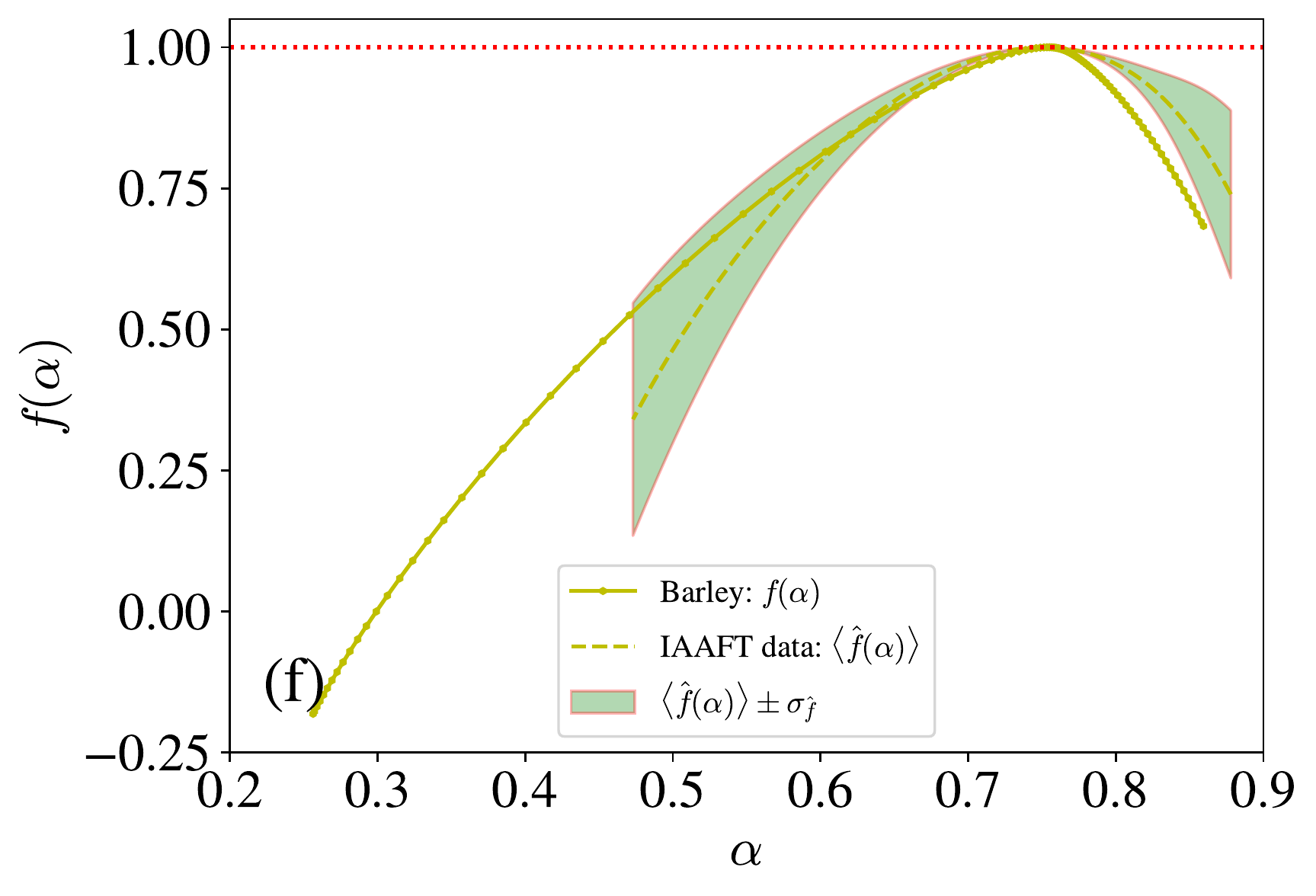}
    \caption{Singularity spectrum $f(\alpha)$ with respect to the singularity $\alpha$ for the GOI index (a), the wheat sub-index (b), the maize sub-index (c), the soyabeans sub-index (d), the rice sub-index (e), and the barley sub-index (f). The polynomial used to detrend the indices is a linear function with $\ell=1$. For each GOI and sub-indices, we generate 1000 surrogate time series using the IAAFT algorithm and calculate the mean $\langle\hat{f}\rangle(\alpha)$ and standard deviation $\sigma_{\hat{f}}$.}
    \label{Fig:GOI:Return:MFDFA:f:alpah}
\end{figure}

For each GOI and sub-indices, the singularity exponents $\alpha(q)$ and the singularity spectrum $f(q)$ can be numerically calculated using Eq.~(\ref{Eq:MF:tau:2:alpha}) and Eq.~(\ref{Eq:MF:alpha:tau:2:f}). For each index, we generate 1000 surrogate time series using the IAAFT algorithm and calculate the mean $\langle\hat{f}\rangle(\alpha)$ and standard deviation $\sigma_{\hat{f}}$. The resulting multifractal singularity spectra are shown in  Fig.~\ref{Fig:GOI:Return:MFDFA:f:alpah}.

We find that all the the mean multifractal singularity spectra $\langle\hat{f}\rangle(\alpha)$ of the IAAFT surrogates have a nice bell-like shape as the singularity spectra of classic mathematical models \cite{Jiang-Xie-Zhou-Sornette-2019-RepProgPhys}. Moreover, their singularity widths $\Delta\hat{\alpha}$ are significantly greater than 0, which highlights the potential power of the broad return distribution and possible linear long-term correlations to generate illusionary multifractality for multifractality-free IAAFT surrogates. For GOI, maize, soyabeans and barley, the singularity spectra $f(\alpha)$ are also bell-shaped, and their singularity $\Delta\alpha$ widths are significantly greater than those of the null models $\langle\Delta\hat{\alpha}\rangle$,
\begin{equation}
    \Delta\alpha > \langle\Delta\hat{\alpha}\rangle.
\end{equation}
In contrast, for wheat and rice, we have
\begin{equation}
    \Delta\alpha < \langle\Delta\hat{\alpha}\rangle,
\end{equation}
and $\Delta\alpha <0$ for rice. The singularity spectra of wheat and rice have a knot. The results obtained from MF-DFA with $\ell=2$ are very similar, as shown in Fig.~\ref{FigA:GOI:Return:MFDFA:f:alpah}, except for wheat whose singularity spectrum becomes bell-shaped and the singularity width is significantly greater than the IAAFT surrogates.

\subsubsection{Statistical tests based on Singularity width}

The strength of multifractality is one of the most important quantities of multifractal time series, which is characterized by the singularity width
\begin{equation}
    \Delta\alpha = \alpha(-\infty)-\alpha(+\infty) \triangleq \alpha_{\max}-\alpha_{\min}
    \label{Eq:MF:Delta:alpha}
\end{equation}
where $\alpha_{\max}$ and $\alpha_{\min}$ are not the asymptotic values for $q = \pm\infty$ but for $q=5$ in our empirical analysis \cite{Jiang-Xie-Zhou-Sornette-2019-RepProgPhys}. In practice, we try to test if $\Delta\alpha$ is greater than $\Delta\hat{\alpha}$ of the IAAFT surrogates \cite{Gao-Shao-Yang-Zhou-2022-ChaosSolitonsFractals}. Speaking differently, we need to calculate the probability that $\Delta\alpha$ is greater than $\Delta\hat{\alpha}$:
\begin{equation}
    p{\mathrm{-value}}=\Pr(\Delta\alpha<\Delta\hat{\alpha}).
\end{equation}
If the $p$-value is smaller than a preset significance level, say 5\%, we reject the hypothesis that the original time series is monofractal. Otherwise, the original time series cannot be distinguished from the IAAFT surrogates which contain only linear correlations.

\begin{figure}[!ht]
    \centering
    \includegraphics[width=0.325\linewidth]{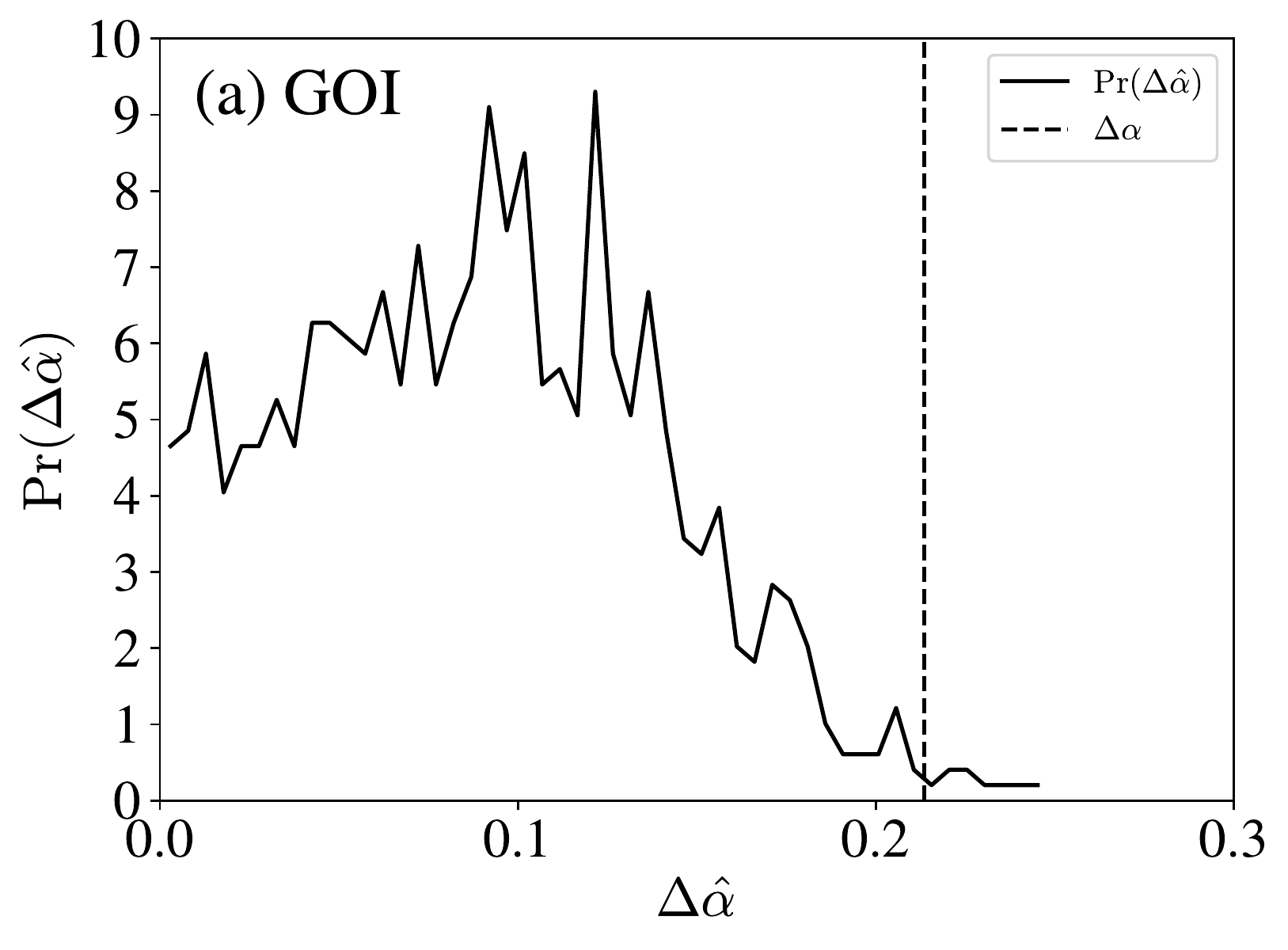}
    \includegraphics[width=0.325\linewidth]{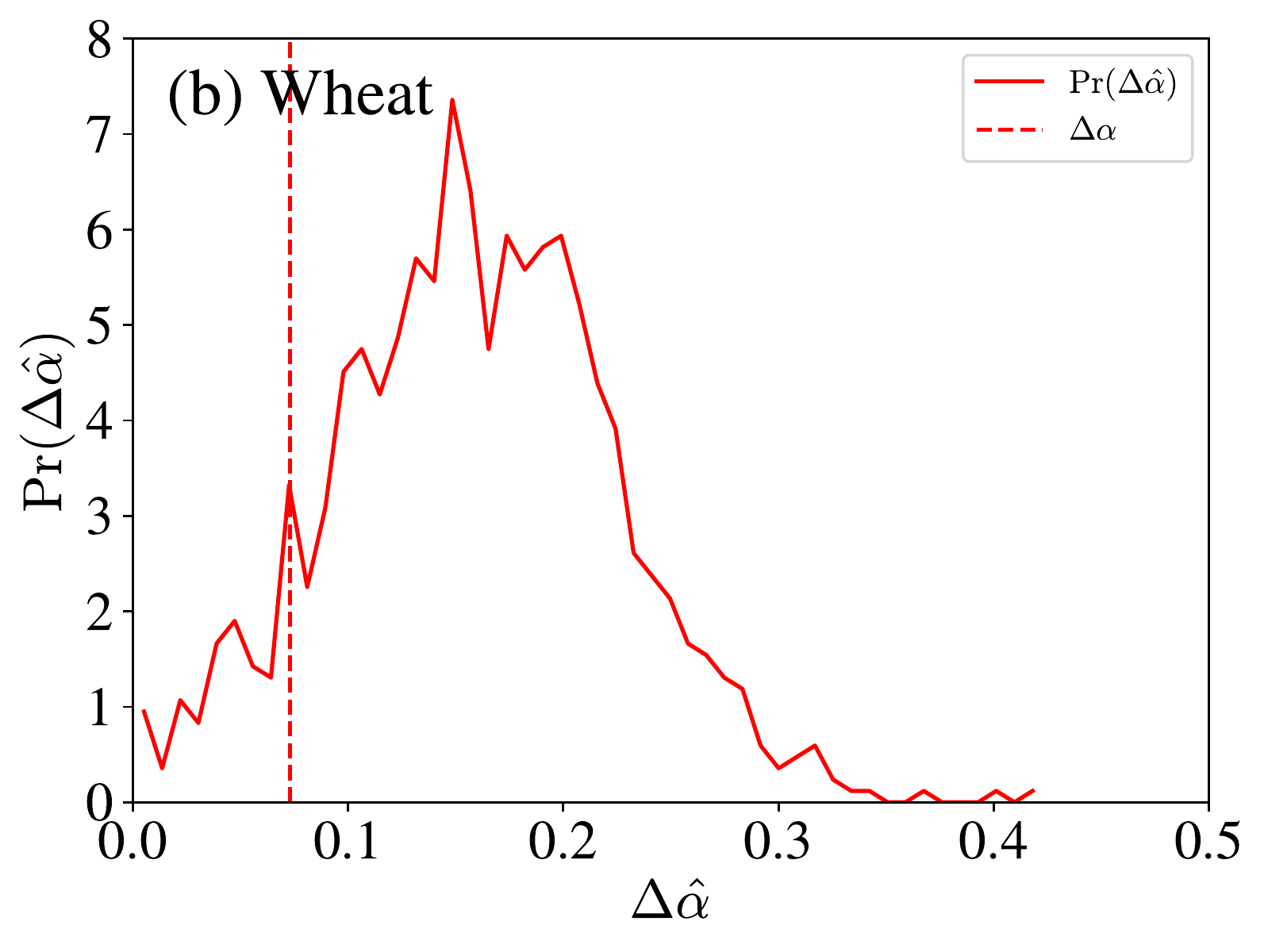}
    \includegraphics[width=0.325\linewidth]{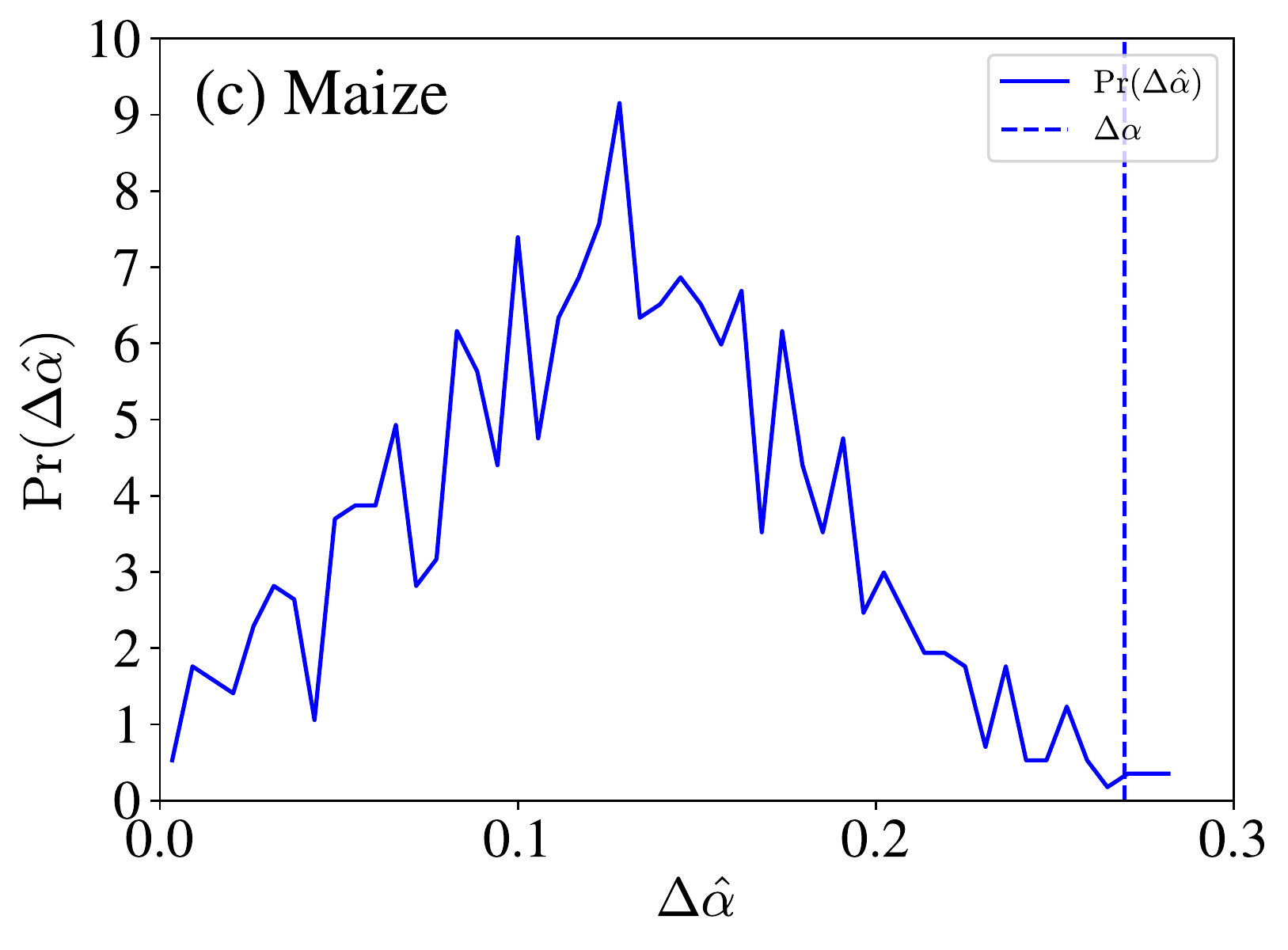}
    \includegraphics[width=0.325\linewidth]{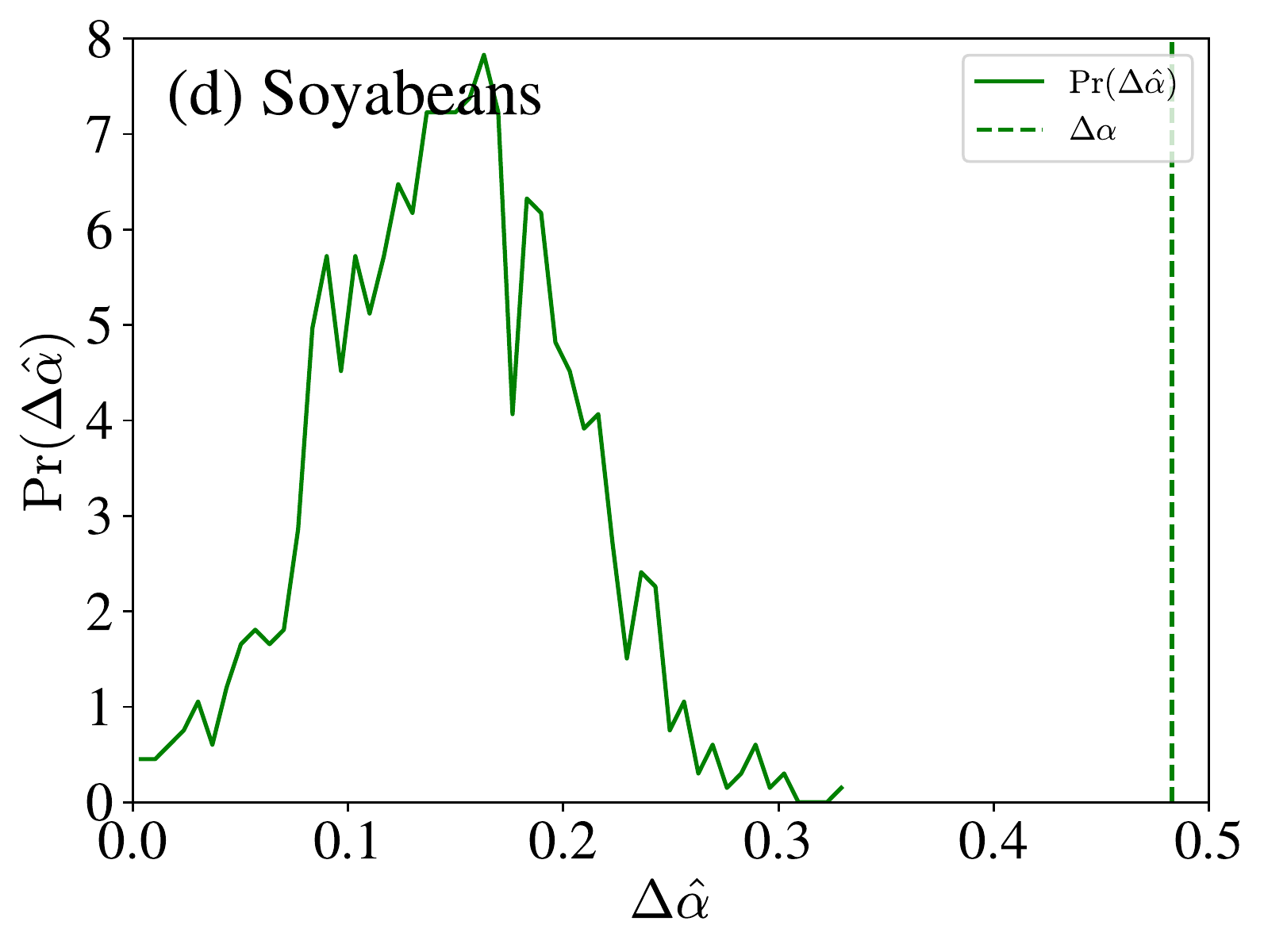}
    \includegraphics[width=0.325\linewidth]{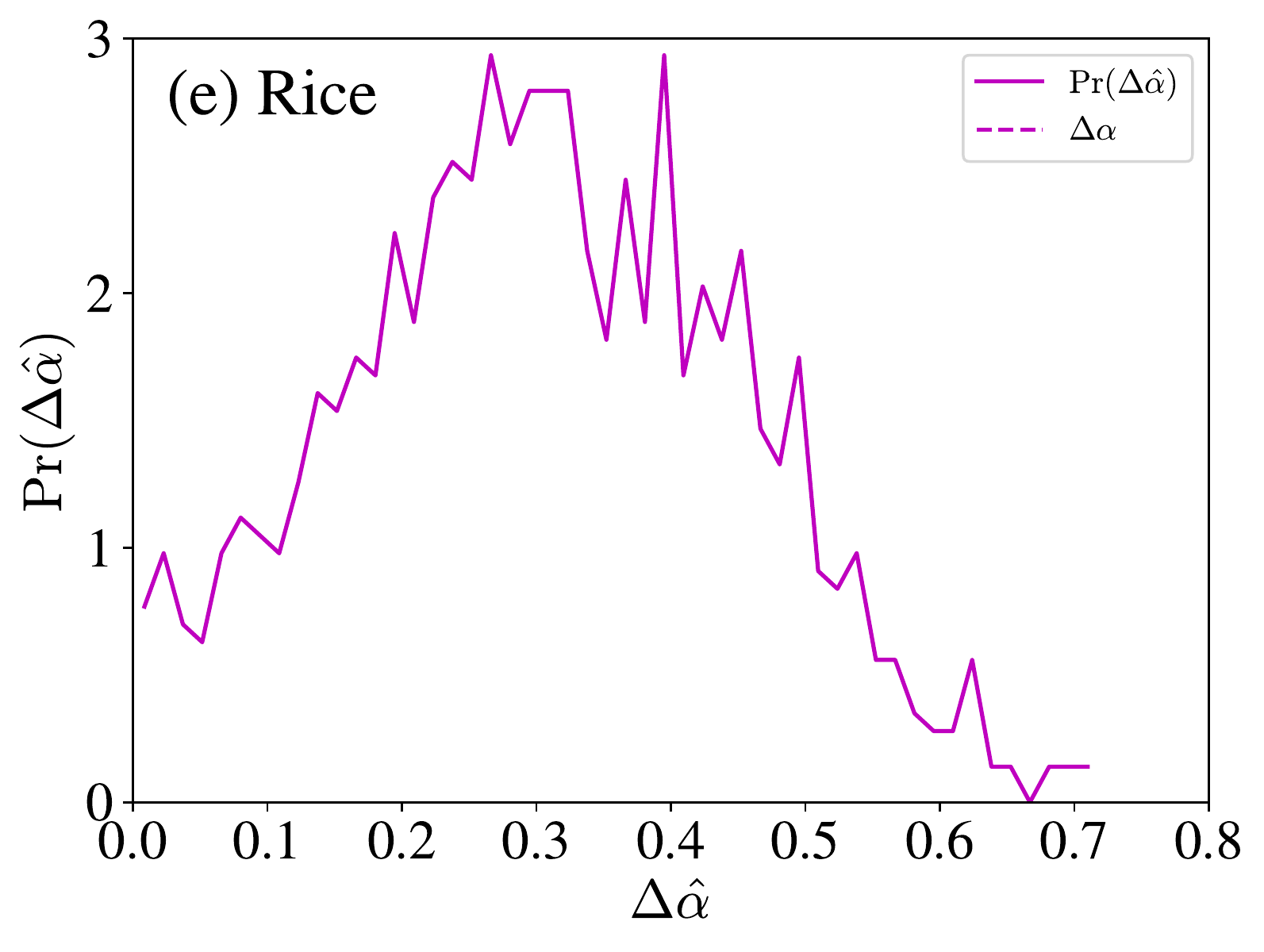}
    \includegraphics[width=0.325\linewidth]{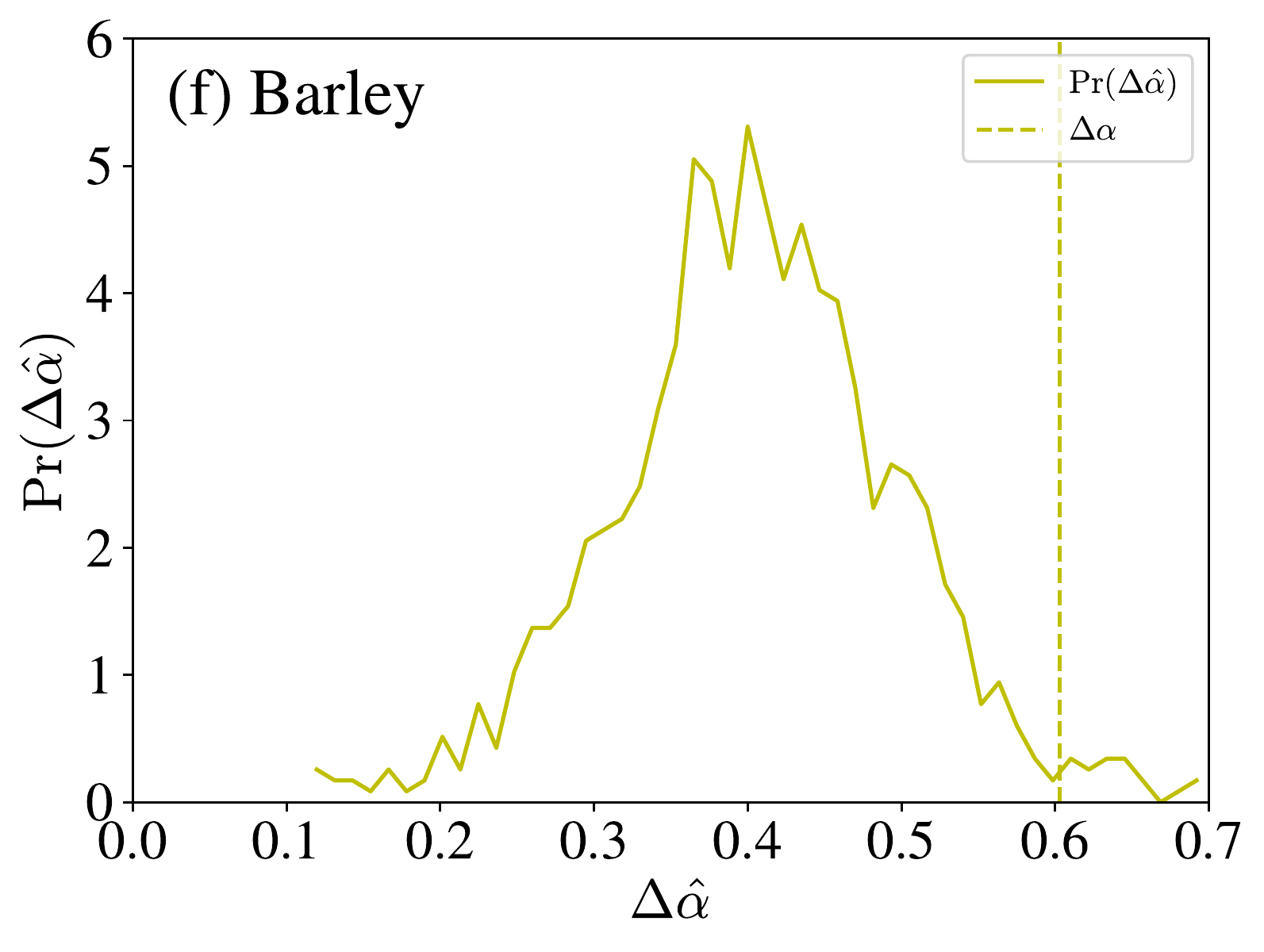}
    \caption{Empirical distribution of the singularity widths $\Delta\hat{\alpha}$ of the 1000 IAAFT surrogates for the GOI index (a), the wheat sub-index (b), the maize sub-index (c), the soyabeans sub-index (d), the rice sub-index (e), and the barley sub-index (f). The polynomial used to detrend the indices is a linear function with $\ell=1$. The vertical dashed lines are the corresponding singularity widths $\Delta \alpha$ of the original time series. Note that $\Delta \alpha<0$ for rice, which is not shown in plot (e).}
    \label{Fig:GOI:Return:MFDFA:dalpah:test}
\end{figure}

For each spot price index, we generate 1000 IAAFT surrogates and determine their singularity widths $\Delta\hat{\alpha}$. The empirical distributions of singularity widths $\Delta\hat{\alpha}$ of the six spot price indices are illustrated in  Fig.~\ref{Fig:GOI:Return:MFDFA:dalpah:test}. We find that wheat in  Fig.~\ref{Fig:GOI:Return:MFDFA:dalpah:test}(b) and rice in  Fig.~\ref{Fig:GOI:Return:MFDFA:dalpah:test}(e) have large $p$-values, while the rest four spot price indices have very small $p$-values. As shown in Fig.~\ref{FigA:GOI:Return:MFDFA:dalpah:test}, the results obtained from MF-DFA with $\ell=2$ are qualitatively the same, except for wheat whose $p$-value becomes much smaller.

\begin{table}[!h]
    \centering
    \caption{Testing multifractality based on singularity width. Here, $\Delta\alpha$ is the singularity width of the original time series, $\langle\Delta\hat{\alpha}\rangle$ is the average singularity width of the 1000 IAAFT surrogates, $\sigma_{\Delta\hat{\alpha}}$ is the standard deviation of the singularity widths of the 1000 IAAFT surrogates, and $p$-value is the proportion of IAAFT surrogates with $\Delta\hat{\alpha} >\Delta\alpha$. }
    \smallskip
    \renewcommand\tabcolsep{9pt}
    \begin{tabular}{cccccccccccccc}
        \toprule
         \multirow{2}{2em}{Index}   & \multicolumn{4}{c}{MF-DFA with $\ell=1$} &&  \multicolumn{4}{c}{MF-DFA with $\ell=2$} \\
               \cline{2-5} \cline{7-10}
         & $\Delta\alpha$ & $\langle\Delta\hat{\alpha}\rangle$ &  $\sigma_{\Delta\hat{\alpha}}$ & $p$-value && $\Delta\alpha$ & $\langle\Delta\hat{\alpha}\rangle$ &  $\sigma_{\Delta\hat{\alpha}}$ & $p$-value  \\
        \midrule
         GOI & 0.2136 & 0.0891 & 0.0502 & 0.0090 && 0.3361 & 0.1140 & 0.0468 & 0.0000   \\
         Wheat & 0.0730 & 0.1594 & 0.0649 & 0.9040 && 0.2385 & 0.1513 & 0.0572 & 0.0640   \\
         Maize & 0.2695 & 0.1266 & 0.0560 & 0.0040  && 0.2124 & 0.1286 & 0.0463 & 0.0310   \\
         Soyabeans & 0.4828 & 0.1474 & 0.0545 & 0.0000  && 0.7253 & 0.1682 & 0.0458 & 0.0000   \\
         Rice & -0.0139 & 0.3060 & 0.1442 & 1.0000  && 0.1494 & 0.3546 & 0.1304 & 0.9320   \\
         Barley & 0.6030 & 0.4050 & 0.0910 & 0.0200  && 0.7183 & 0.4141 & 0.0817 & 0.0010   \\
        \bottomrule
    \end{tabular}
    \label{Tab:MFtest:dalpha}
\end{table}

For each global grain spot price index, We calculate the singularity width $\delta\alpha$, the average singularity width $\langle\Delta\hat{\alpha}\rangle$ of the 1000 IAAFT surrogates, the standard deviation of the singularity widths $\sigma_{\Delta\hat{\alpha}}$ of the 1000 IAAFT surrogates, and the $p$-value that $\delta\alpha$ is smaller than $\Delta\hat{\alpha}$. These characteristic values are presented in Table~\ref{Tab:MFtest:dalpha} for $\ell=1$ and $\ell=2$. It shows that rice and barley have much larger $\sigma_{\Delta\hat{\alpha}}$ values, quantifying the broader distributions in Fig.~\ref{Fig:GOI:Return:MFDFA:dalpah:test}(e,f). For GOI, maize, soyabeans, and barley, $\langle\Delta\hat{\alpha}\rangle < \Delta\alpha$ and their $p$-values are very small, suggesting that these time series may have an intrinsic multifractal nature that cannot be explained by the fat-tailedness of returns and possible long-term liner correlations in returns. In contrast, for wheat and rice, the $p$-values are large, which indicates the absence of intrinsic multifractality.

\begin{figure}[!ht]
    \centering
    \includegraphics[width=0.325\linewidth]{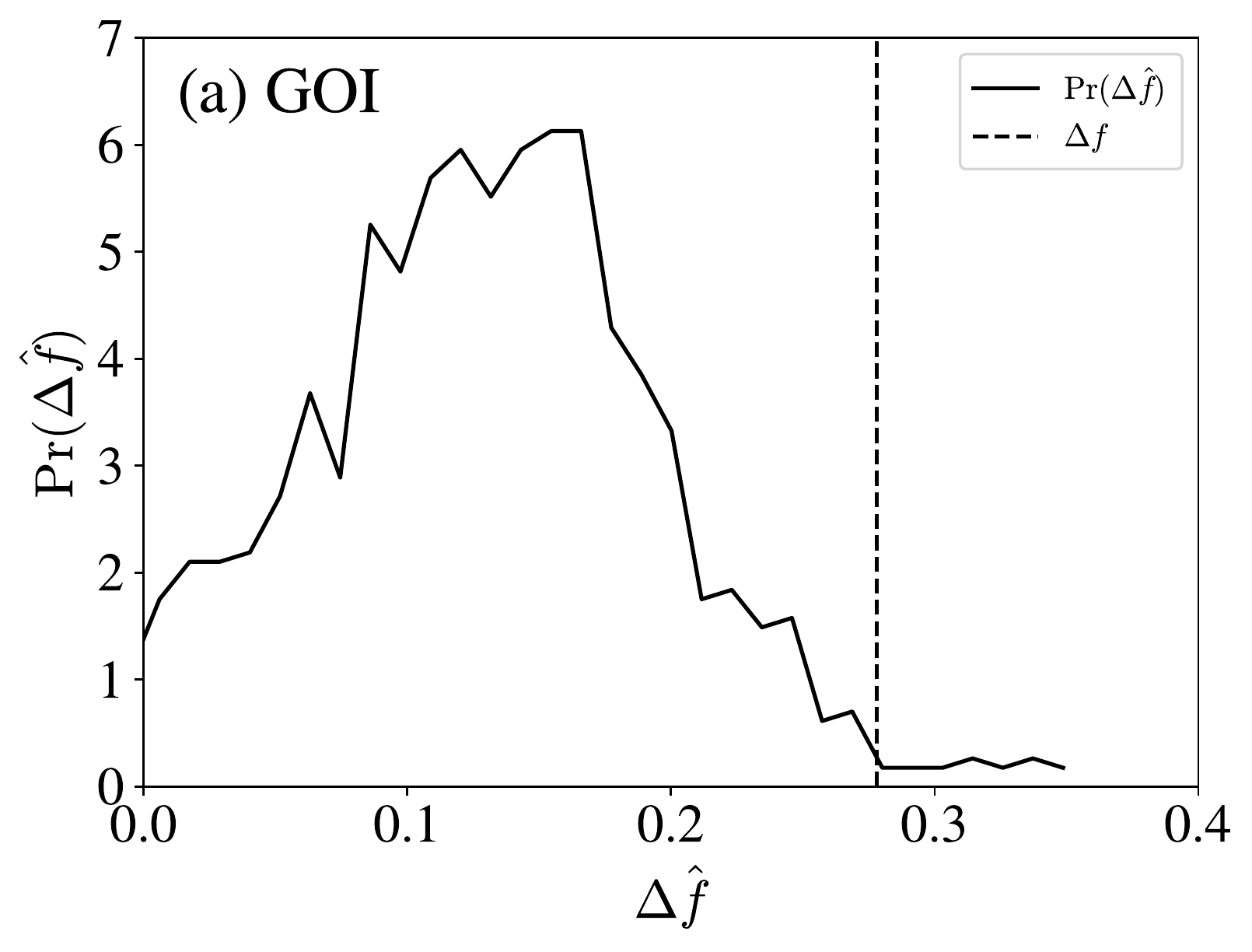}
    \includegraphics[width=0.325\linewidth]{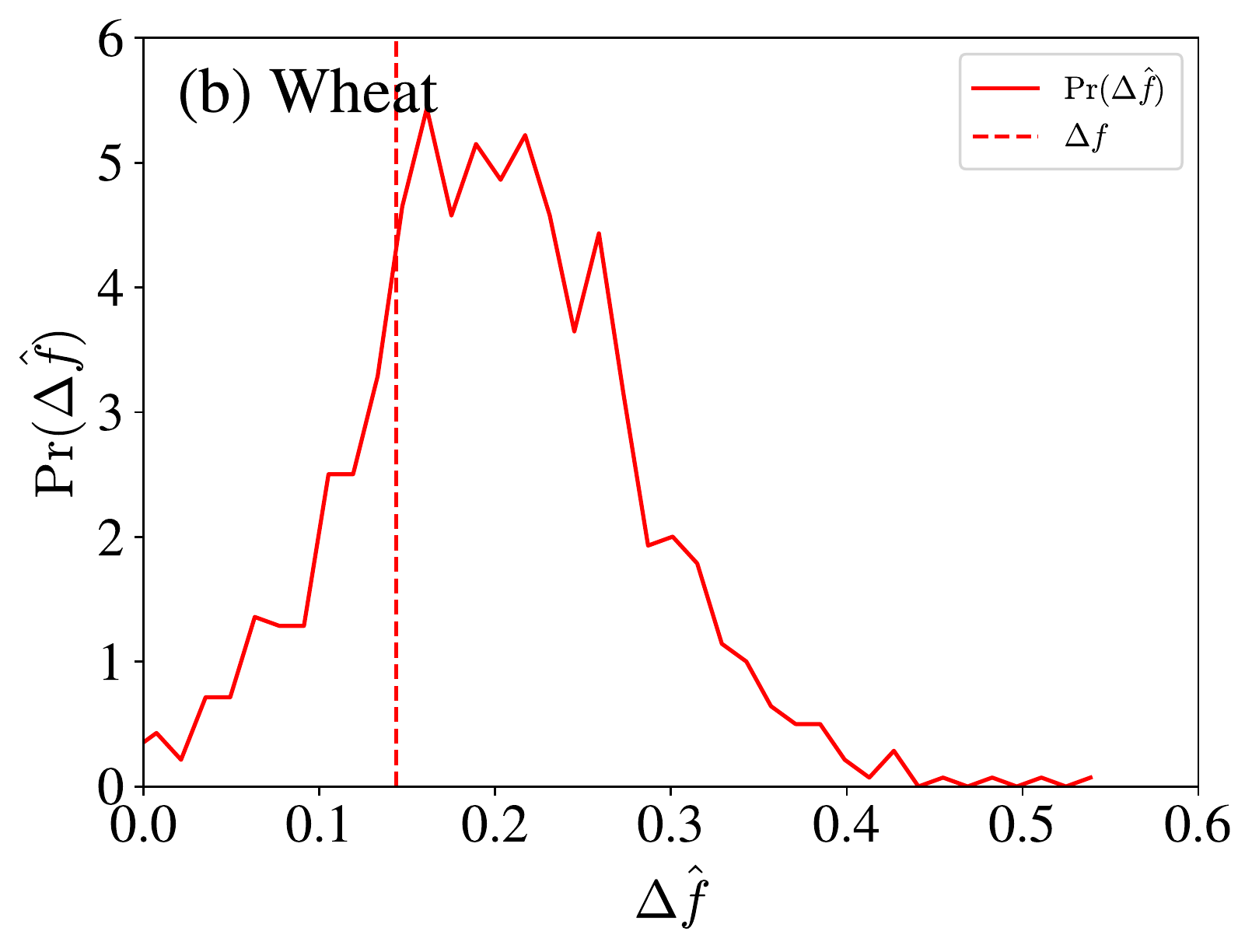}
    \includegraphics[width=0.325\linewidth]{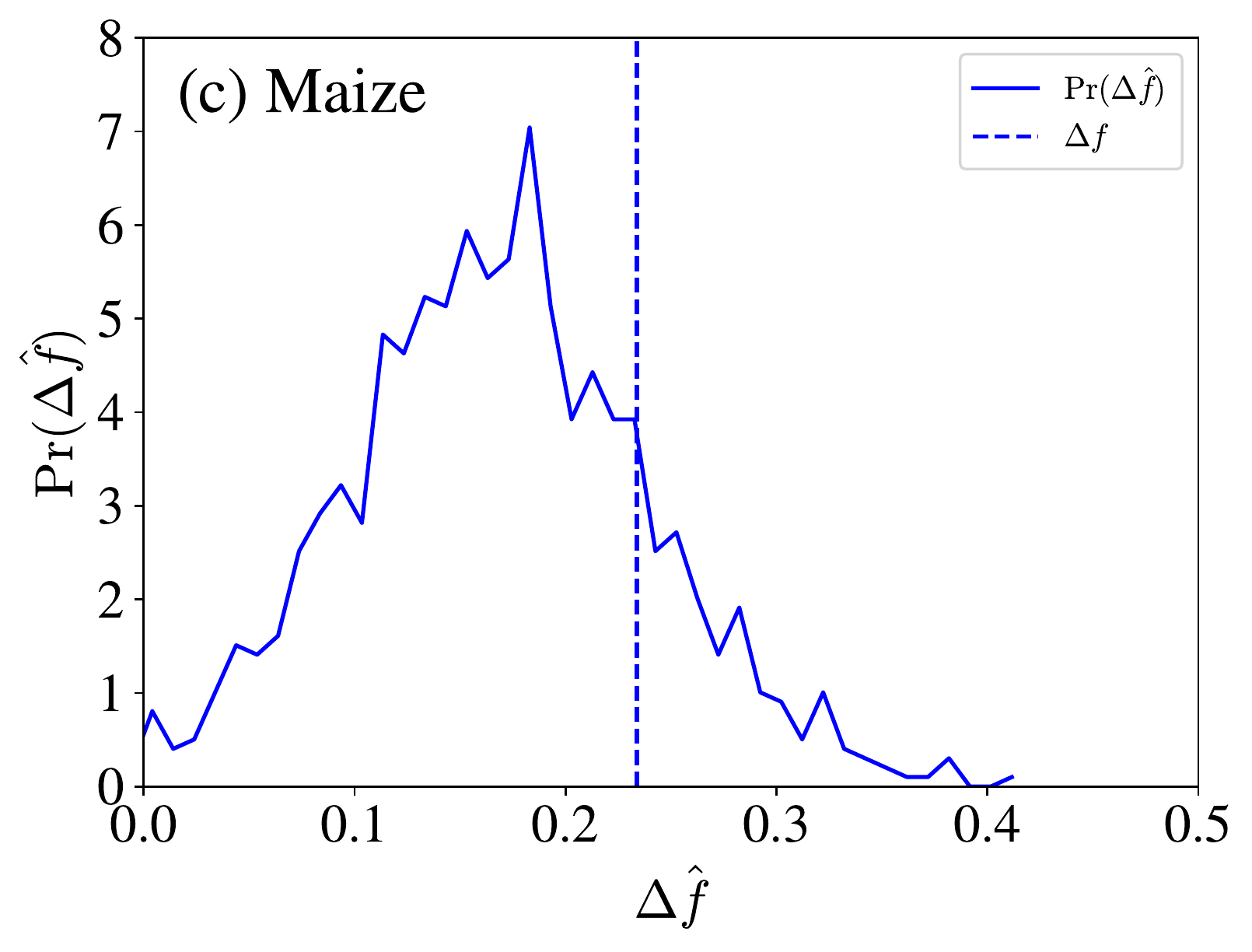}
    \includegraphics[width=0.325\linewidth]{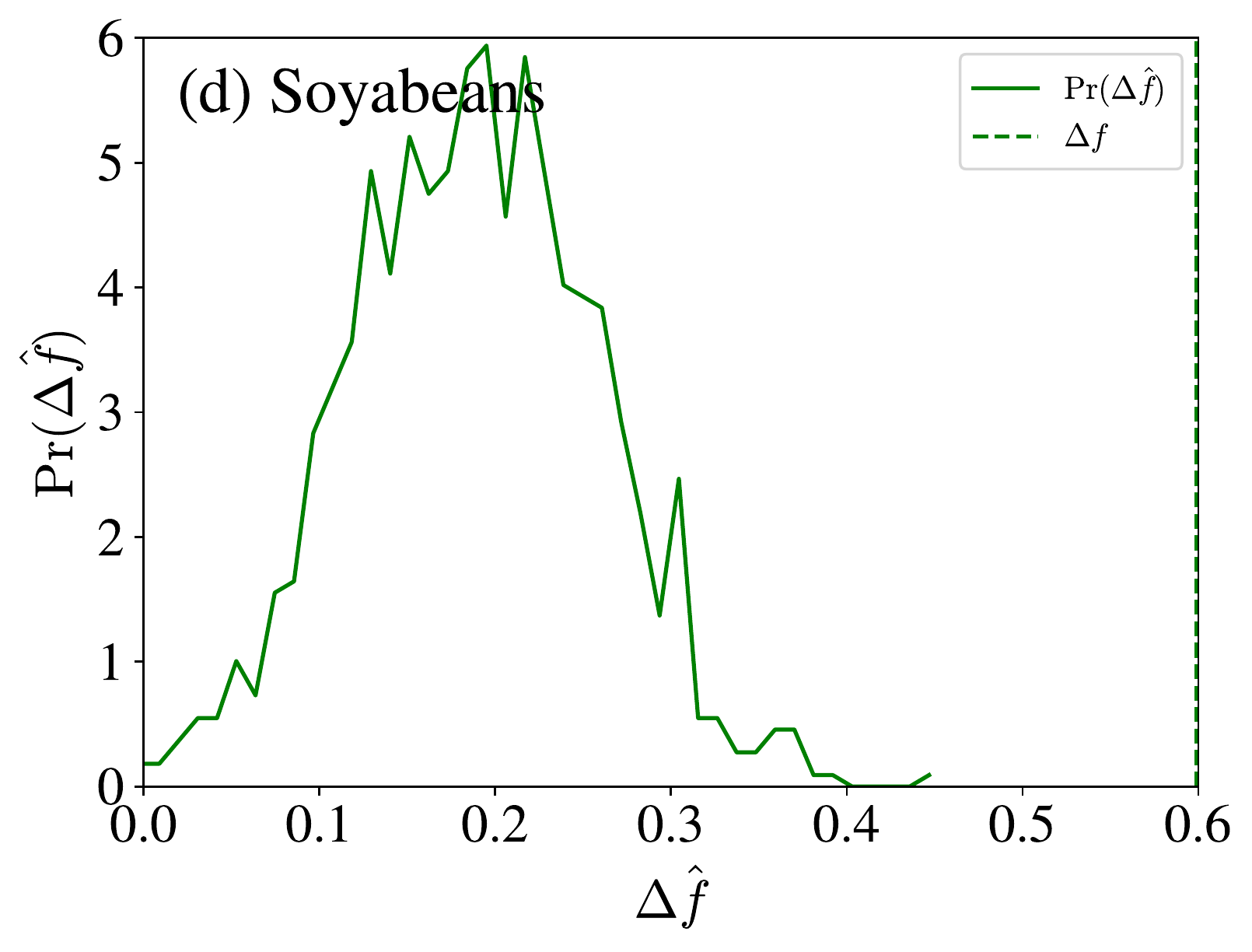}
    \includegraphics[width=0.325\linewidth]{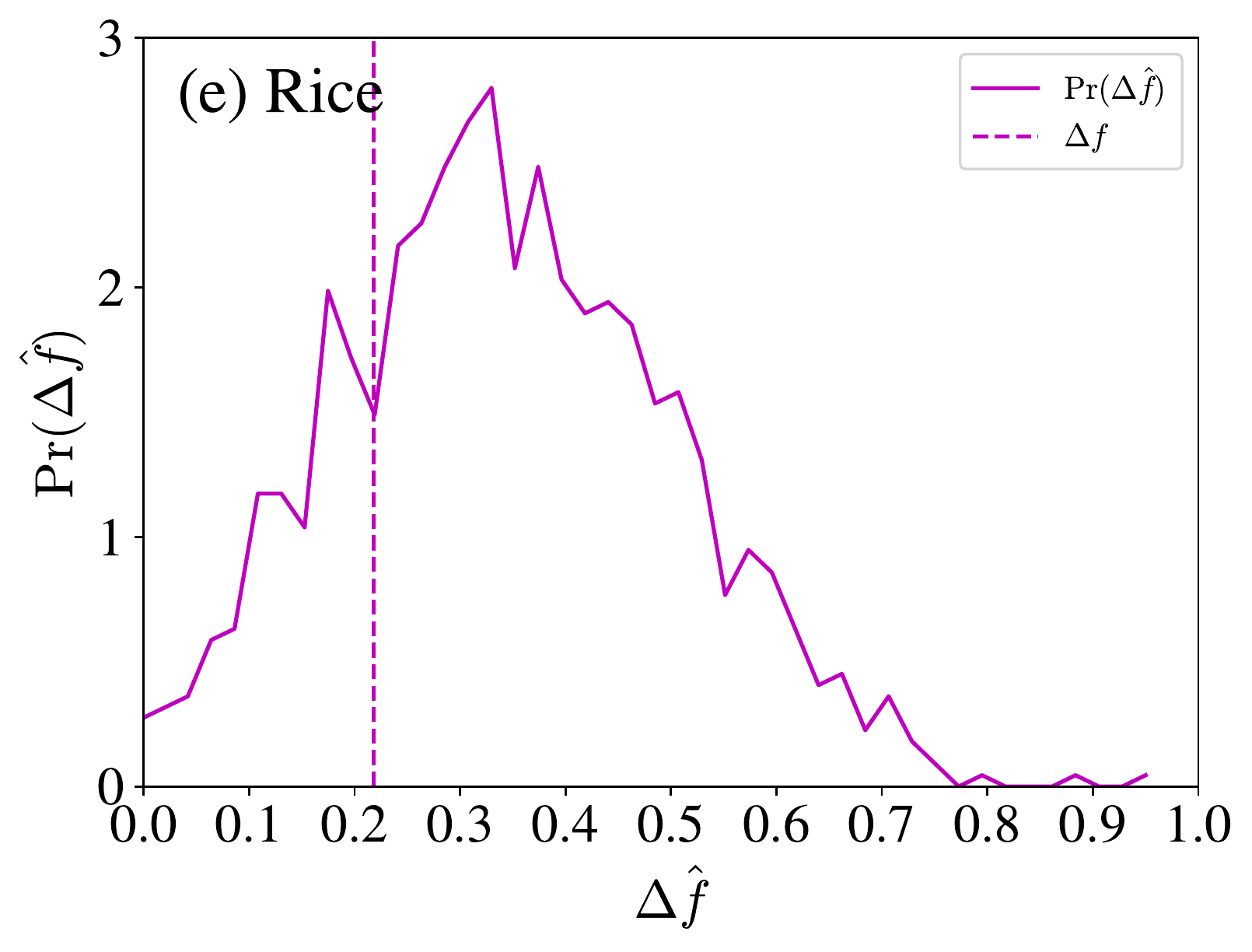}
    \includegraphics[width=0.325\linewidth]{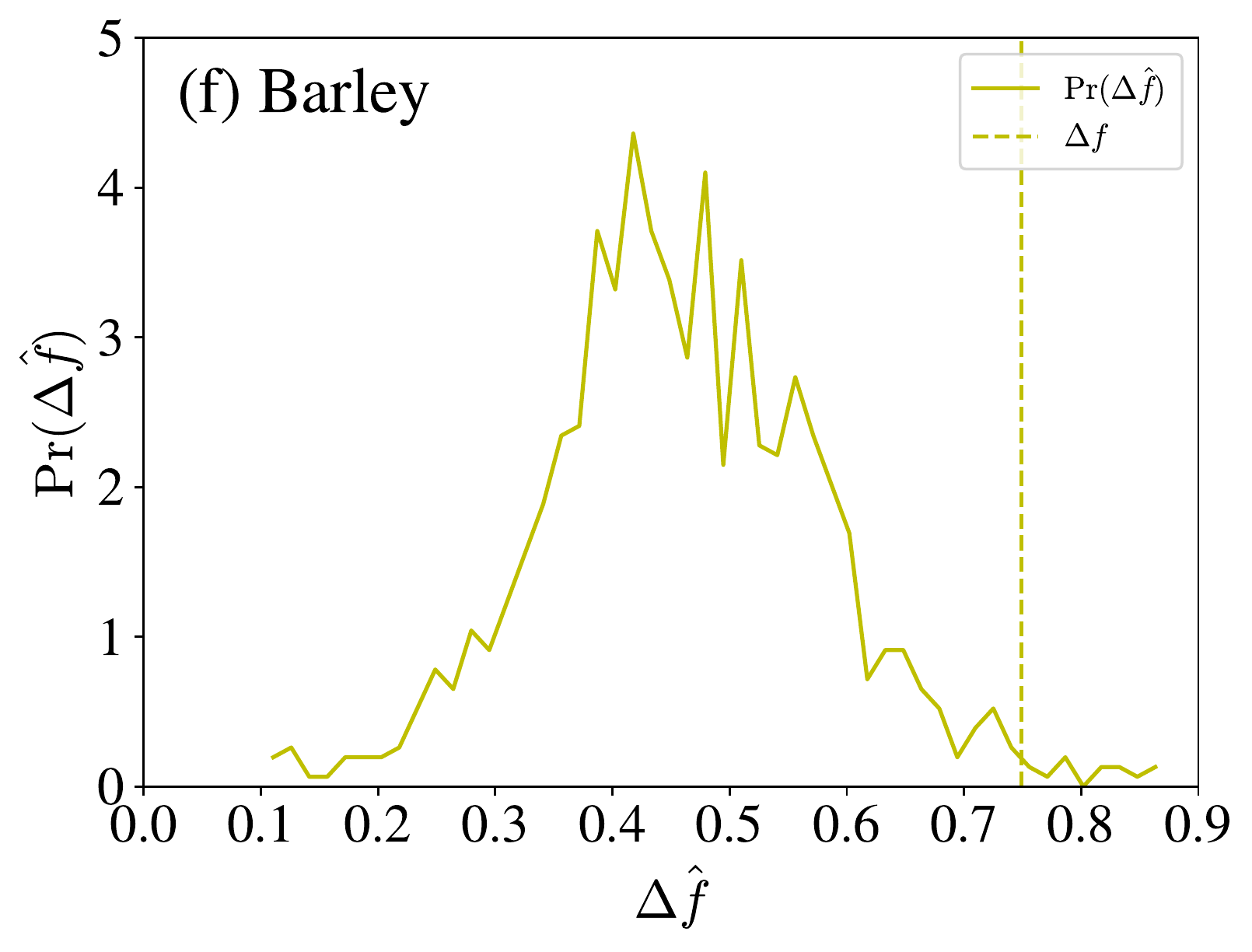}
    \caption{Empirical distribution of the spectrum differences $\Delta\hat{f}$ of the 1000 IAAFT surrogates for the GOI index (a), the wheat sub-index (b), the maize sub-index (c), the soyabeans sub-index (d), the rice sub-index (e), and the barley sub-index (f). The polynomial used to detrend the indices is a linear function with $\ell=1$. The vertical dashed lines are the corresponding spectrum differences $\Delta{f}$ of the original time series.}
    \label{Fig:GOI:Return:MFDFA:df:test}
\end{figure}

\subsubsection{Statistical tests based on spectrum difference}

Following Refs.~\cite{Jiang-Zhou-2008a-PhysicaA,Jiang-Zhou-2008b-PhysicaA,Gao-Shao-Yang-Zhou-2022-ChaosSolitonsFractals}, we next perform statistical tests based on the spectrum difference $\Delta{f}$, which is defined as follows,
\begin{equation}
    \Delta f = 1 - \left[f(\alpha_{\min})+f(\alpha_{\max})\right] / 2.
    \label{Eq:MF:Delta:f}
\end{equation}
We aim to test if $\Delta{f}$ is significantly greater than $\Delta\hat{f}$ of the IAAFT surrogates \cite{Gao-Shao-Yang-Zhou-2022-ChaosSolitonsFractals}. Speaking differently, we calculate the probability that $\Delta{f}$ is greater than $\Delta\hat{f}$:
\begin{equation}
    p{\mathrm{-value}}=\Pr(\Delta{f}<\Delta\hat{f}).
\end{equation}
We note that, this test alone is not capable of rejecting multifractality; Rather, it is used to confirm the conclusion of the statistical tests based on singularity width.

For each spot price index, we generate 1000 IAAFT surrogates and determine their spectrum differences $\Delta\hat{f}$. The empirical distributions of spectrum differences  $\Delta\hat{f}$ of the six spot price indices are illustrated in  Fig.~\ref{Fig:GOI:Return:MFDFA:df:test}. We find that wheat in  Fig.~\ref{Fig:GOI:Return:MFDFA:df:test}(b), maize in  Fig.~\ref{Fig:GOI:Return:MFDFA:df:test}(c) and rice in  Fig.~\ref{Fig:GOI:Return:MFDFA:df:test}(e) have large $p$-values, while the rest three spot price indices have very small $p$-values. As shown in Fig.~\ref{FigA:GOI:Return:MFDFA:df:test}, the results obtained from MF-DFA with $\ell=2$ are qualitatively the same, except again for wheat whose $p$-value becomes much smaller.

For each global grain spot price index, we calculate the spectrum difference $\delta{f}$, the average spectrum difference $\langle\Delta\hat{f}\rangle$ of the 1000 IAAFT surrogates, the standard deviation of spectrum differences $\sigma_{\Delta\hat{f}}$ of the 1000 IAAFT surrogates, and the $p$-value that $\delta{f}$ is smaller than $\Delta\hat{f}$. These characteristic values are presented in Table~\ref{Tab:MFtest:df} for $\ell=1$ and $\ell=2$. The large $p$-values for wheat and maize echo the results for $\Delta\alpha$ in Table~\ref{Tab:MFtest:dalpha}. However, the large $p$-value for maize does not deny the conclusion extracted from Table~\ref{Tab:MFtest:dalpha}.

\begin{table}[!h]
\centering
    \caption{Testing multifractality based on singularity spectrum difference $\Delta{f}$.  Here, $\Delta{f}$ is the spectrum difference of the original time series, $\langle\Delta\hat{f}\rangle$ is the average spectrum difference of the 1000 IAAFT surrogates, $\sigma_{\Delta\hat{f}}$ is the standard deviation of the spectrum differences of the 1000 IAAFT surrogates, and $p$-value is the proportion of IAAFT surrogates with $\Delta\hat{f} >\Delta{f}$. }
    \smallskip
    \renewcommand\tabcolsep{9pt}
    \begin{tabular}{cccccccccccccc}
        \toprule
         \multirow{2}{2em}{Index}    & \multicolumn{4}{c}{MF-DFA with $\ell=1$} &&  \multicolumn{4}{c}{MF-DFA with $\ell=2$} \\
               \cline{2-5} \cline{7-10}
            & $\Delta{f}$ & $\langle\Delta\hat{f}\rangle$ &  $\sigma_{\Delta\hat{f}}$ & $p$-value && $\Delta{f}$ & $\langle\Delta\hat{f}\rangle$ &  $\sigma_{\Delta\hat{f}}$ & $p$-value  \\
        \midrule
         GOI & 0.2780 & 0.1245 & 0.0724 & 0.0150  &&  0.3819 & 0.1474 & 0.0601 & 0.0000   \\
         Wheat & 0.1435 & 0.1977 & 0.0848 & 0.7660 && 0.3523 & 0.1885 & 0.0745 & 0.0140   \\
         Maize & 0.2338 & 0.1665 & 0.0727 & 0.1680 && 0.1812 & 0.1649 & 0.0596 & 0.3880   \\
         Soyabeans & 0.5991 & 0.1870 & 0.0711 & 0.0000 && 0.8569 & 0.2069 & 0.0597 & 0.0000   \\
         Rice & 0.2182 & 0.3426 & 0.1603 & 0.7720 && 0.2431 & 0.3634 & 0.1476 & 0.7960   \\
         Barley & 0.7493 & 0.4602 & 0.1190 & 0.0130 && 0.9276 & 0.4841 & 0.1066 & 0.0000   \\
        \bottomrule
    \end{tabular}
    \label{Tab:MFtest:df}
\end{table}


\section{Summary and conclusion}
\label{S1:Summary}

We have performed multifractal detrended fluctuation analysis on six spot price indices of the global grain markets. To check if the extracted apparent multifractality is caused by the intrinsic nonlinear correlations in the original time series, we performed extensive statistical tests using surrogate time series generated by the IAAFT algorithm that preserve the values and power spectra of the original time series but destroy any nonlinear long-range correlations. We summarize and compare the MF-DMA and MF-DFA results of statistical tests for multifractal nature in Table~\ref{Tab:MFtest:summary}.
The column ``$\mathrm{d}H(q)/\mathrm{d}q<0$'' checks if $H(q)$ is a monotonically decreasing function. For a multifractal time series, we should have $\mathrm{d}H(q)/\mathrm{d}q<0$. The column ``Bell-shaped'' checks if the singularity spectrum is bell-shaped without a knotted tie on the top. The singularity spectrum $f(\alpha)$ of a multifractal time series should be bell-shaped without a knotted tie.
The column ``$a_2\neq0$'' checks if $\tau(q)$ is a nonlinear function with $\Pr(a_2\neq0)<0.05$. A multifractal time series should have a nonlinear $\tau(q)$ function so that $a_2\neq 0$.
The column of ``$a_2<0$'' checks if the coefficient $a_2<0$. A multifractal time series should have a concave $\tau(q)$ function so that $a_2<0$. We would like to point out that these criteria are used to check if the time series exhibits apparent multifractality.
The column ``$\Pr(\Delta\alpha<\Delta\hat{\alpha})$ '' shows the $p$-values of $\Delta\alpha<\Delta\hat{\alpha}$. For a multifractal time series, its singularity width $\Delta\alpha$ should be wider than the singularity width $\Delta\hat{\alpha}$ of its IAAFT surrogates.
The column ``$\Pr(\Delta{f}<\Delta\hat{f})$'' shows the $p$-values of $\Delta{f}<\Delta\hat{f}$. For a multifractal time series, we expect that its spectrum difference $\Delta{f}$ should be greater than the spectrum difference $\Delta\hat{f}$ of its IAAFT surrogates. However, we note that it is also possible to have $\Delta{f}<\Delta\hat{f}$ for multifractal and non-multifractal time series. Hence, this last column is to provide further supporting evidence for the presence of multifractal nature in time series, but not to reject the presence of multifractality.

\begin{table}[!h]
\centering
    \caption{Summary of the results of statistical tests for multifractal nature.
    The column of ``$\mathrm{d}H(q)/\mathrm{d}q$'' checks if $H(q)$ is a monotonically decreasing function.
    The column of ``Bell-shaped'' checks if the singularity spectrum is bell-shaped without a knotted tie on the top.
    The column of ``$a_2\neq0$'' checks if $\tau(q)$ is a nonlinear function with $\Pr(a_2\neq0)<0.05$.
    The column of ``$a_2<0$'' checks if the coefficient $a_2<0$.
    The column of ``$\Pr(\Delta\alpha<\Delta\hat{\alpha})$ '' shows the $p$-values of $\Delta\alpha<\Delta\hat{\alpha}$.
    The column of ``$\Pr(\Delta{f}<\Delta\hat{f})$'' shows the $p$-values of $\Delta{f}<\Delta\hat{f}$. }
    \smallskip
    \renewcommand\tabcolsep{6.5pt}
    \begin{tabular}{ccccccccccc}
        \toprule
        Index  & Method & $\mathrm{d}H(q)/\mathrm{d}q<0$ & Bell-shaped &  $a_2\neq0$ &  $a_2<0$ &   $\Pr(\Delta\alpha<\Delta\hat{\alpha})$ &  $\Pr(\Delta{f}<\Delta\hat{f})$  \\
        \midrule
        \multirow{3}{2em}{Maize}
         & MF-DMA            & \checkmark & \checkmark  & \checkmark & \checkmark & 0.0010 & 0.0010   \\
         & MF-DFA ($\ell=1$) & \checkmark & \checkmark  & \checkmark & \checkmark & 0.0040 & 0.1680   \\
         & MF-DFA ($\ell=2$) & \checkmark & \checkmark  & \checkmark & \checkmark & 0.0310 & 0.3880   \\
        \midrule
        \multirow{3}{2em}{Barley}
         & MF-DMA            & \checkmark & \checkmark  & \checkmark & \checkmark & 0.0970 & 0.0690   \\
         & MF-DFA ($\ell=1$) & \checkmark & \checkmark  & \checkmark & \checkmark & 0.0200 & 0.0130   \\
         & MF-DFA ($\ell=2$) & \checkmark & \checkmark  & \checkmark & \checkmark & 0.0010 & 0.0000   \\
        \midrule
        \multirow{3}{2em}{GOI}
         & MF-DMA            & \ding{55}  & \ding{55}   & \checkmark & \ding{55}  & 0.7140 & 0.2040   \\
         & MF-DFA ($\ell=1$) & \checkmark & \checkmark  & \checkmark & \checkmark & 0.0090 & 0.0150   \\
         & MF-DFA ($\ell=2$) & \checkmark & \checkmark  & \checkmark & \checkmark & 0.0000 & 0.0000   \\
        \midrule
        \multirow{3}{4em}{Soyabeans}
         & MF-DMA            & \ding{55}  & \ding{55}   & \checkmark & \checkmark & 0.2110 & 0.0380   \\
         & MF-DFA ($\ell=1$) & \checkmark & \checkmark  & \checkmark & \checkmark & 0.0000 & 0.0000   \\
         & MF-DFA ($\ell=2$) & \checkmark & \checkmark  & \checkmark & \checkmark & 0.0000 & 0.0000   \\
        \midrule
        \multirow{3}{2em}{Wheat}
         & MF-DMA            & \ding{55}  & \ding{55}   & \ding{55}  & \ding{55}  & 0.8970 & 0.6770   \\
         & MF-DFA ($\ell=1$) & \ding{55}  & \ding{55}   & \checkmark & \checkmark & 0.9040 & 0.7660   \\
         & MF-DFA ($\ell=2$) & \checkmark & \checkmark  & \checkmark & \checkmark & 0.0640 & 0.0140   \\
        \midrule
        \multirow{3}{2em}{Rice}
         & MF-DMA            & \checkmark & \checkmark  & \checkmark & \checkmark & 0.6740 & 0.4020   \\
         & MF-DFA ($\ell=1$) & \ding{55}  & \ding{55}   & \checkmark & \ding{55}  & 1.0000 & 0.7720   \\
         & MF-DFA ($\ell=2$) & \ding{55}  & \ding{55}   & \checkmark & \checkmark & 0.9320 & 0.7960   \\
        \bottomrule
    \end{tabular}
    \label{Tab:MFtest:summary}
\end{table}

We find that, for the maize spot price index and all three methods (MF-DMA, MF-DFA with $\ell=1$, and MF-DFA with $\ell=2$), the generalized Hurst index functions $H(q)$ are monotonically decreasing ($\mathrm{d}H(q)/\mathrm{d}q<0$), the singularity spectra $f(\alpha)$ are concave and bell-shaped, and the mass exponent functions $\tau(q)$ are nonlinear and concave. Moreover, all the three $p$-values of $\Delta\alpha<\Delta\hat{\alpha}$ are less than 5\%. Therefore, the maize spot price index exhibits an undoubted intrinsic multifractal nature.
For the barley spot price index, the generalized Hurst index functions $H(q)$ are monotonically decreasing ($\mathrm{d}H(q)/\mathrm{d}q<0$), the singularity spectra $f(\alpha)$ are concave and bell-shaped, and the mass exponent functions $\tau(q)$ are nonlinear and concave. we argue that there is also an intrinsic multifractal nature, which is supported by all the information except that the $p$-value of $\Delta\alpha<\Delta\hat{\alpha}$ is 0.0970 when the MF-DMA method is utilized.
For the GOI and the soyabeans sub-index, MF-DMA denies the possible presence of apparent multifractality since $H(q)$ is not a monotonically decreasing function and $f(\alpha)$ is not bell-shaped. MF-DMA also denies the presence of intrinsic multifractality since the $p$-values of $\Delta\alpha<\Delta\hat{\alpha}$ is large. However, MF-DFA with $\ell=1$ and MF-DFA with $\ell=2$ support the presence of an intrinsic multifractal nature with high probability.
For the wheat spot price index, MF-DMA and MF-DFA with $\ell=1$ deny the possible presence of apparent multifractality due to the incorrect shapes of $H(q)$ and $f(\alpha)$ and intrinsic multifractality due to the large $p$-values of $\Delta\alpha<\Delta\hat{\alpha}$. However, the results obtained by using MF-DFA with $\ell=2$ show that the generalized Hurst index functions $H(q)$ are monotonically decreasing ($\mathrm{d}H(q)/\mathrm{d}q<0$), the singularity spectra $f(\alpha)$ are concave and bell-shaped, and the mass exponent functions $\tau(q)$ are nonlinear and concave, which favors the possibility that there is intrinsic multifractality at the significance level of 0.0640.
Finally, the rice spot price index suffers the most convincing evidence for the absence of an intrinsic multifractal nature. Particularly, the three $p$-values of $\Delta\alpha<\Delta\hat{\alpha}$ are all very large. Moreover, MF-DFA with $\ell=1$ and MF-DFA with $\ell=2$ suggest that there is even no apparent multifractality because $H(q)$ is not monotonically decreasing and $f(\alpha)$ is not bell-shaped (not concave).

The summarized results in Table~\ref{Tab:MFtest:summary} highlight the very complex nonlinear behavior of the global grain spot markets. Previous studies on other markets often reported the presence of multifractality, although many of them did not perform rigorous statistical tests \cite{Jiang-Xie-Zhou-Sornette-2019-RepProgPhys}. On the one hand, we argue that, as what has been done in this work and in Ref.~\cite{Gao-Shao-Yang-Zhou-2022-ChaosSolitonsFractals}, any multifractal analysis of time series should perform similar statistical tests to confirm or deny the presence of multifractality. On the other hand, it is of vital importance to understand the underlying internal and external mechanisms that generate the multifractal or non-multifractal nature.

We note that, although the IAAFT technique is widely applied in nonlinear time series analysis, its asymptotic properties about statistical size and power (and hence consistency) have not been properly established. Theoretical and simulation studies reported that, in the context of IAAFT surrogates, the critical values used (or, equivalently, the $p$-values calculated) might not be appropriate \cite{Kugiumtzis-1999-PhysRevE,Kugiumtzis-2000-PhysRevE,Kugiumtzis-2001-IntJBifurcationChaos,Mammen-Nandi-2004-PhysRevE,Kugiumtzis-2008-StudNonlinearDynEconom}. As an alternative to the IAAFT method, the statically transformed autoregressive process (STAP) method was proposed \cite{Kugiumtzis-2000-PhysRevE,Kugiumtzis-2002-PhysRevE}, which was found to give consistently good results in terms of the size and power of the test \cite{Kugiumtzis-2008-StudNonlinearDynEconom}. Therefore, it would be necessary to apply the STAP surrogates to check if the IAAFT algorithm over-rejects the null hypothesis of the absence of multifractality.





\section*{Acknowledgments}
\label{S1:Funding}

This work was partly supported by the National Natural Science Foundation of China (72171083), the Shanghai Outstanding Academic Leaders Plan, and the Fundamental Research Funds for the Central Universities.


\clearpage
\appendix
\renewcommand\thefigure{\Alph{section}\arabic{figure}}
\section{Comparison of MF-DFA results}
\setcounter{figure}{0}

\begin{figure}[!ht]
    \centering
    \includegraphics[width=0.325\linewidth]{Fig_GOI_Return_MFDFA1_Fs_GOI.pdf}
    \includegraphics[width=0.325\linewidth]{Fig_GOI_Return_MFDFA1_Fs_Wheat.pdf}
    \includegraphics[width=0.325\linewidth]{Fig_GOI_Return_MFDFA1_Fs_Maize.pdf}
    \includegraphics[width=0.325\linewidth]{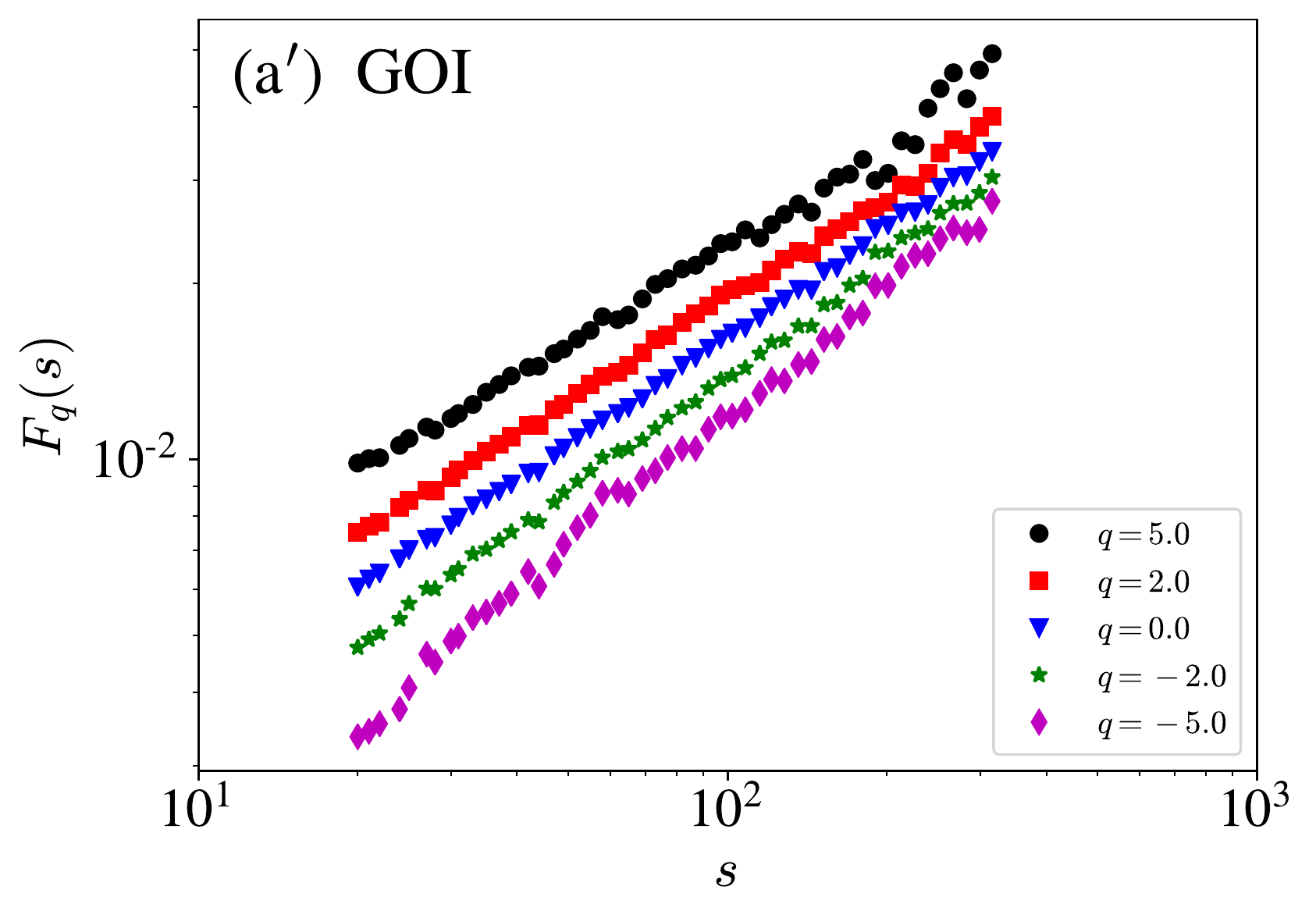}
    \includegraphics[width=0.325\linewidth]{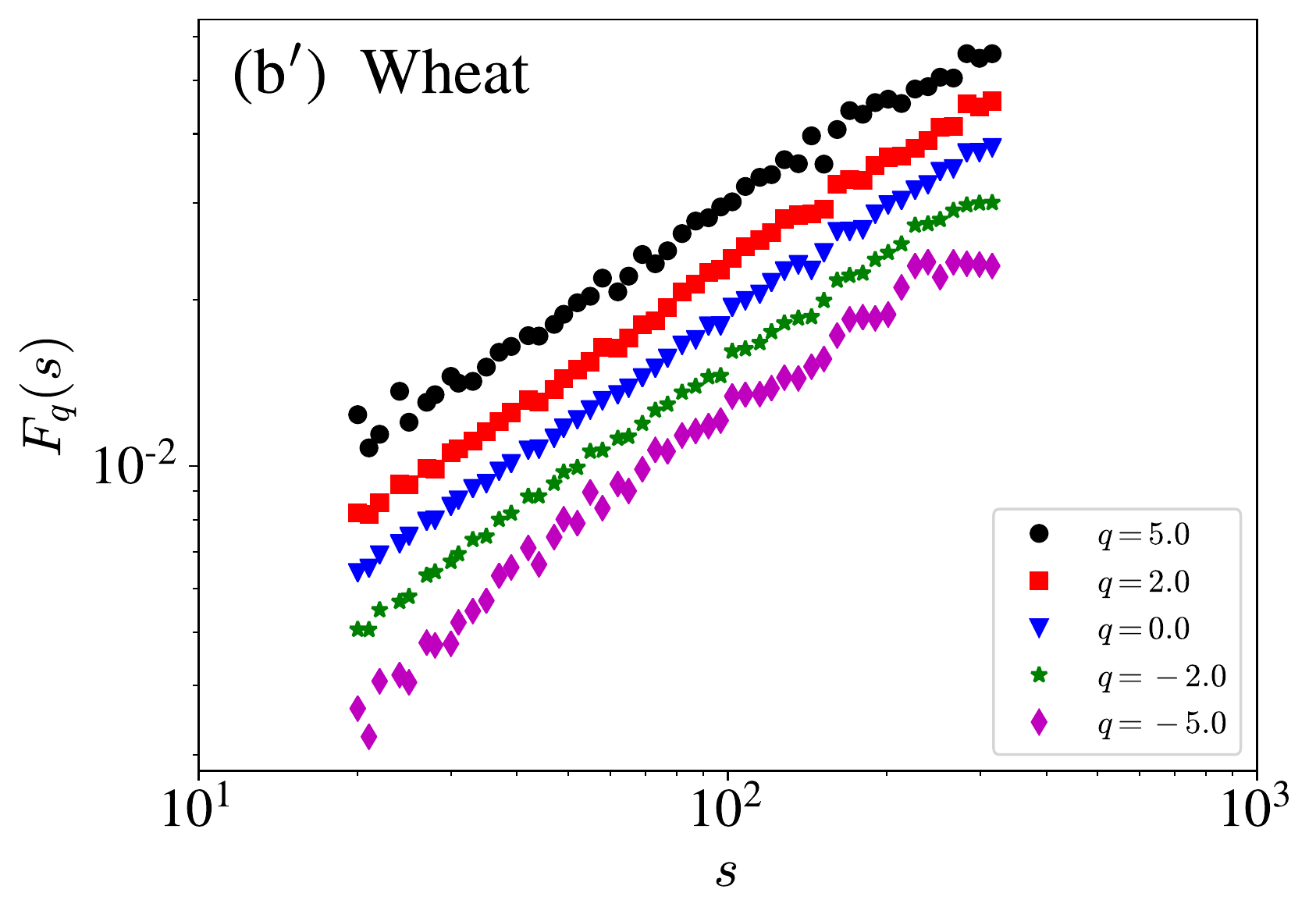}
    \includegraphics[width=0.325\linewidth]{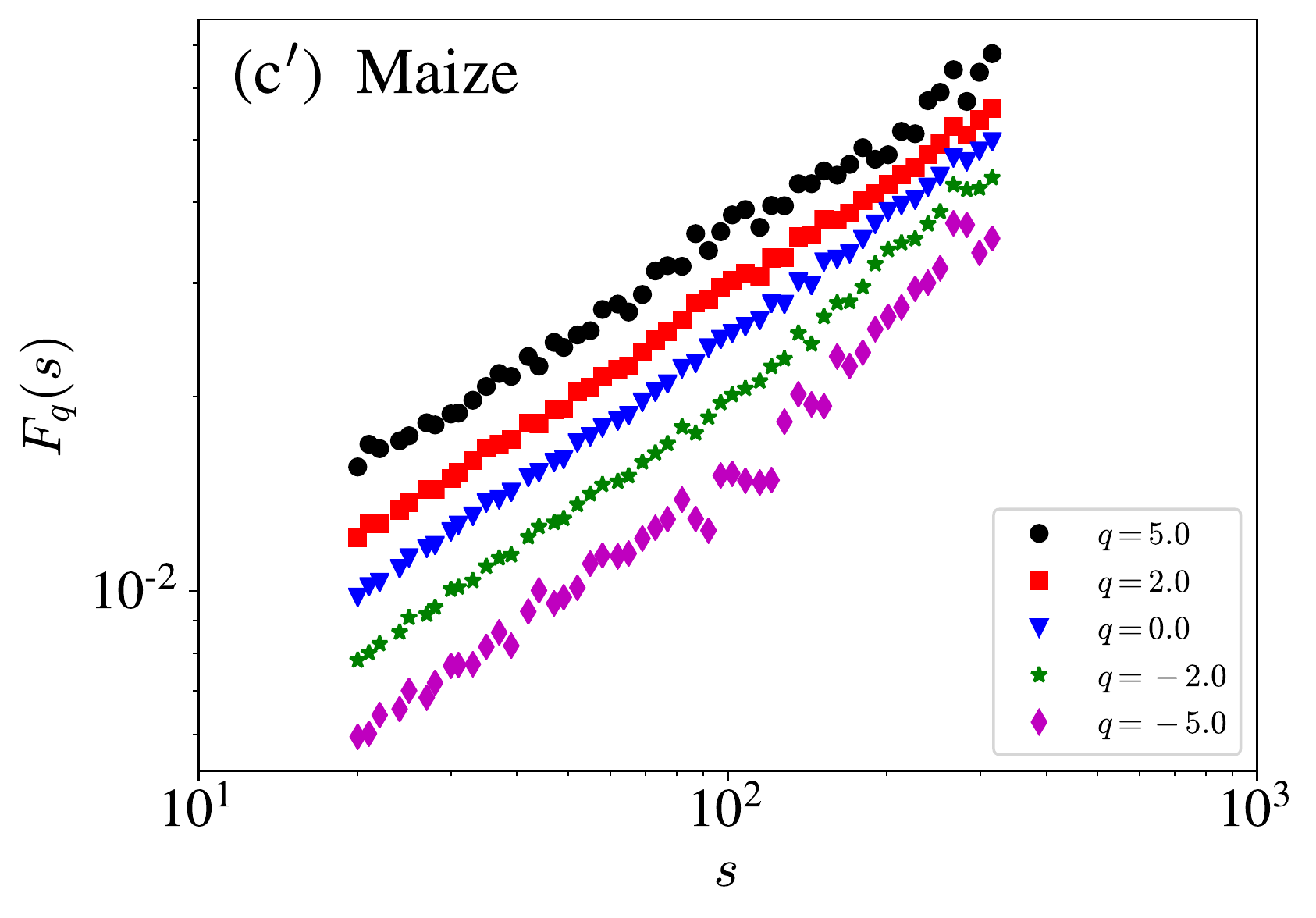}
    \includegraphics[width=0.325\linewidth]{Fig_GOI_Return_MFDFA1_Fs_Soyabeans.pdf}
    \includegraphics[width=0.325\linewidth]{Fig_GOI_Return_MFDFA1_Fs_Rice.pdf}
    \includegraphics[width=0.325\linewidth]{Fig_GOI_Return_MFDFA1_Fs_Barley.pdf}
    \includegraphics[width=0.325\linewidth]{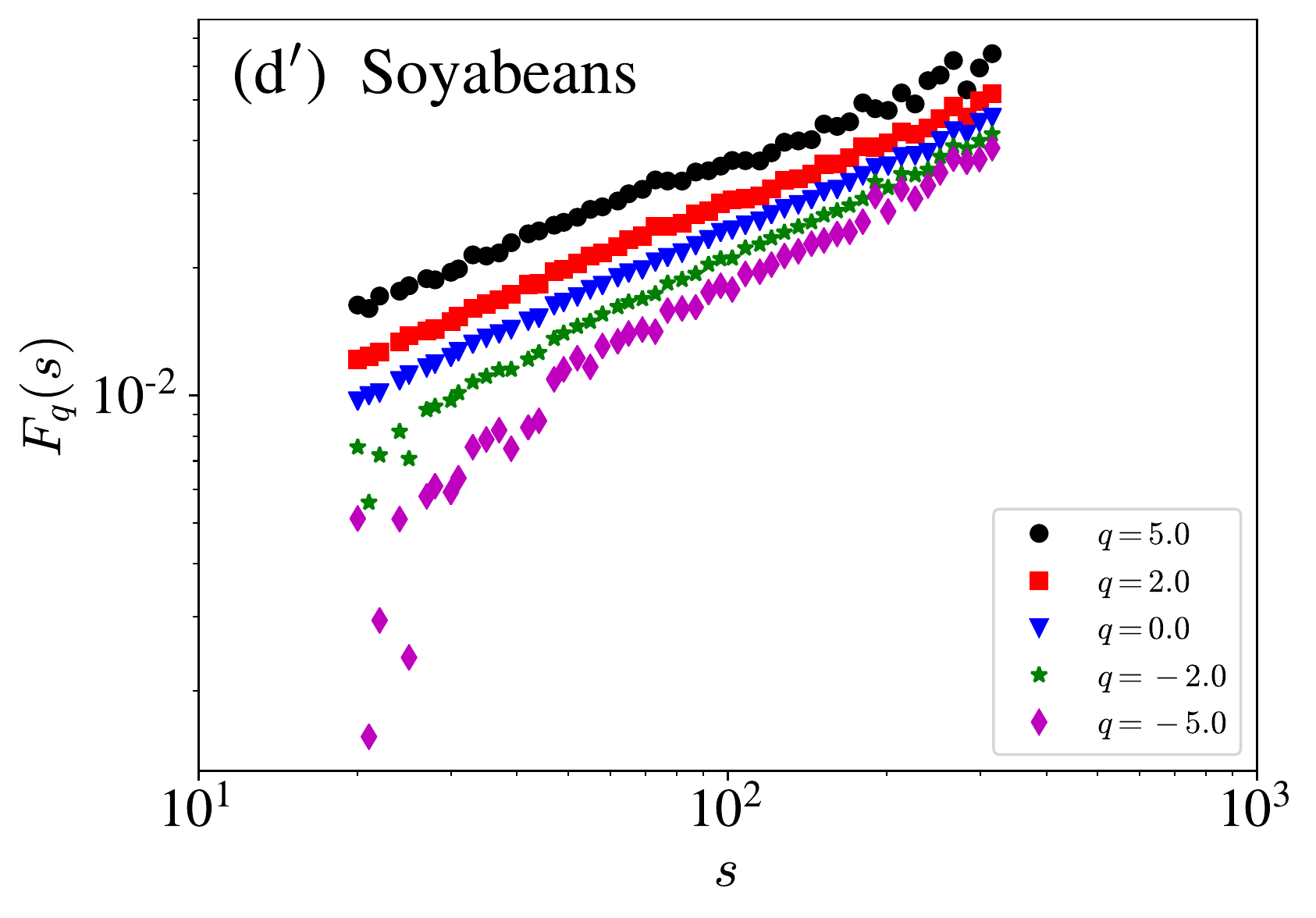}
    \includegraphics[width=0.325\linewidth]{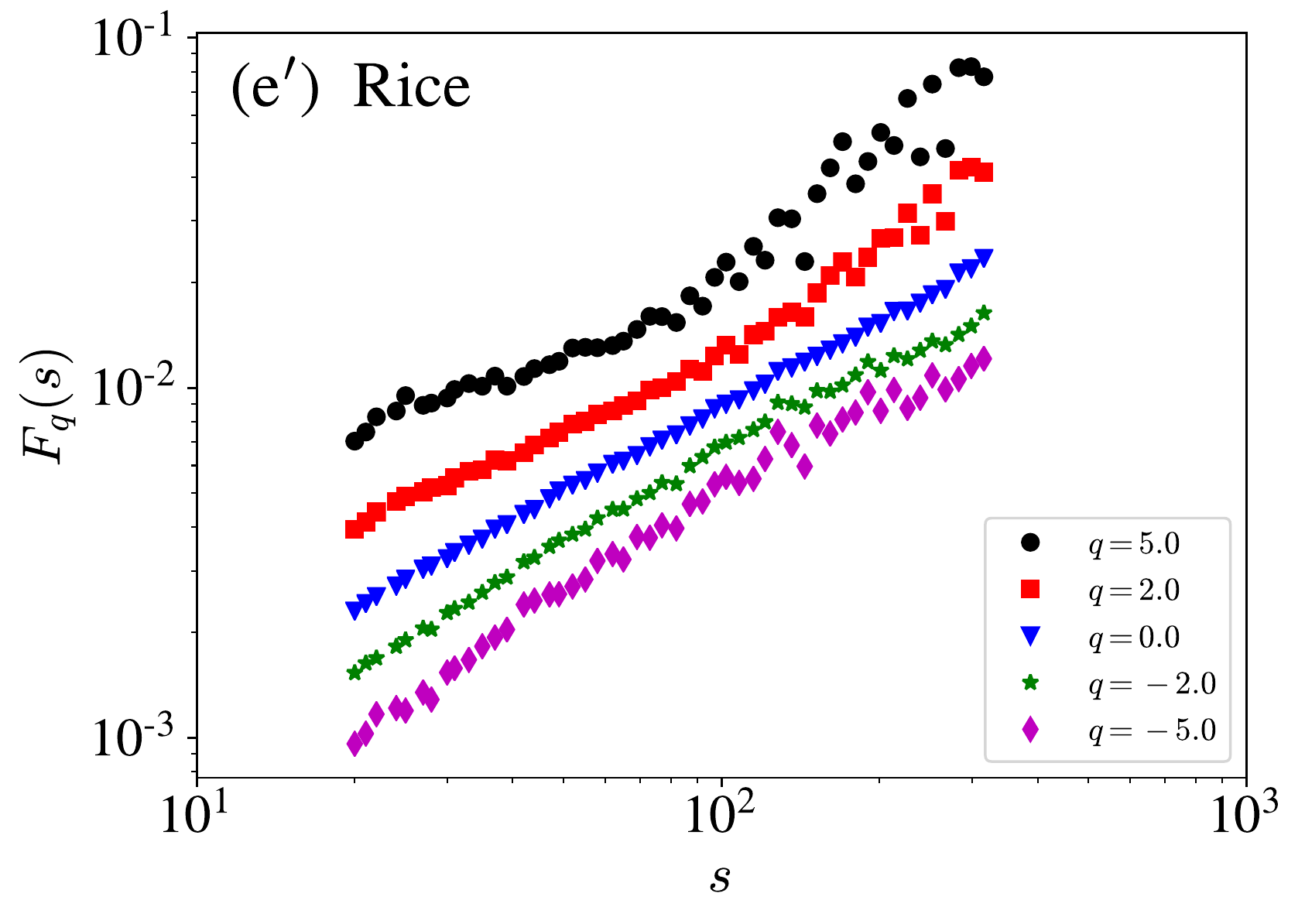}
    \includegraphics[width=0.325\linewidth]{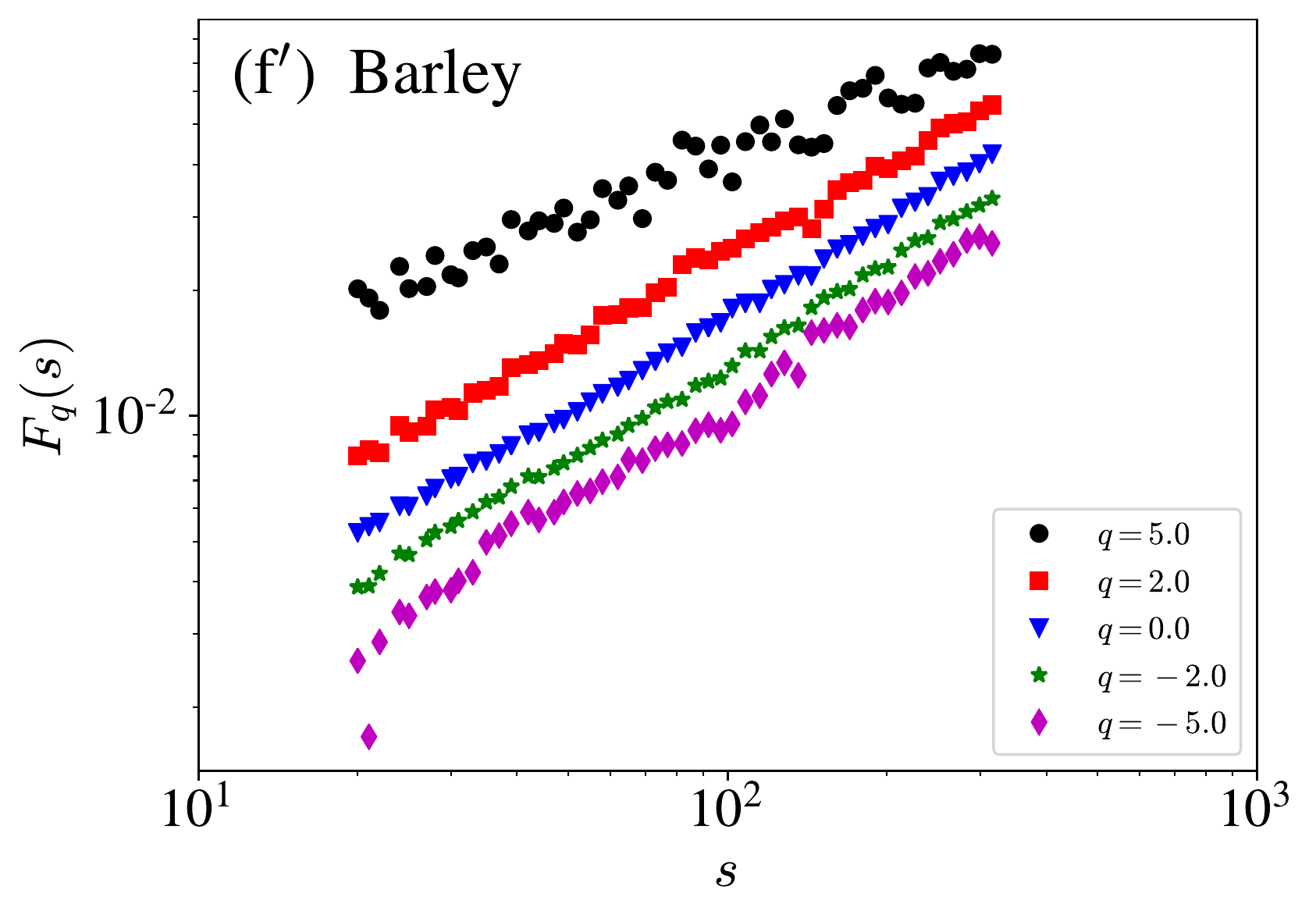}
    \caption{Scaling plots of the MF-DMA fluctuation function $F_q(s)$ with respect to the scale $s$ for the GOI index [(a) for $\ell=1$ and (a$^\prime$) for $\ell=2$], the wheat sub-index [(b) for $\ell=1$ and (b$^\prime$) for $\ell=2$], the maize sub-index [(c) for $\ell=1$ and (c$^\prime$) for $\ell=2$], the soyabeans sub-index [(d) for $\ell=1$ and (d$^\prime$) for $\ell=2$], the rice sub-index [(e) for $\ell=1$ and (e$^\prime$) for $\ell=2$], and the barley sub-index [(f) for $\ell=1$ and (f$^\prime$) for $\ell=2$] released by the International Grains Council.}
    \label{FigA:GOI:Return:MFDFA:Fs}
\end{figure}

\begin{figure}[!ht]
    \centering
    \includegraphics[width=0.325\linewidth]{Fig_GOI_Return_MFDFA1_Hq_GOI.pdf}
    \includegraphics[width=0.325\linewidth]{Fig_GOI_Return_MFDFA1_Hq_Wheat.pdf}
    \includegraphics[width=0.325\linewidth]{Fig_GOI_Return_MFDFA1_Hq_Maize.pdf}
    \includegraphics[width=0.325\linewidth]{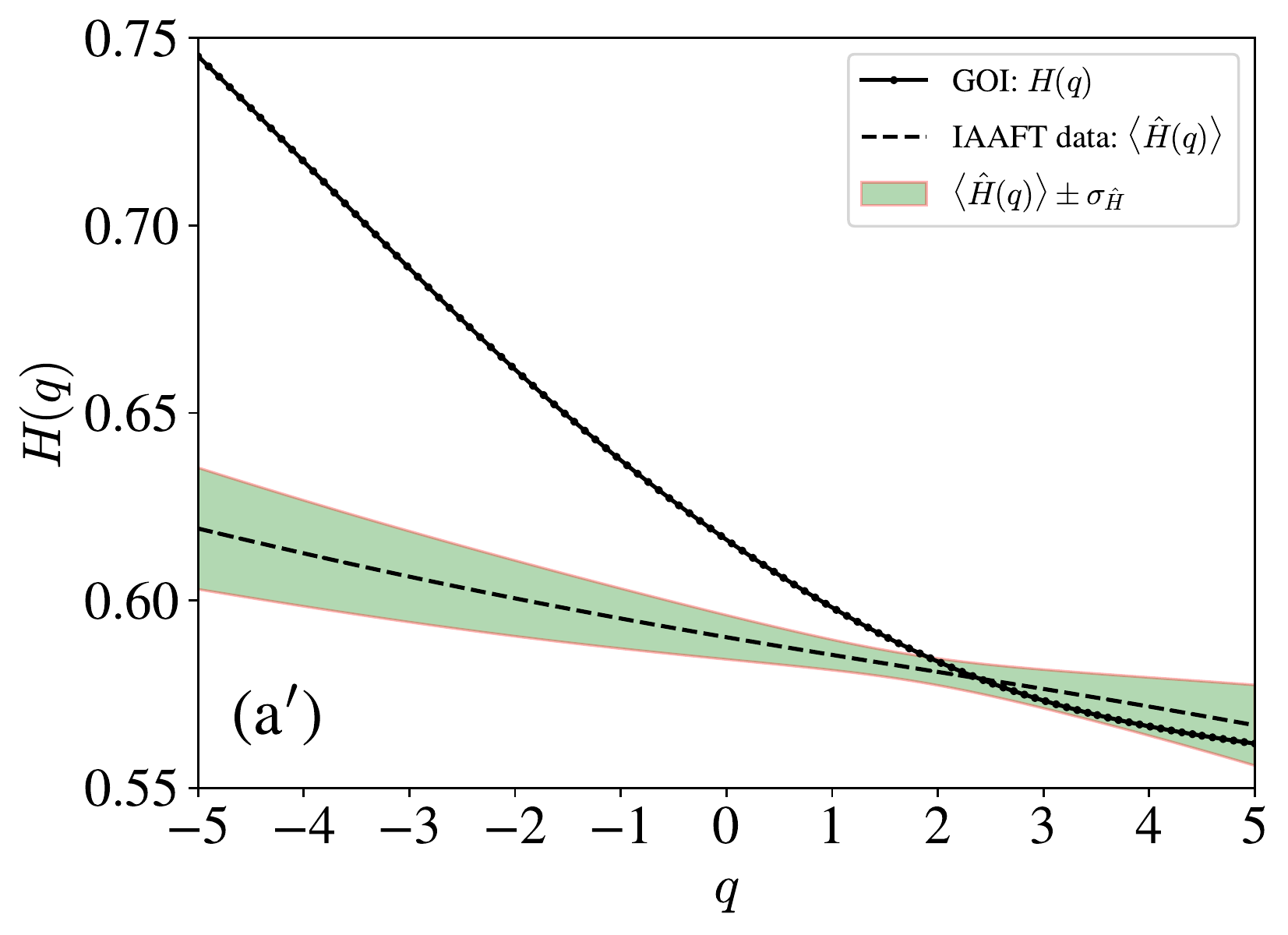}
    \includegraphics[width=0.325\linewidth]{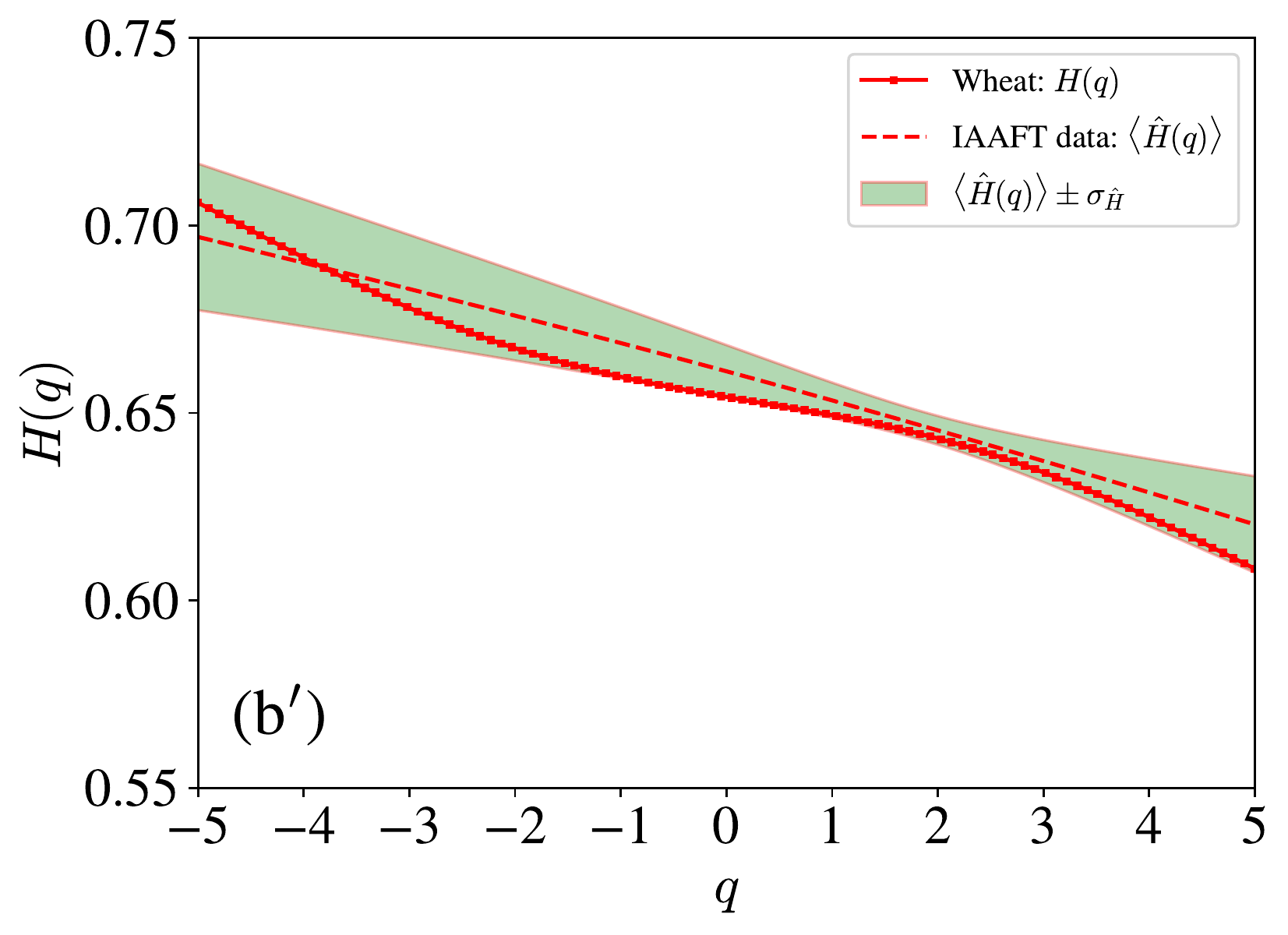}
    \includegraphics[width=0.325\linewidth]{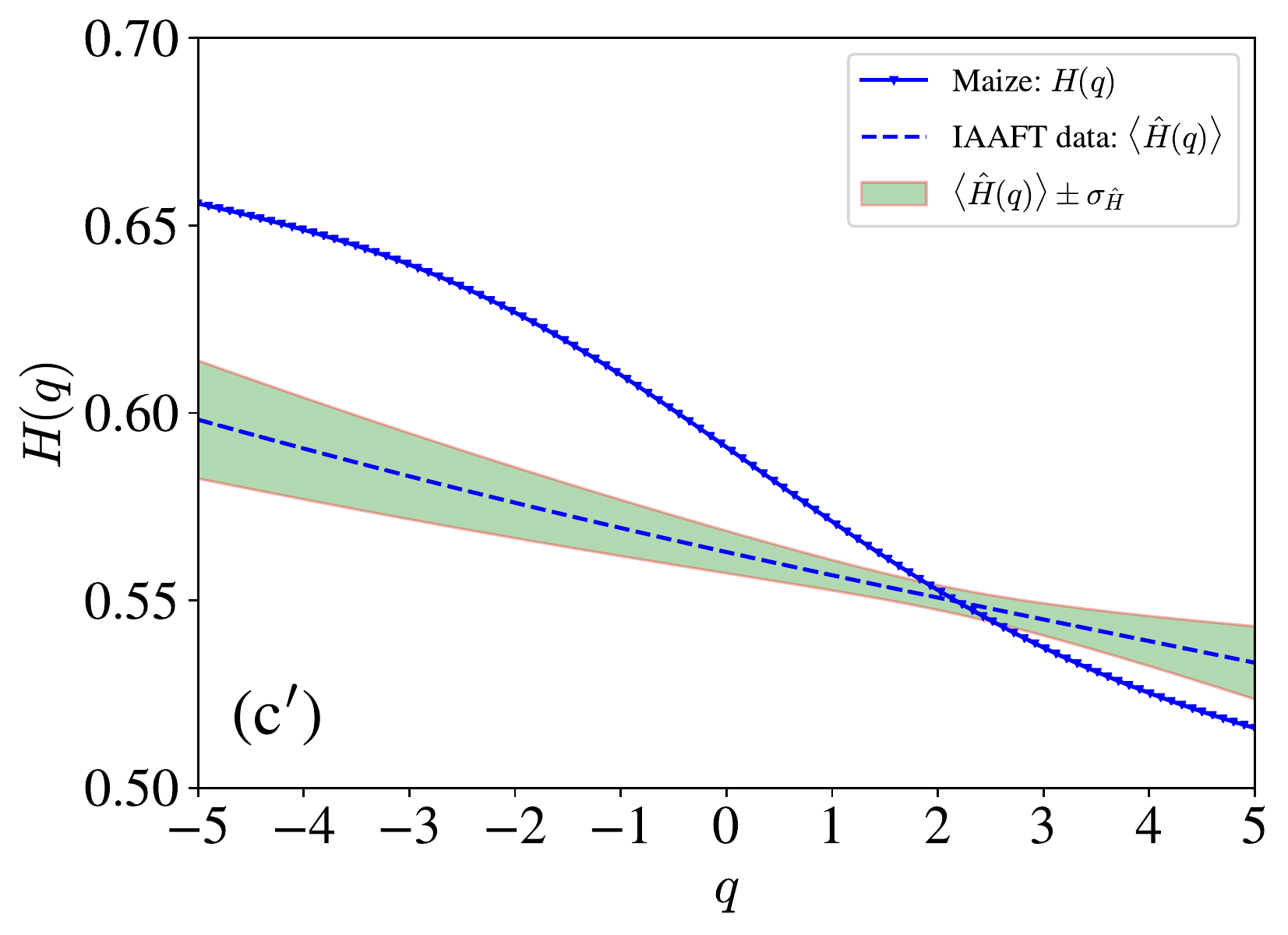}
    \includegraphics[width=0.325\linewidth]{Fig_GOI_Return_MFDFA1_Hq_Soyabeans.pdf}
    \includegraphics[width=0.325\linewidth]{Fig_GOI_Return_MFDFA1_Hq_Rice.pdf}
    \includegraphics[width=0.325\linewidth]{Fig_GOI_Return_MFDFA1_Hq_Barley.pdf}
    \includegraphics[width=0.325\linewidth]{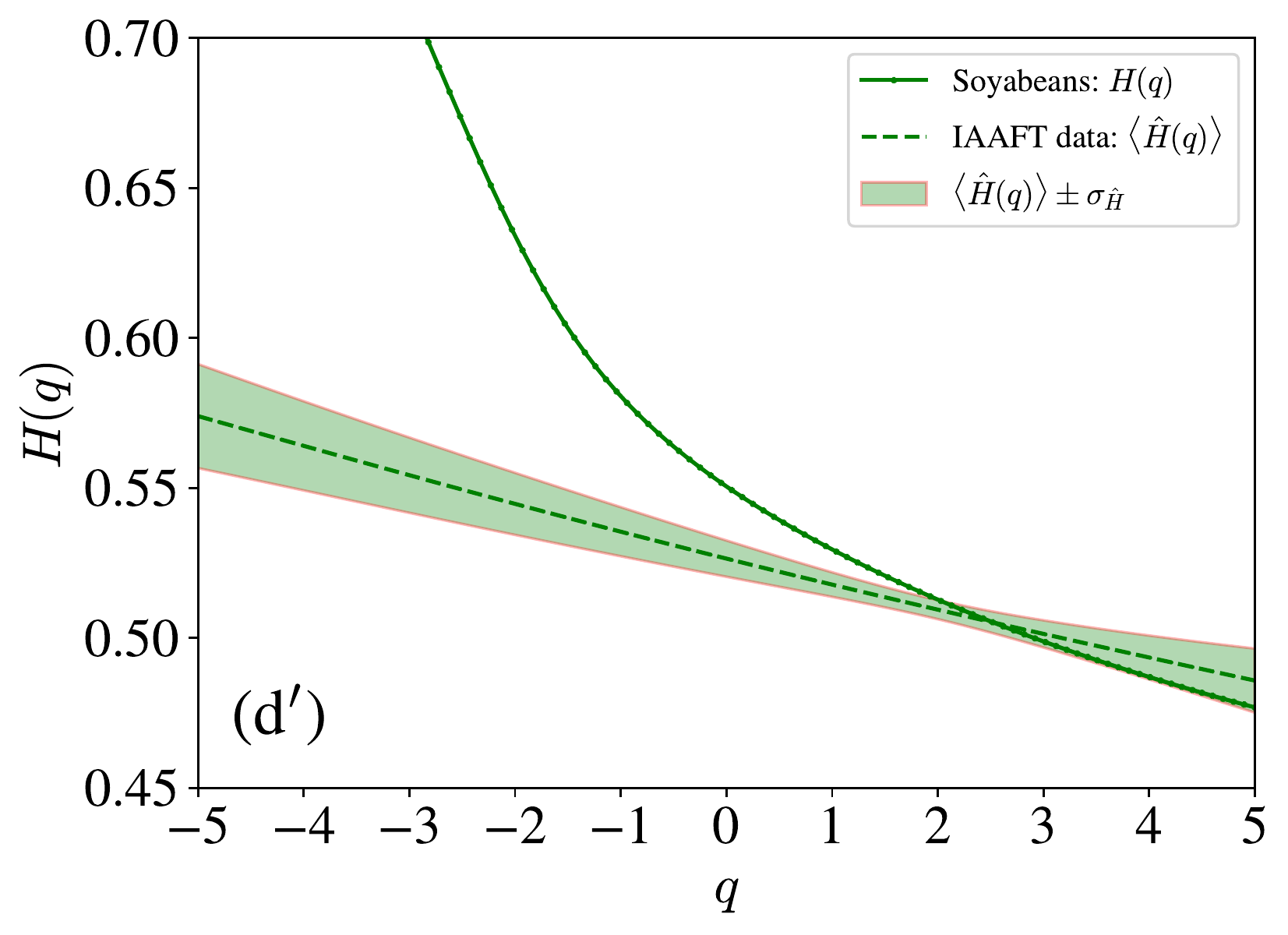}
    \includegraphics[width=0.325\linewidth]{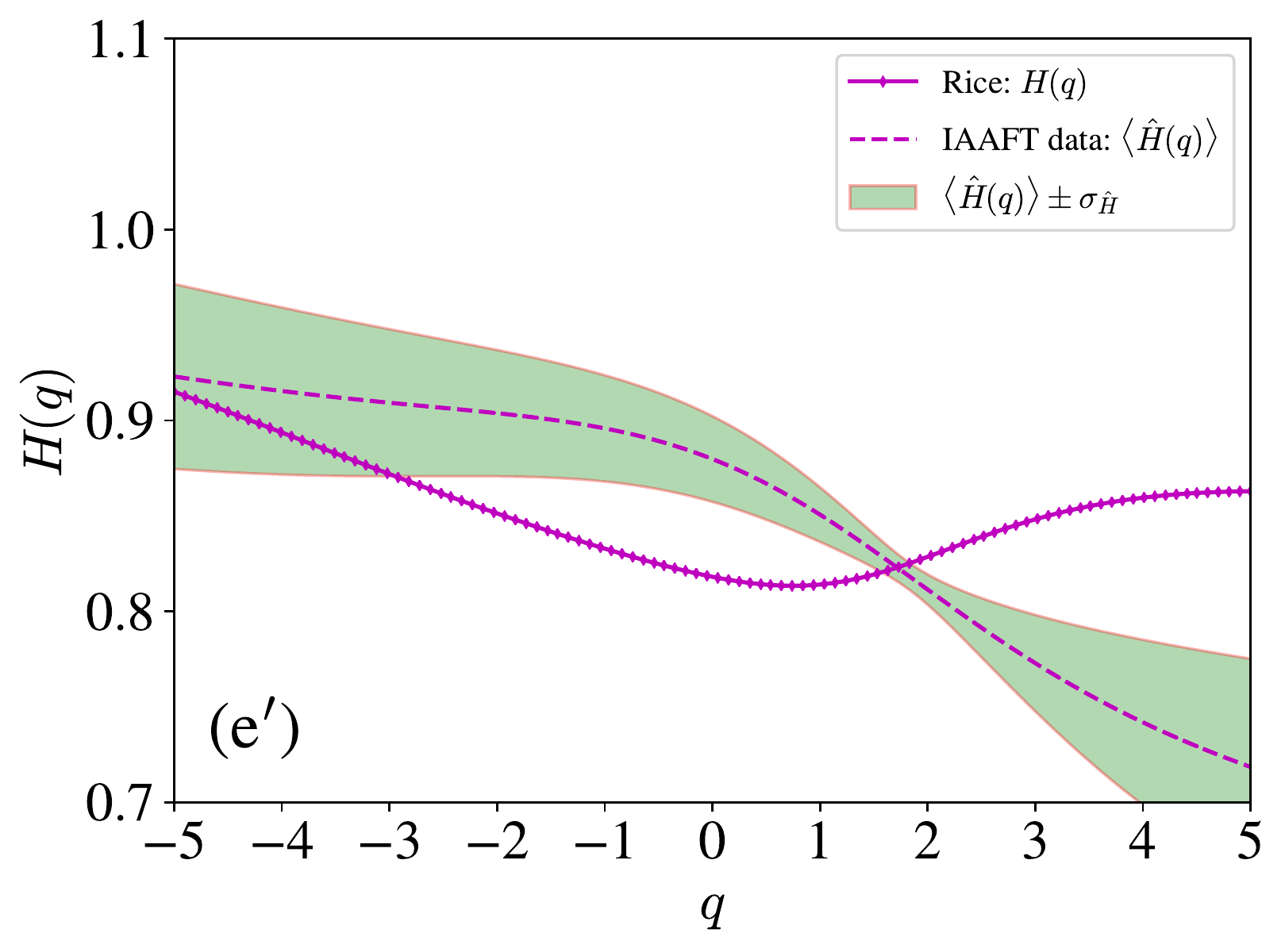}
    \includegraphics[width=0.325\linewidth]{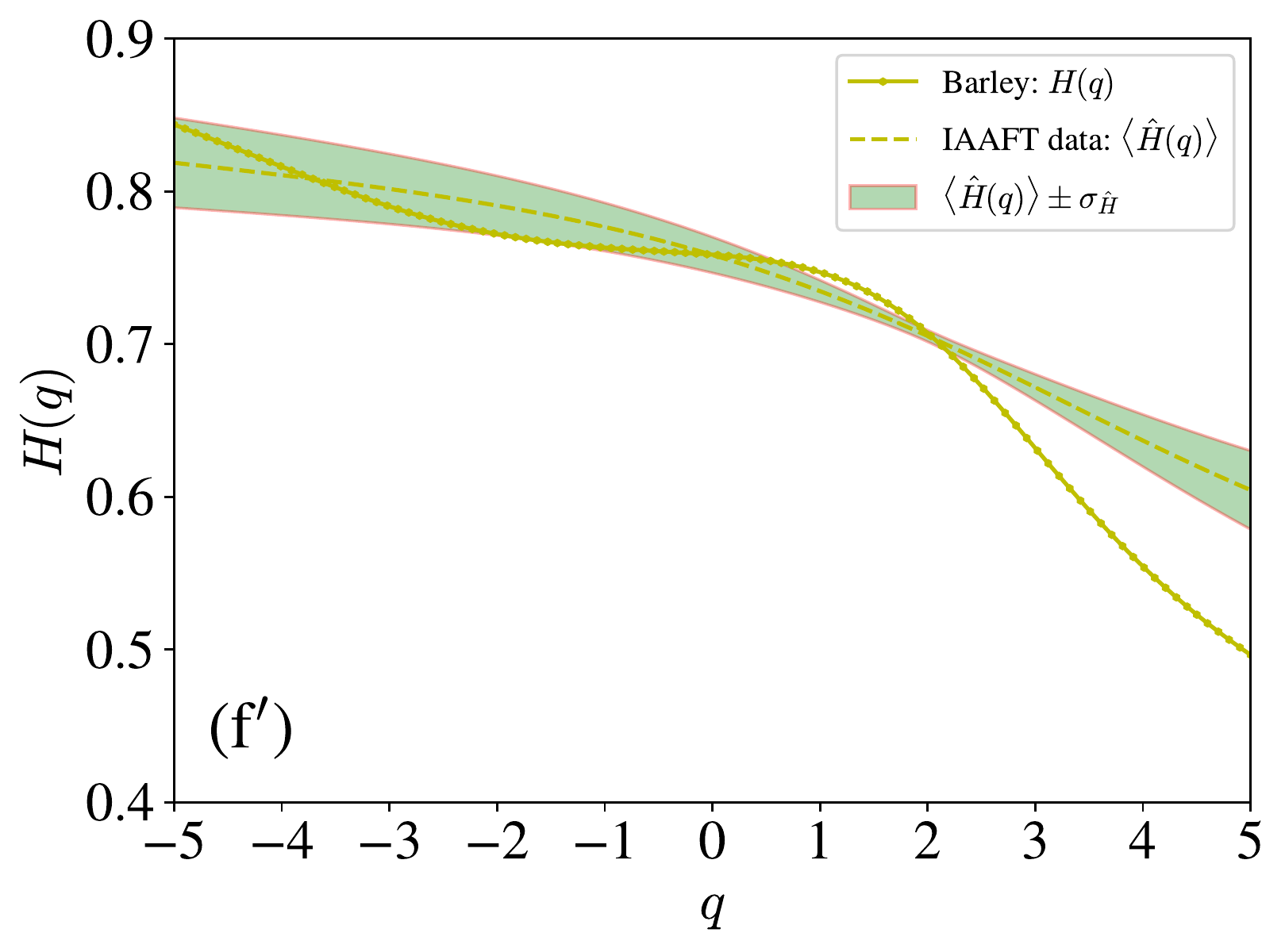}
    \caption{Generalized Hurst indexes $H(q)$ with respect to the order $q$ for the GOI index [(a) for $\ell=1$ and (a$^\prime$) for $\ell=2$], the wheat sub-index [(b) for $\ell=1$ and (b$^\prime$) for $\ell=2$], the maize sub-index [(c) for $\ell=1$ and (c$^\prime$) for $\ell=2$], the soyabeans sub-index [(d) for $\ell=1$ and (d$^\prime$) for $\ell=2$], the rice sub-index [(e) for $\ell=1$ and (e$^\prime$) for $\ell=2$], and the barley sub-index [(f) for $\ell=1$ and (f$^\prime$) for $\ell=2$] released by the International Grains Council. For each GOI and sub-indices, we generate 1000 surrogate time series using the IAAFT algorithm and calculate the mean $\langle\hat{H}\rangle$ and standard deviation $\sigma_{\hat{H}}$.}
    \label{FigA:GOI:Return:MFDFA:Hq}
\end{figure}

\begin{figure}[!ht]
    \centering
    \includegraphics[width=0.325\linewidth]{Fig_GOI_Return_MFDFA1_tau_GOI.pdf}
    \includegraphics[width=0.325\linewidth]{Fig_GOI_Return_MFDFA1_tau_Wheat.pdf}
    \includegraphics[width=0.325\linewidth]{Fig_GOI_Return_MFDFA1_tau_Maize.pdf}
    \includegraphics[width=0.325\linewidth]{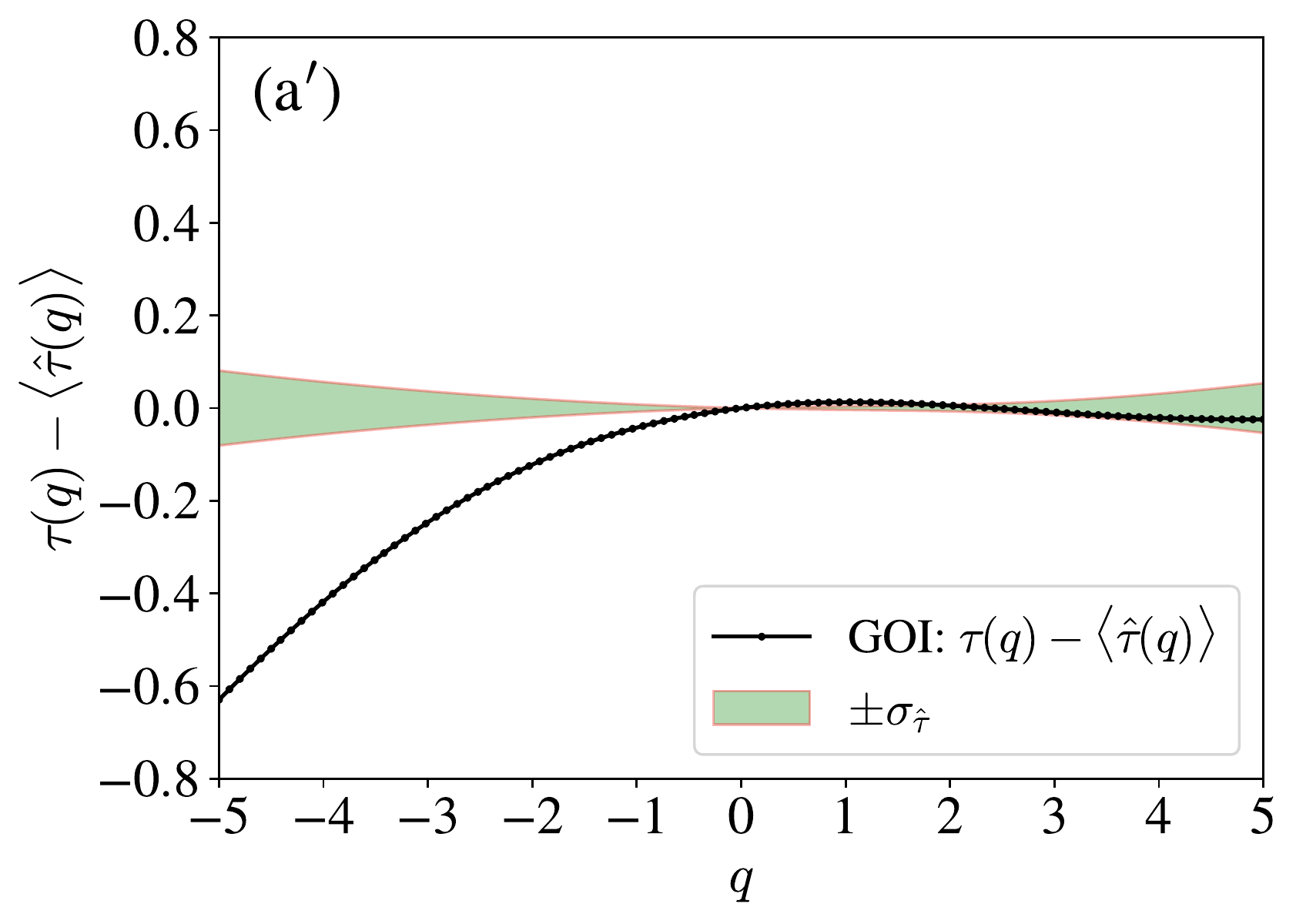}
    \includegraphics[width=0.325\linewidth]{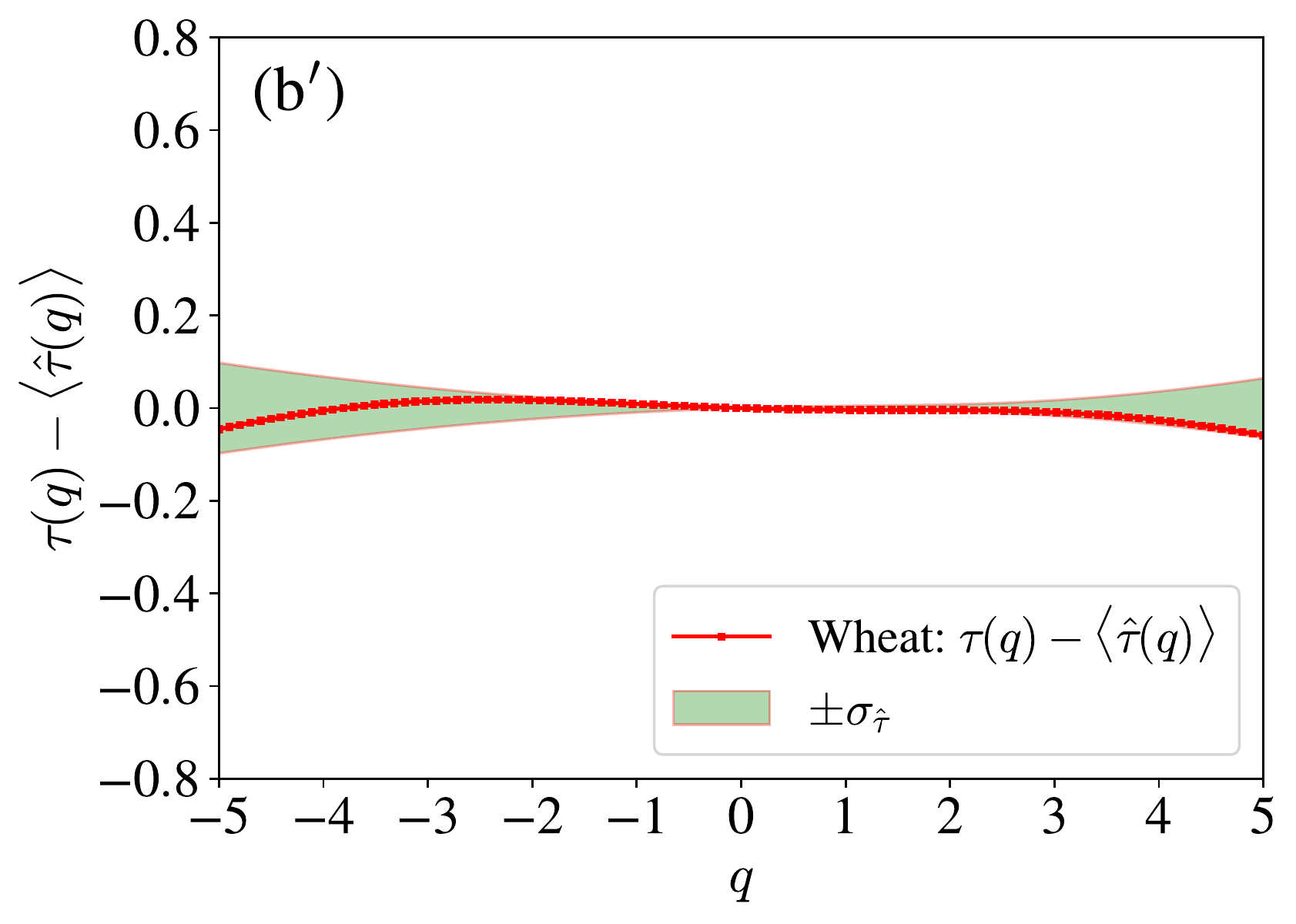}
    \includegraphics[width=0.325\linewidth]{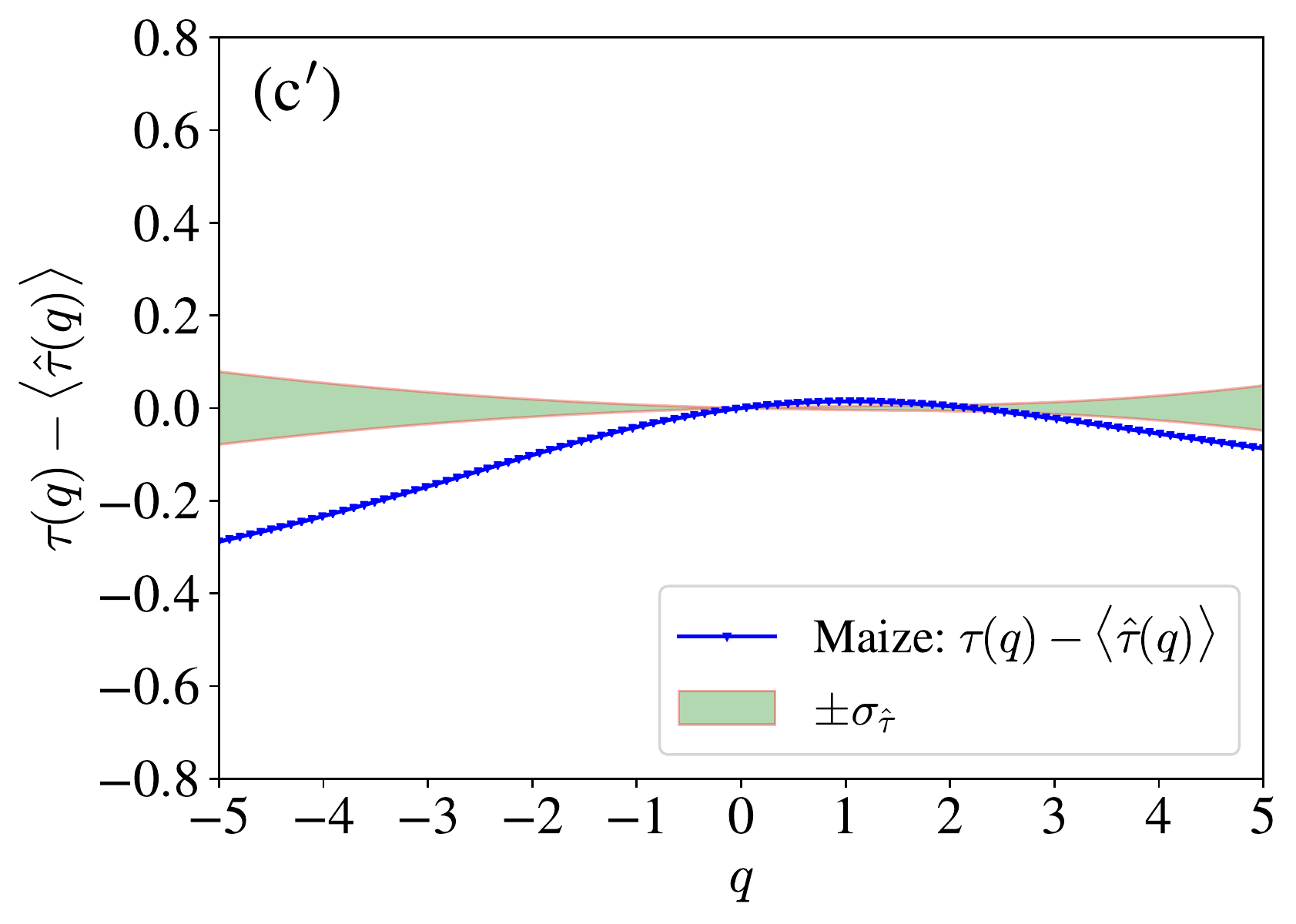}
    \includegraphics[width=0.325\linewidth]{Fig_GOI_Return_MFDFA1_tau_Soyabeans.pdf}
    \includegraphics[width=0.325\linewidth]{Fig_GOI_Return_MFDFA1_tau_Rice.pdf}
    \includegraphics[width=0.325\linewidth]{Fig_GOI_Return_MFDFA1_tau_Barley.pdf}
    \includegraphics[width=0.325\linewidth]{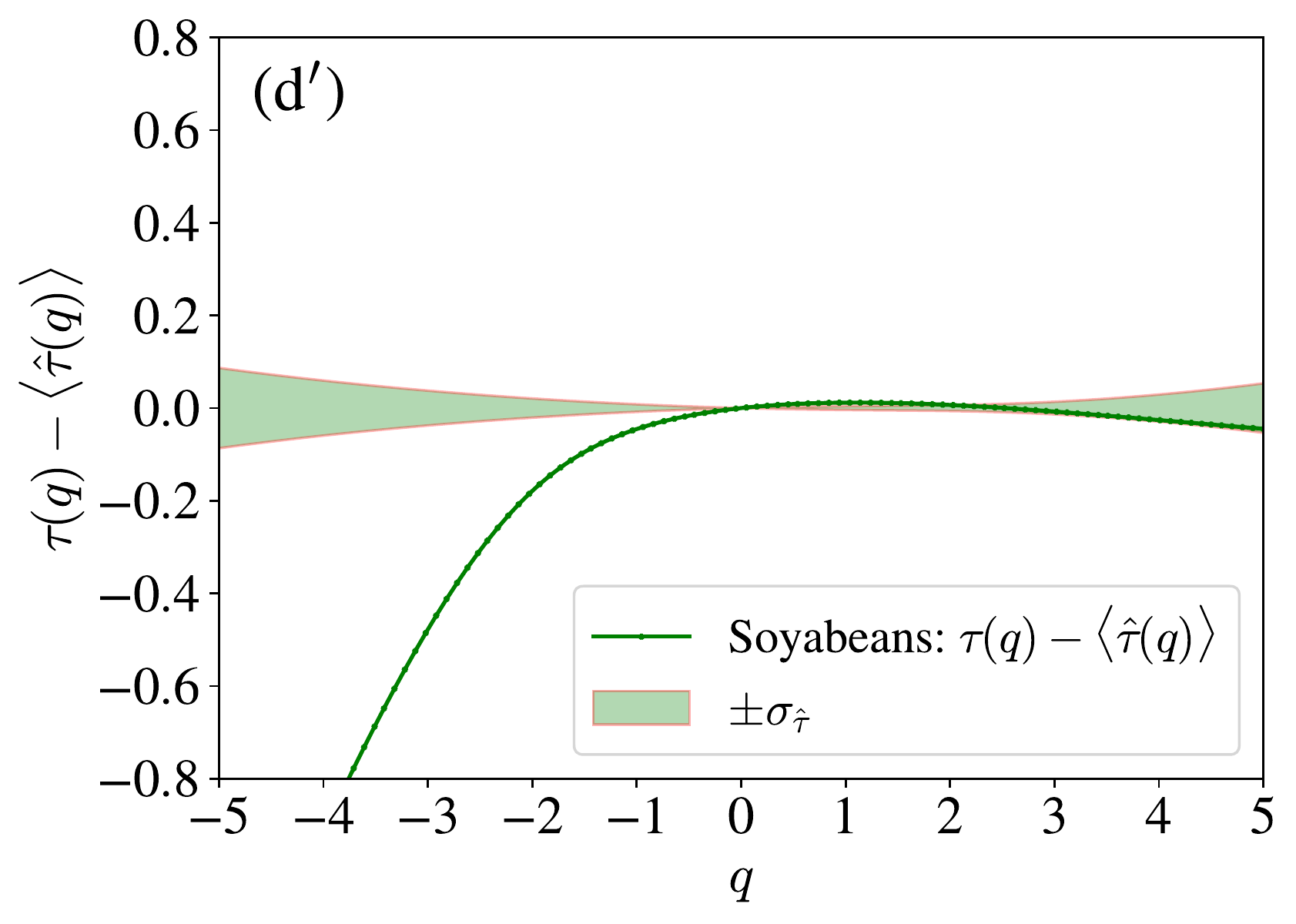}
    \includegraphics[width=0.325\linewidth]{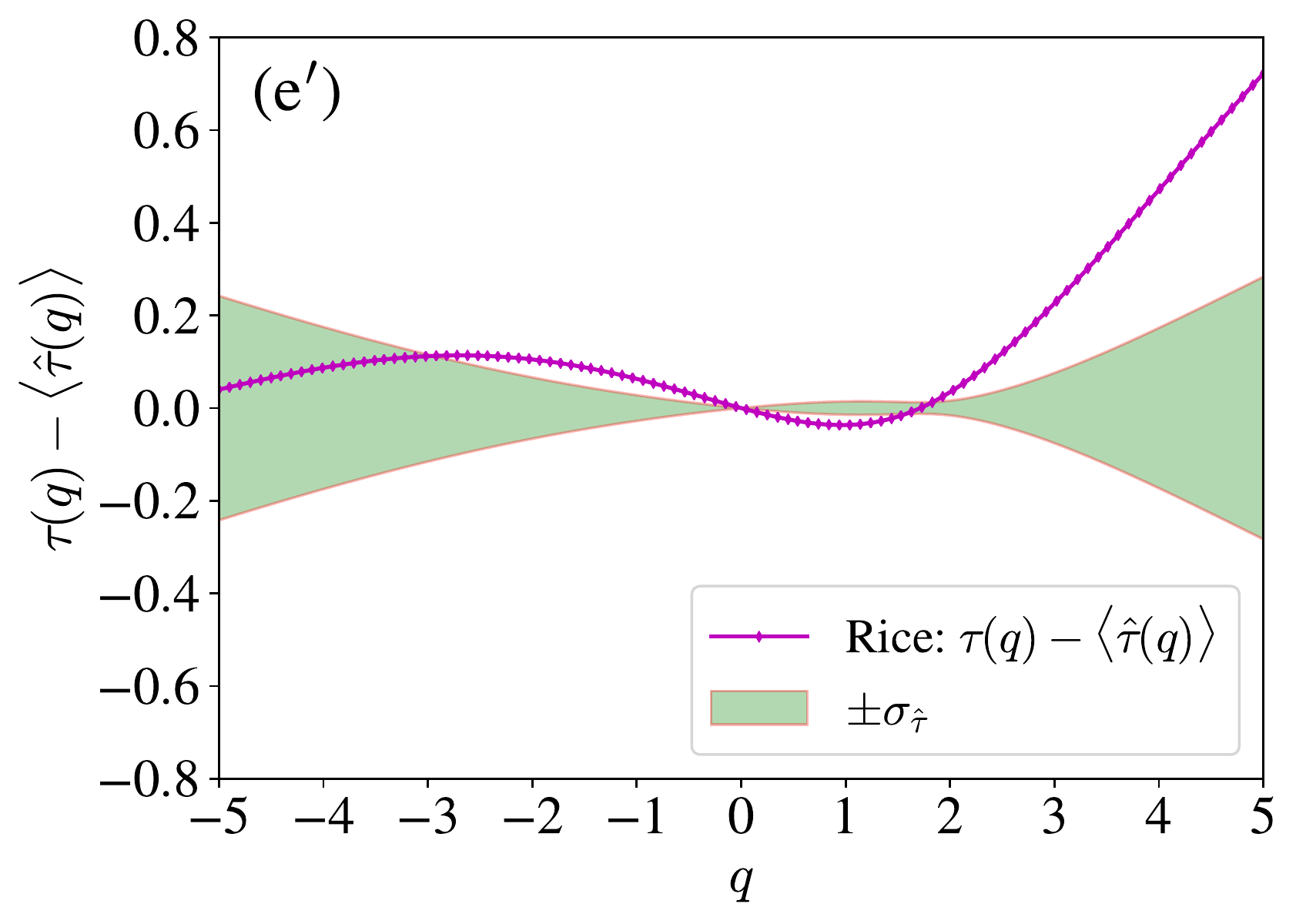}
    \includegraphics[width=0.325\linewidth]{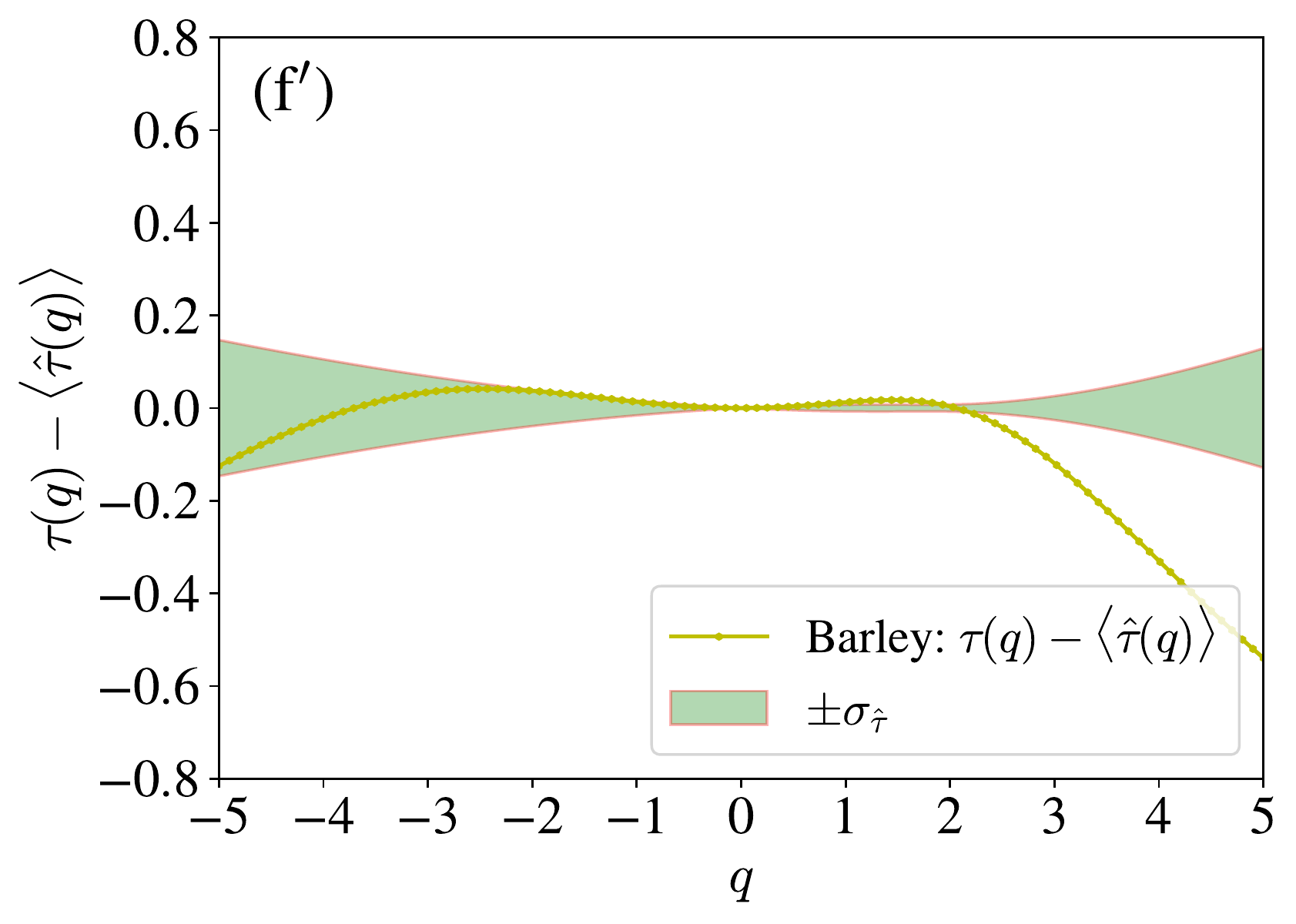}
    \caption{Deviations ($\tau(q)-\langle\hat{\tau}(q)\rangle$) of the mass exponents $\tau(q)$ of the original time series from the average mass exponents $\langle\hat{\tau}(q)\rangle$ of the IAAFT surrogates with respect to the order $q$ for the GOI index [(a) for $\ell=1$ and (a$^\prime$) for $\ell=2$], the wheat sub-index [(b) for $\ell=1$ and (b$^\prime$) for $\ell=2$], the maize sub-index [(c) for $\ell=1$ and (c$^\prime$) for $\ell=2$], the soyabeans sub-index [(d) for $\ell=1$ and (d$^\prime$) for $\ell=2$], the rice sub-index [(e) for $\ell=1$ and (e$^\prime$) for $\ell=2$], and the barley sub-index [(f) for $\ell=1$ and (f$^\prime$) for $\ell=2$] released by the International Grains Council. For each GOI and sub-indices, we generate 1000 surrogate time series using the IAAFT algorithm and calculate the mean $\langle\hat{\tau}\rangle$ and standard deviation $\sigma_{\hat{\tau}}$. }
    \label{FigA:GOI:Return:MFDFA:dtau}
\end{figure}

\begin{figure}[!ht]
    \centering
    \includegraphics[width=0.325\linewidth]{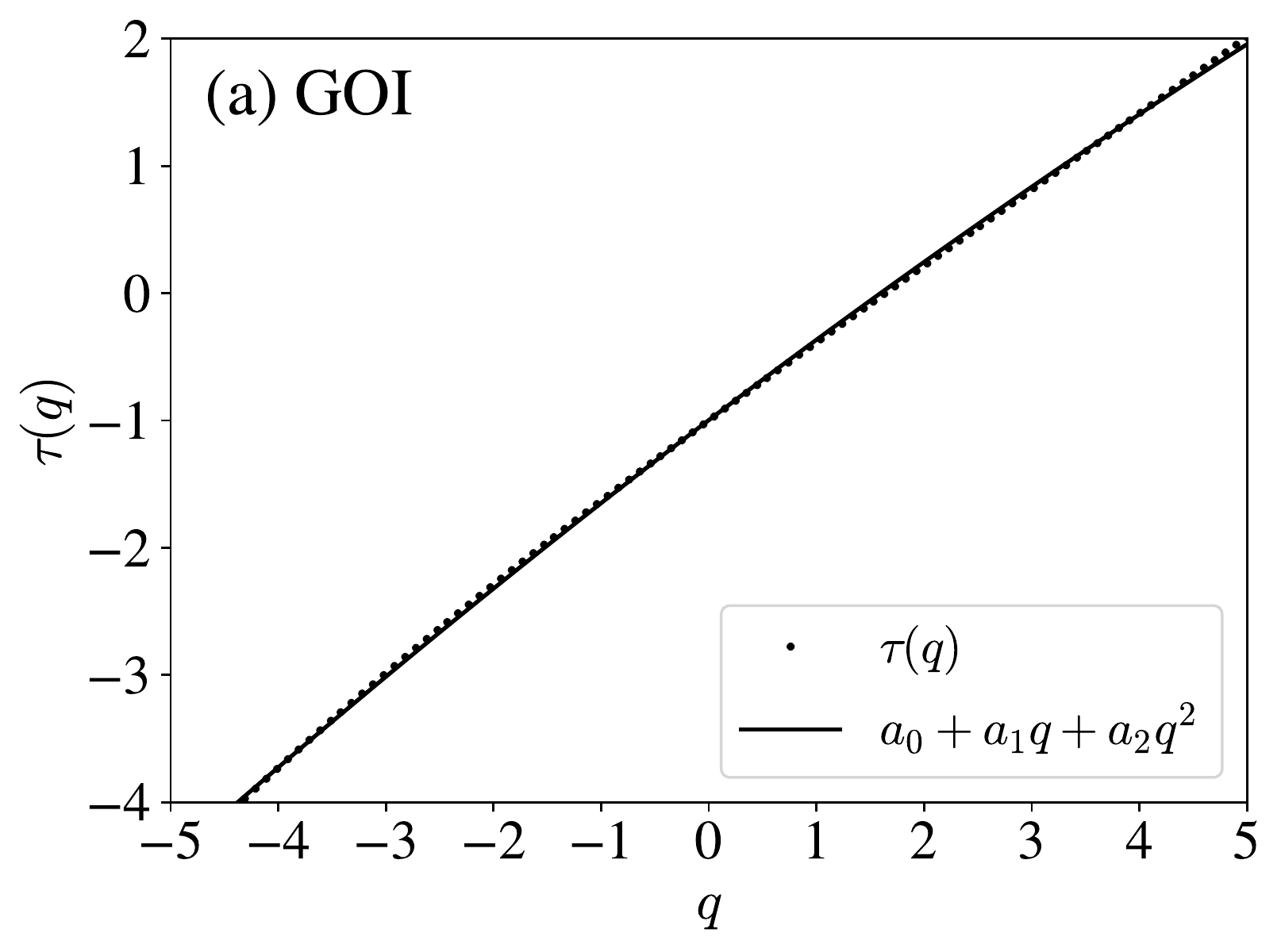}
    \includegraphics[width=0.325\linewidth]{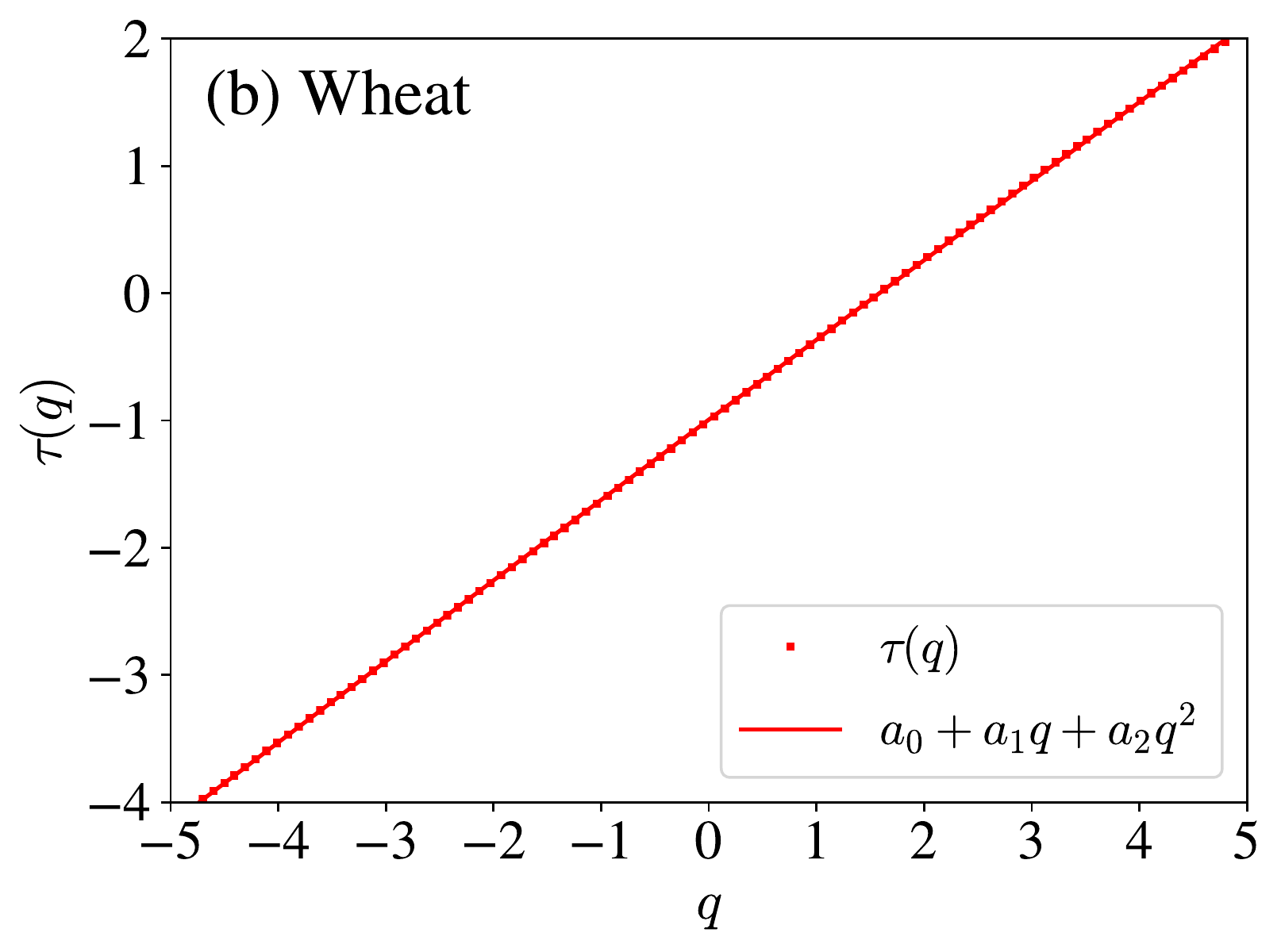}
    \includegraphics[width=0.325\linewidth]{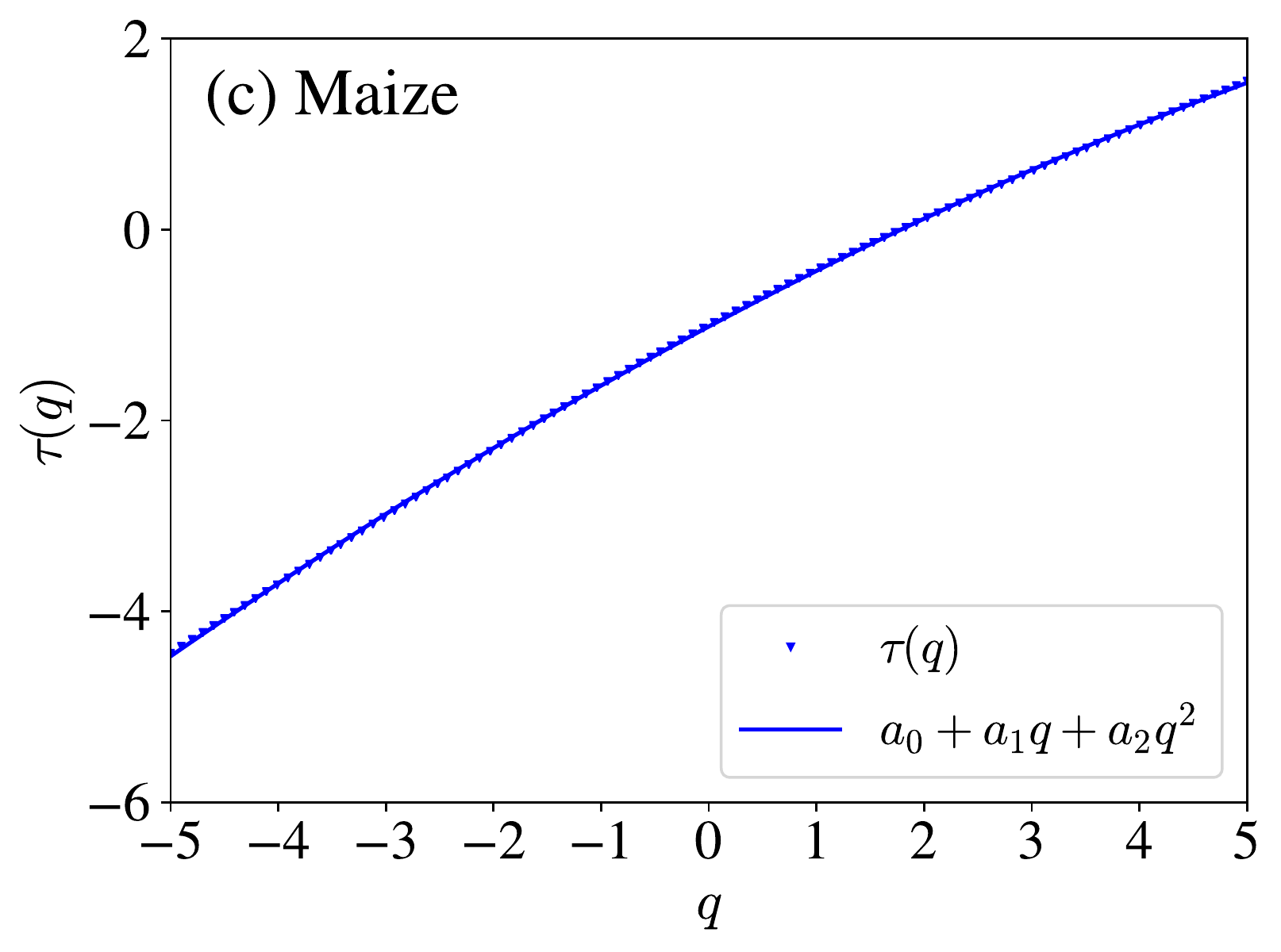}
    \includegraphics[width=0.325\linewidth]{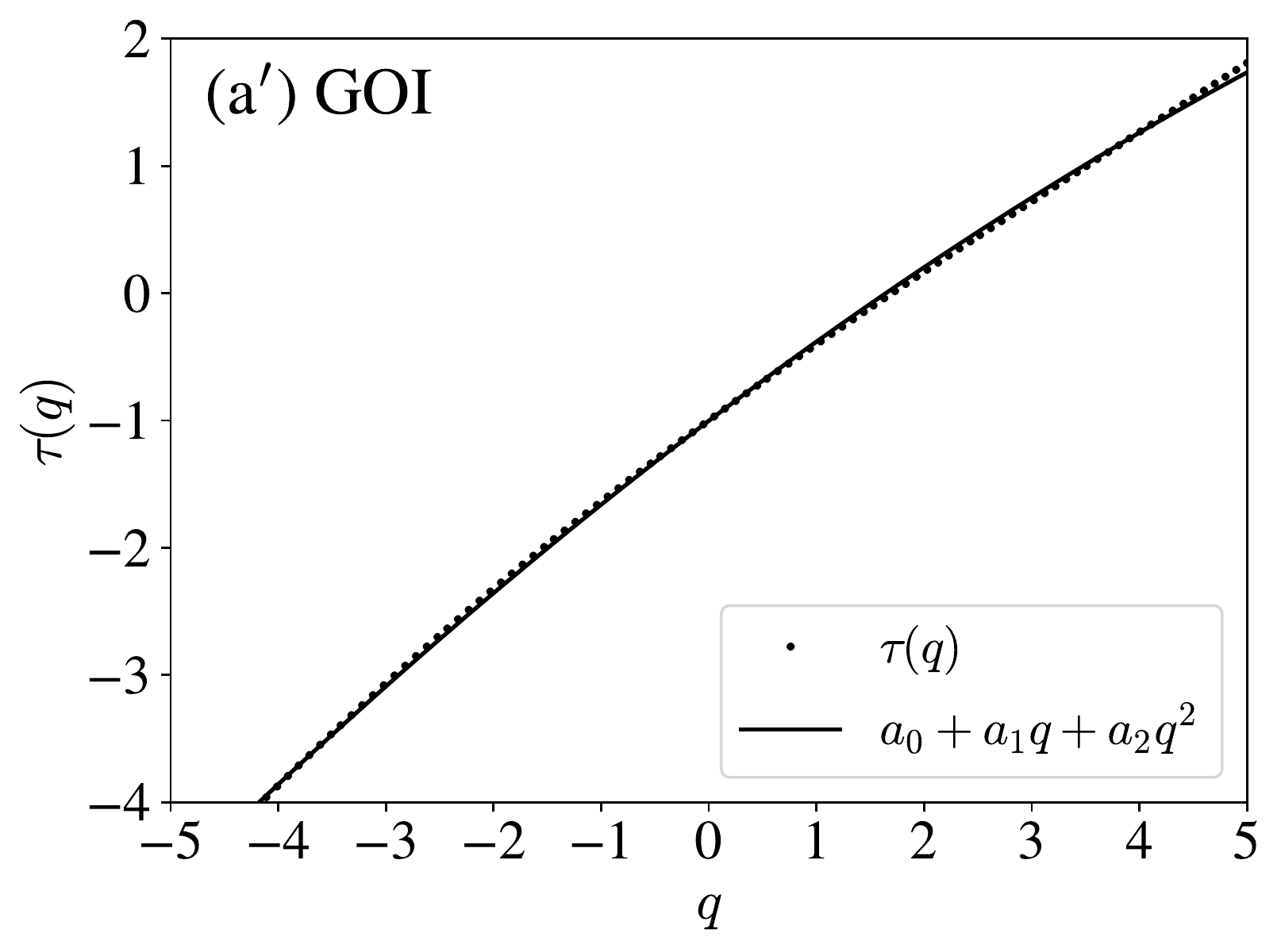}
    \includegraphics[width=0.325\linewidth]{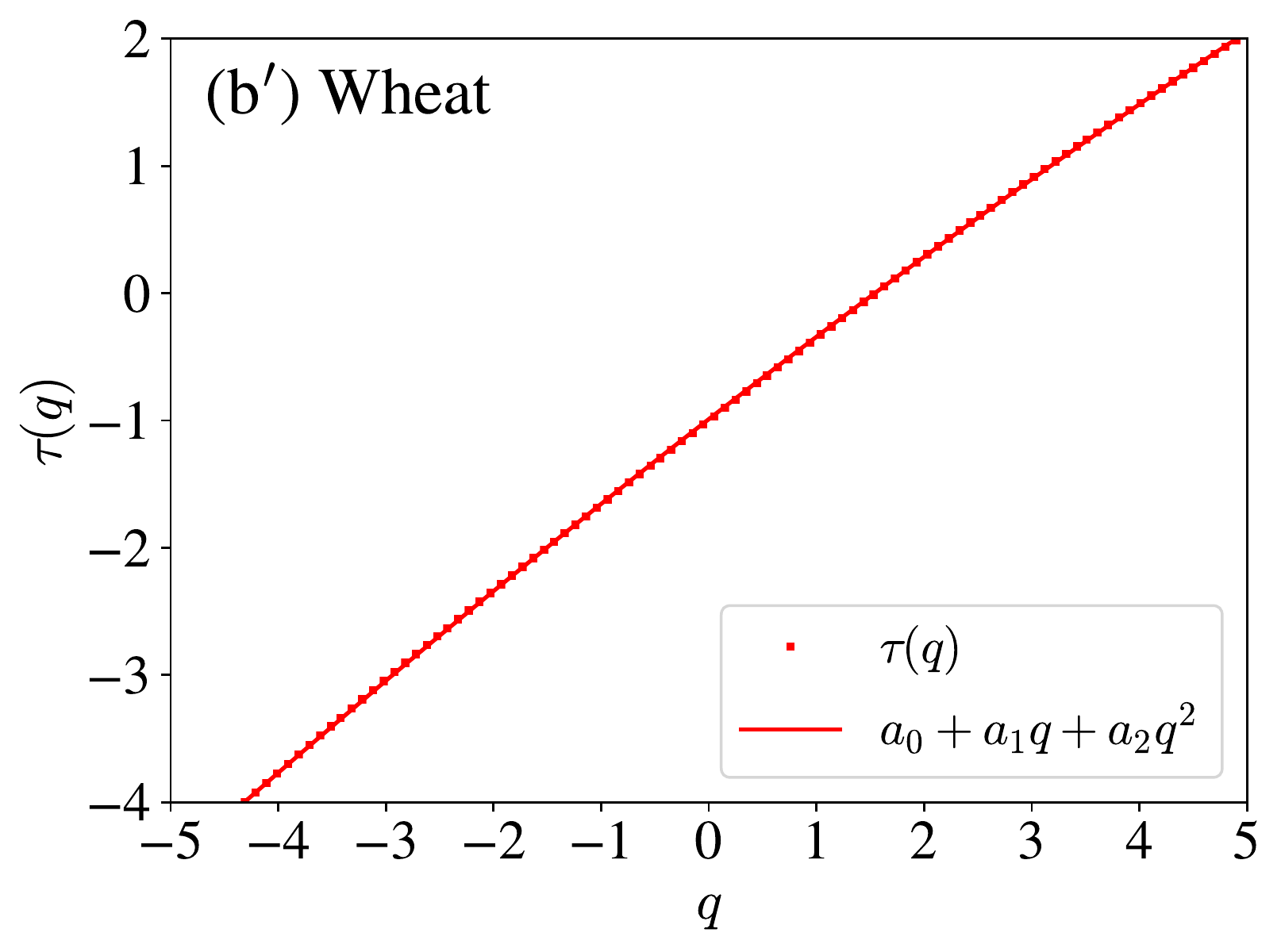}
    \includegraphics[width=0.325\linewidth]{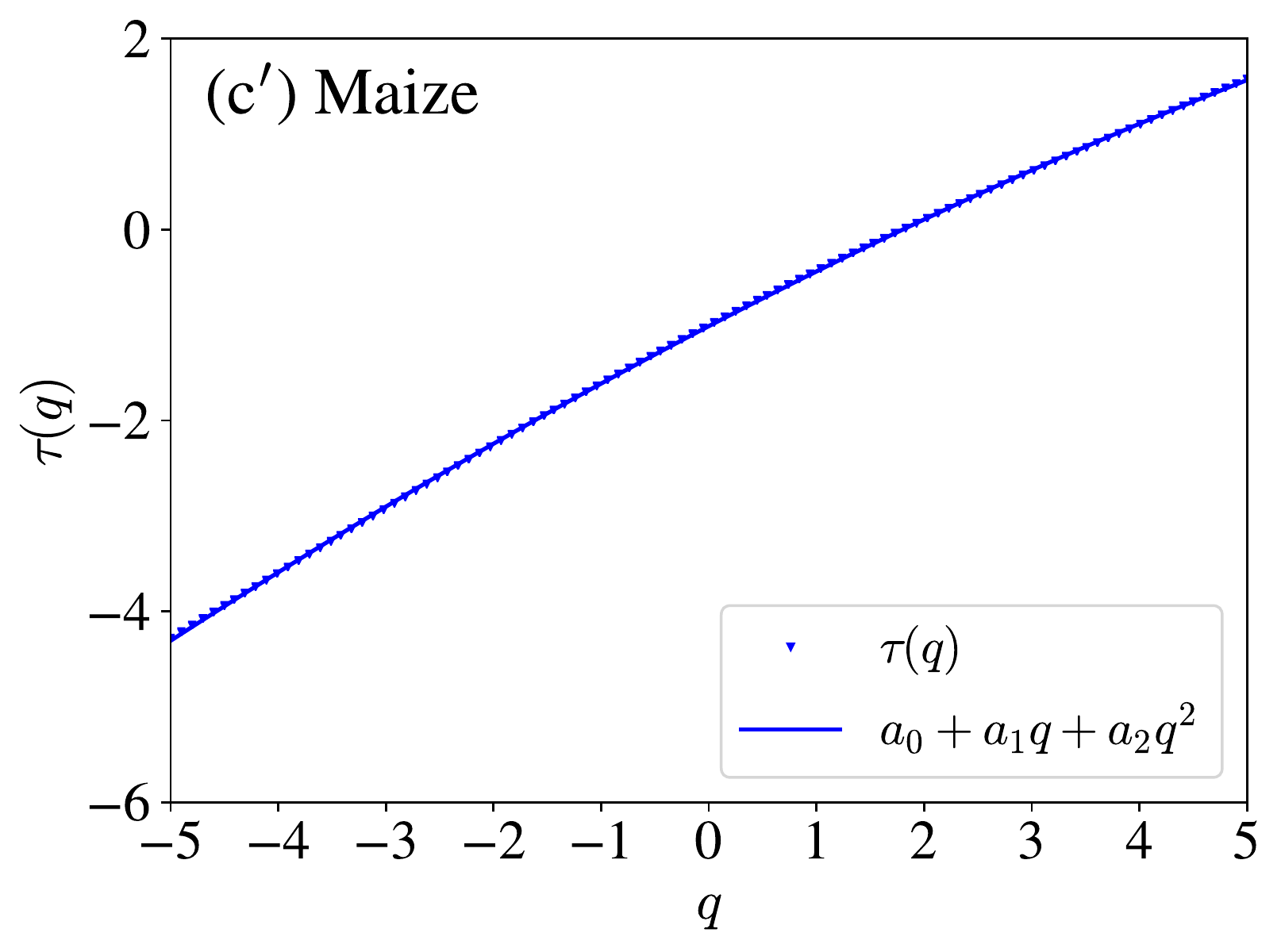}
    \includegraphics[width=0.325\linewidth]{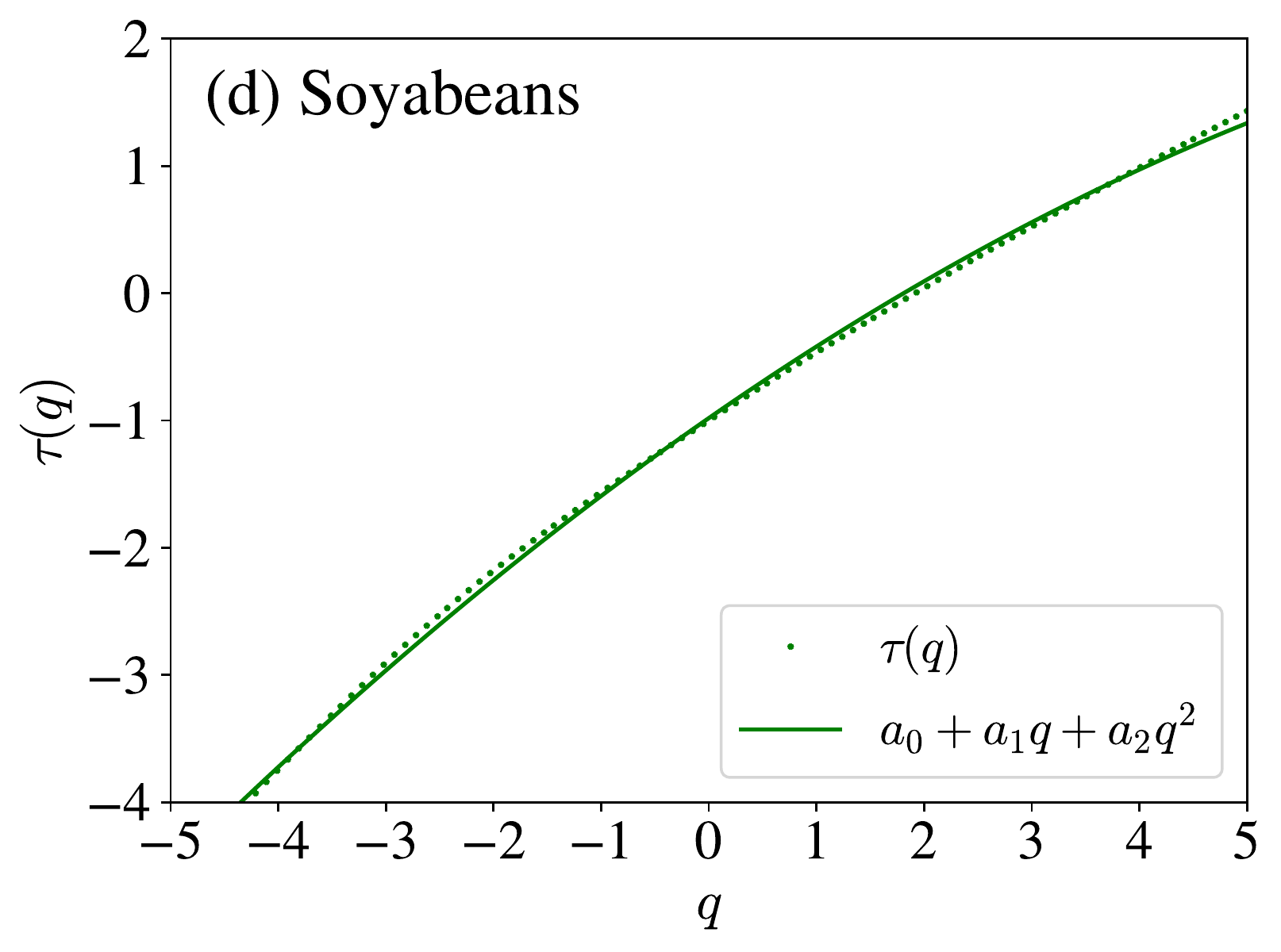}
    \includegraphics[width=0.325\linewidth]{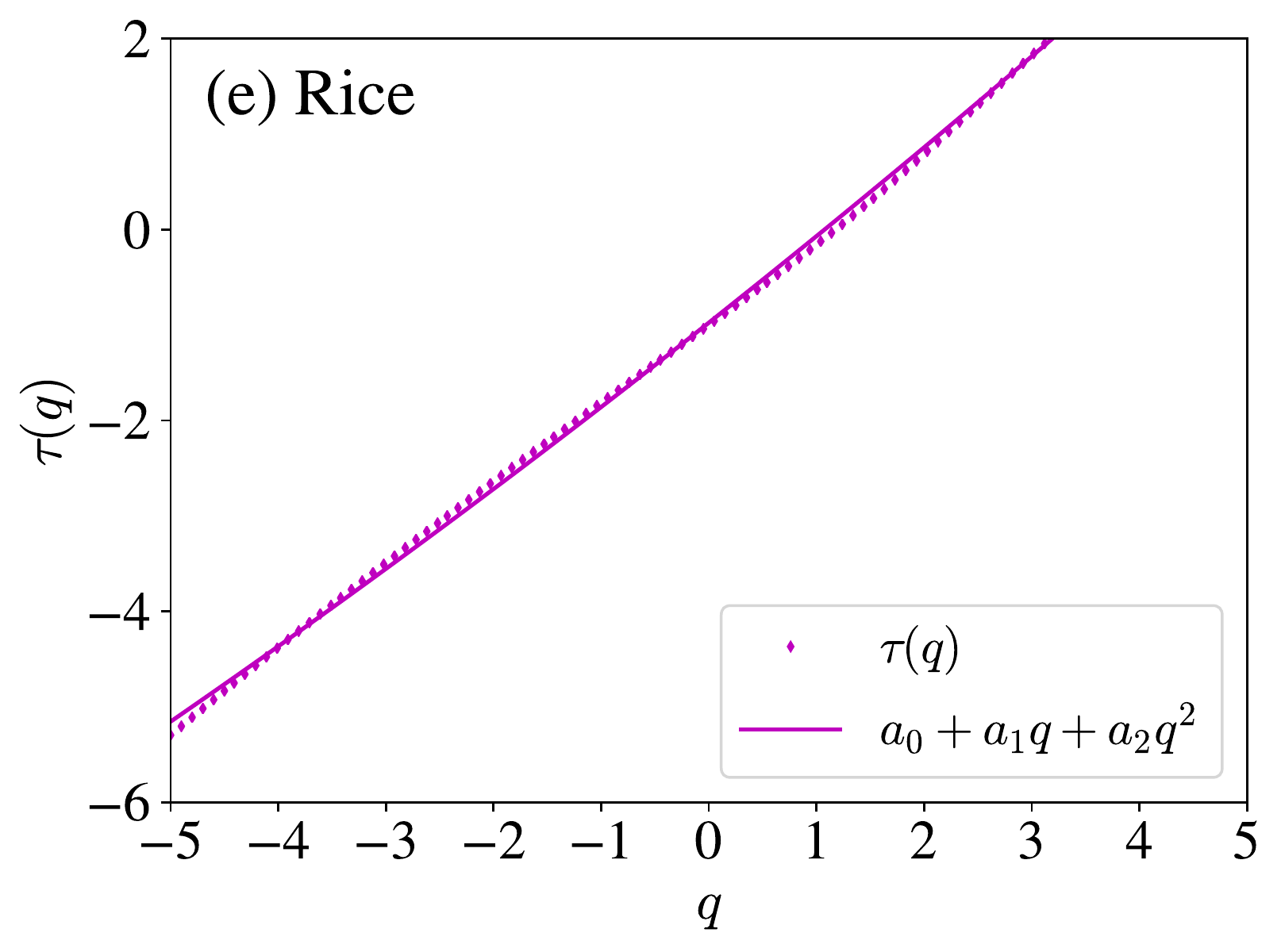}
    \includegraphics[width=0.325\linewidth]{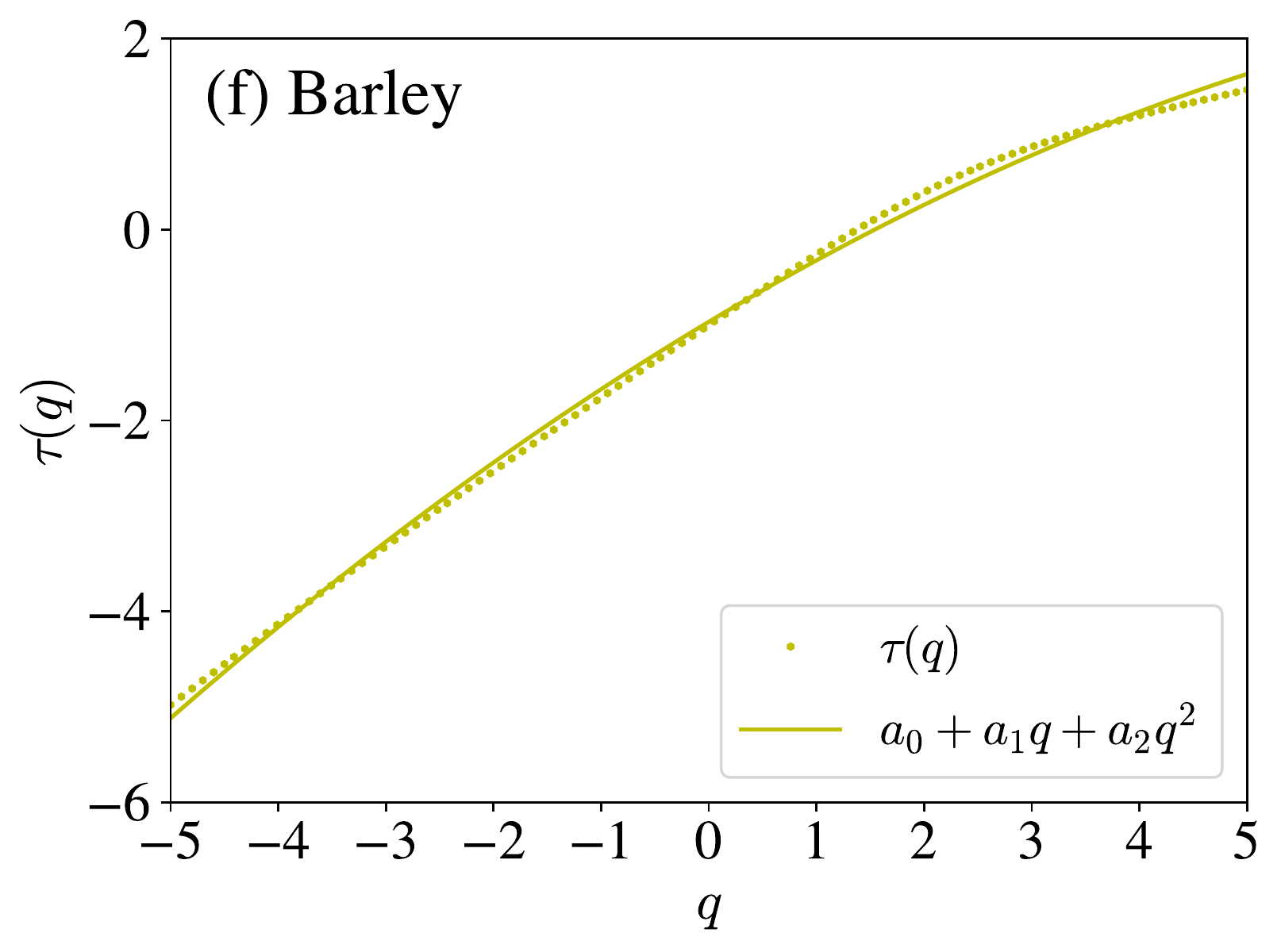}
    \includegraphics[width=0.325\linewidth]{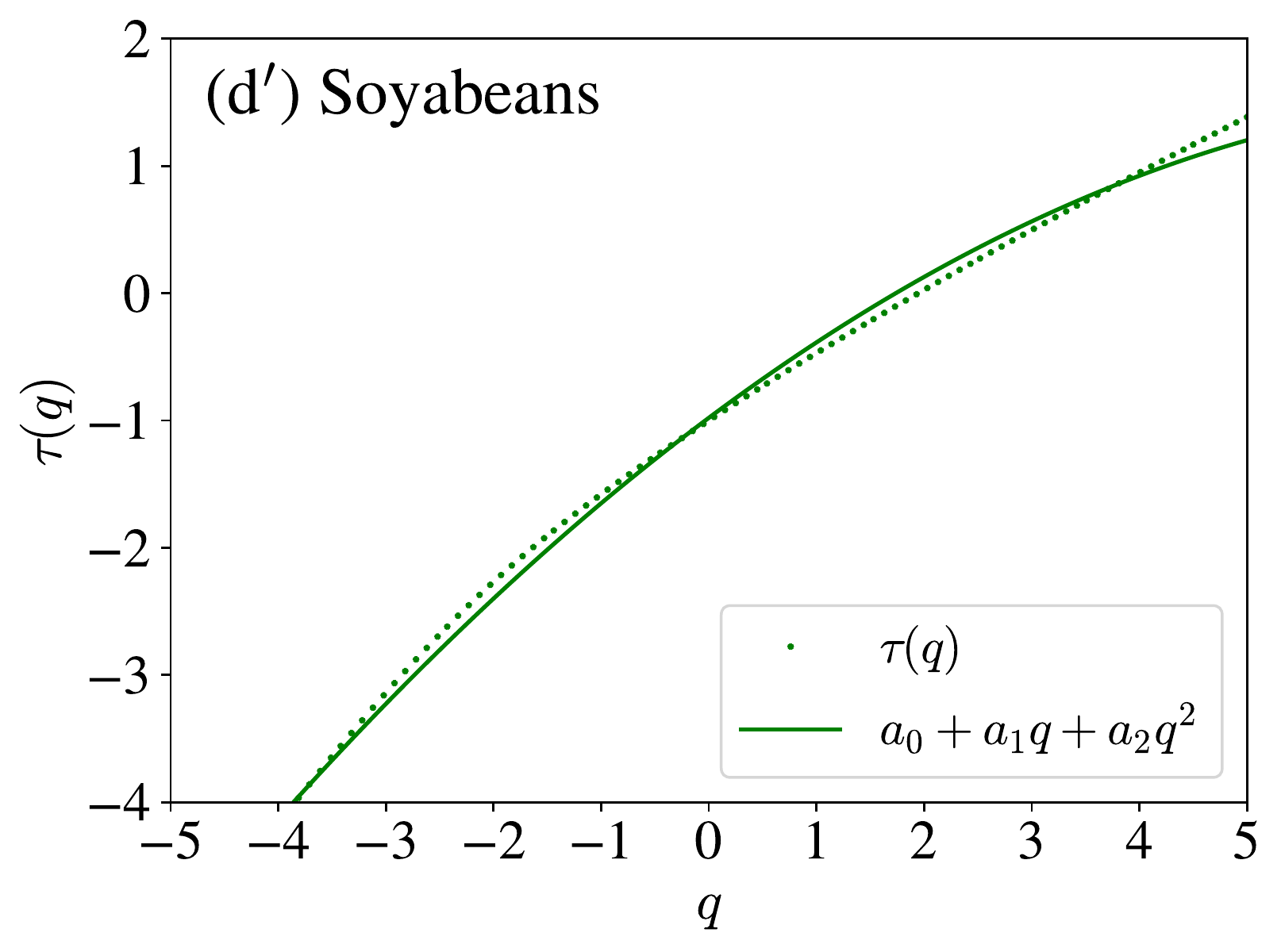}
    \includegraphics[width=0.325\linewidth]{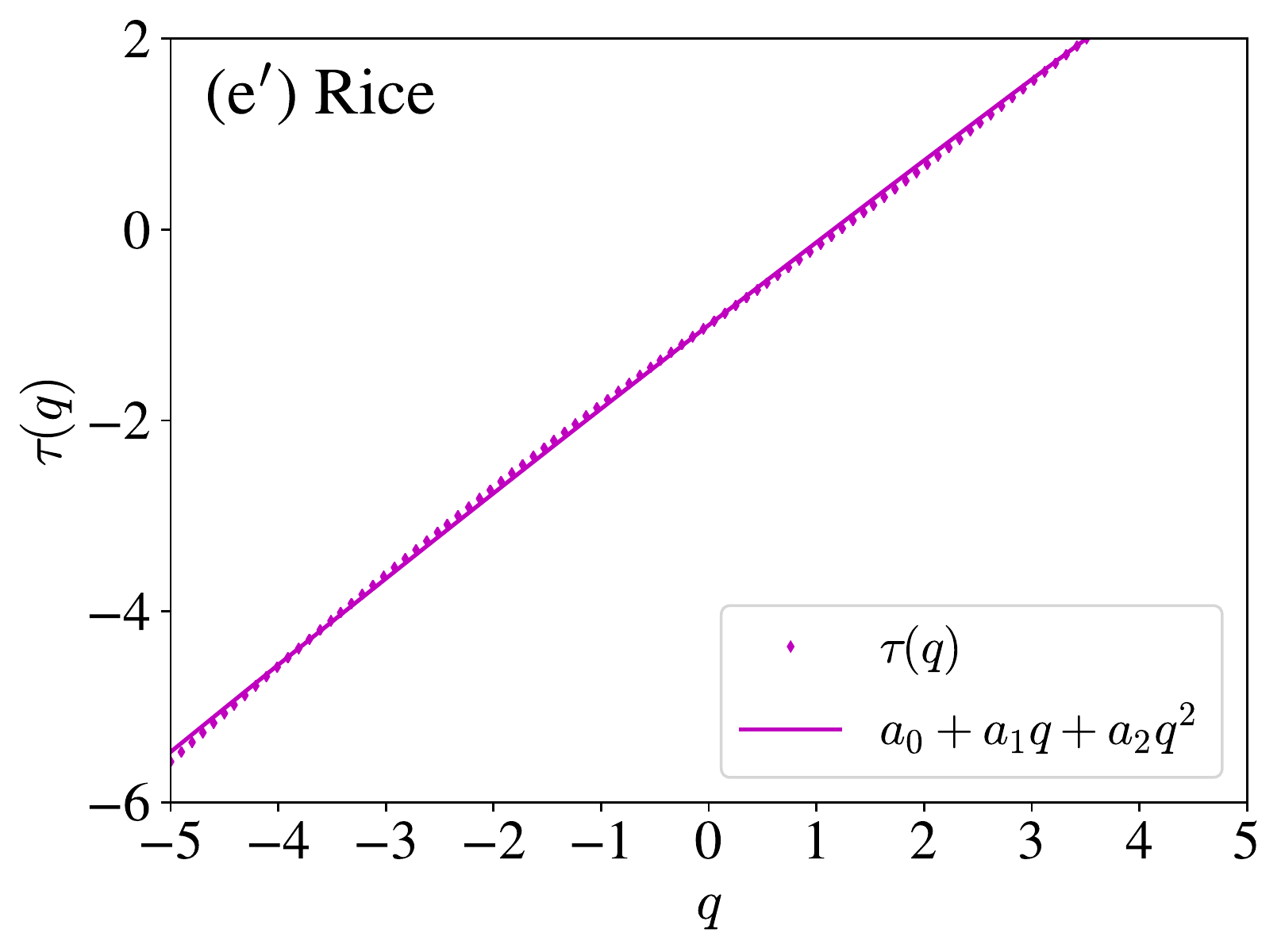}
    \includegraphics[width=0.325\linewidth]{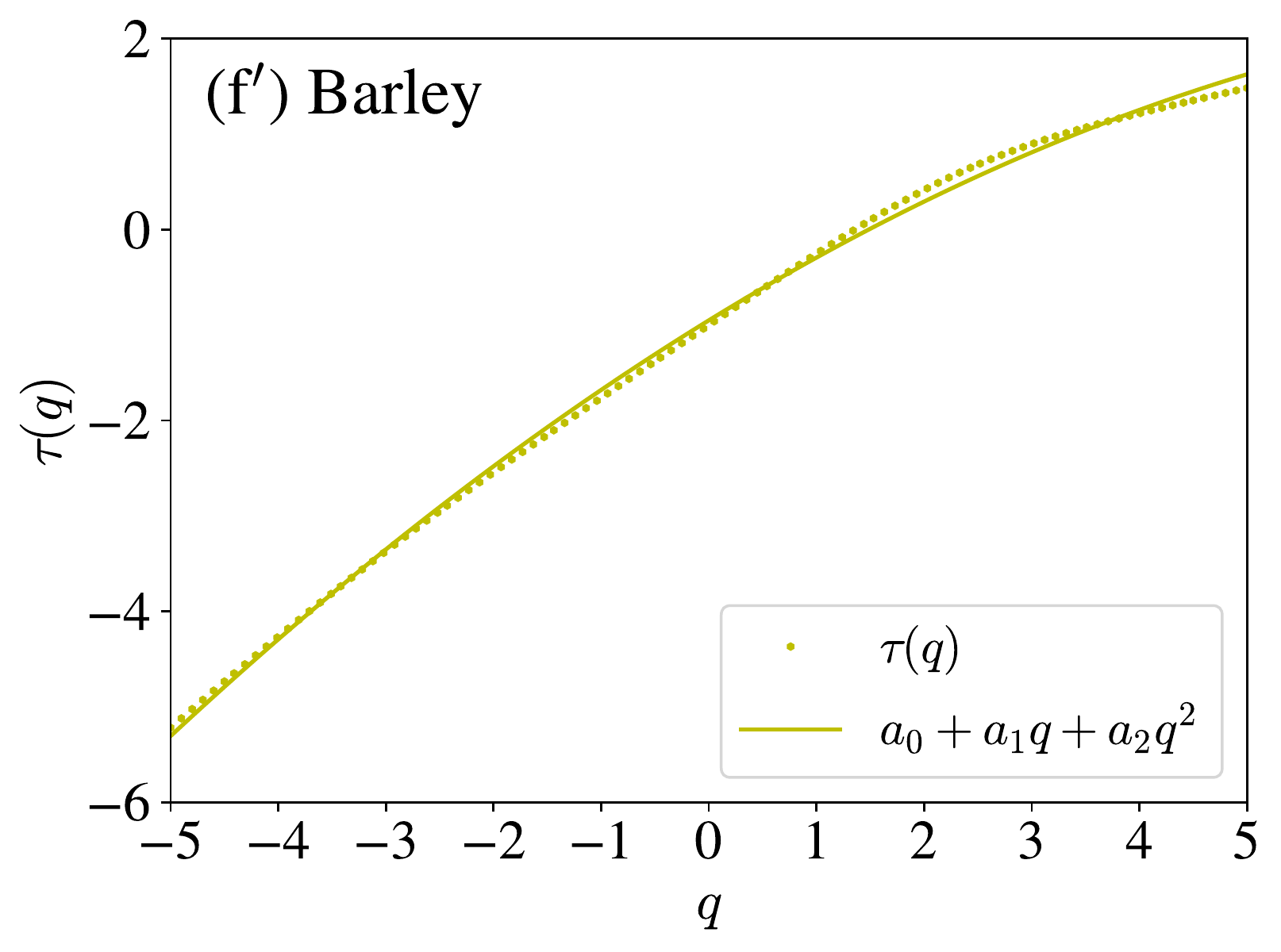}
    \caption{Testing the nonlinearity in the mass exponents $\tau(q)$ for the GOI index [(a) for $\ell=1$ and (a$^\prime$) for $\ell=2$], the wheat sub-index [(b) for $\ell=1$ and (b$^\prime$) for $\ell=2$], the maize sub-index [(c) for $\ell=1$ and (c$^\prime$) for $\ell=2$], the soyabeans sub-index [(d) for $\ell=1$ and (d$^\prime$) for $\ell=2$], the rice sub-index [(e) for $\ell=1$ and (e$^\prime$) for $\ell=2$], and the barley sub-index [(f) for $\ell=1$ and (f$^\prime$) for $\ell=2$].}
    \label{FigA:GOI:Return:MFDFA:tau}
\end{figure}

\begin{figure}[!ht]
    \centering
    \includegraphics[width=0.325\linewidth]{Fig_GOI_Return_MFDFA1_f_alpha_GOI.pdf}
    \includegraphics[width=0.325\linewidth]{Fig_GOI_Return_MFDFA1_f_alpha_Wheat.pdf}
    \includegraphics[width=0.325\linewidth]{Fig_GOI_Return_MFDFA1_f_alpha_Maize.pdf}
    \includegraphics[width=0.325\linewidth]{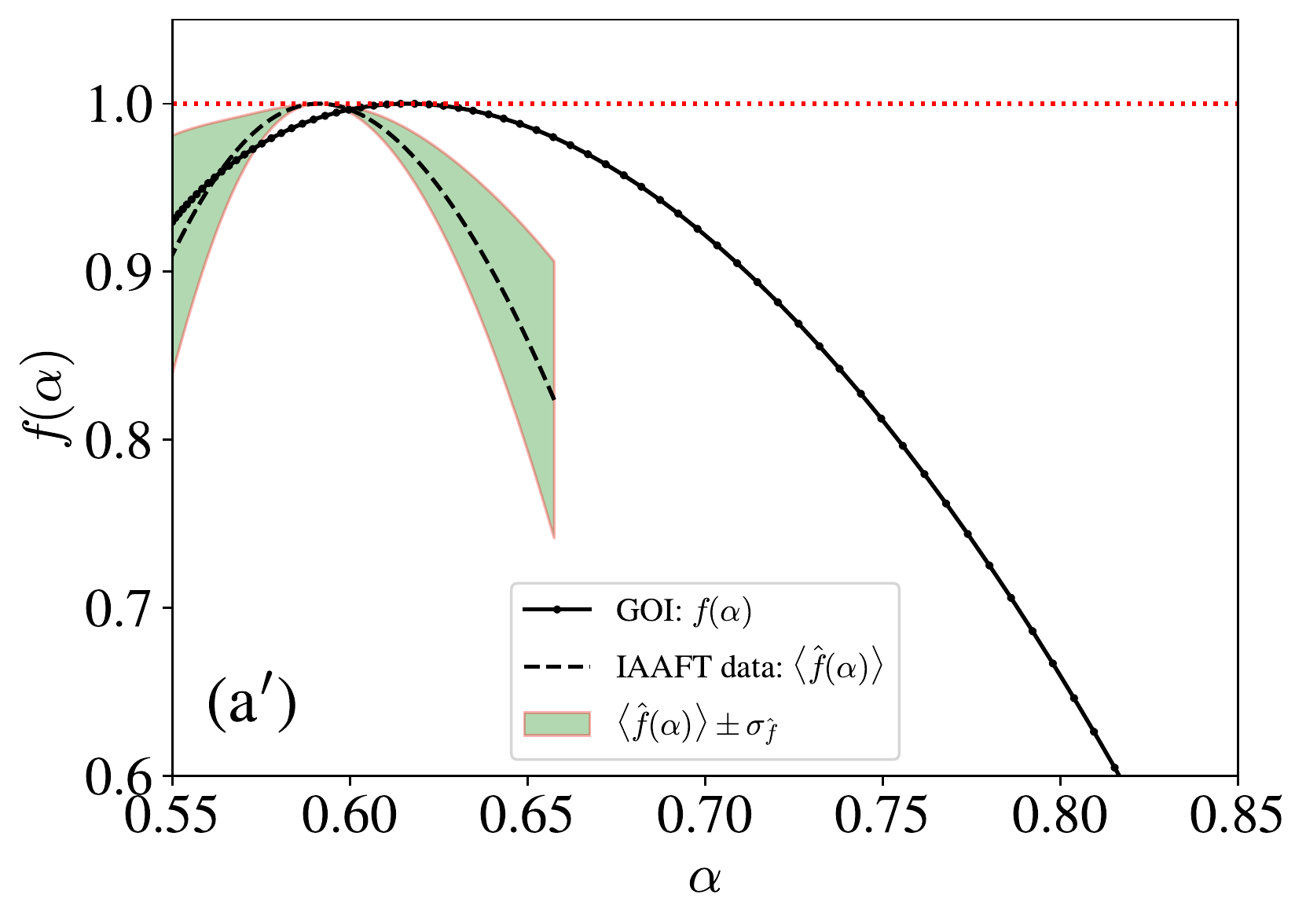}
    \includegraphics[width=0.325\linewidth]{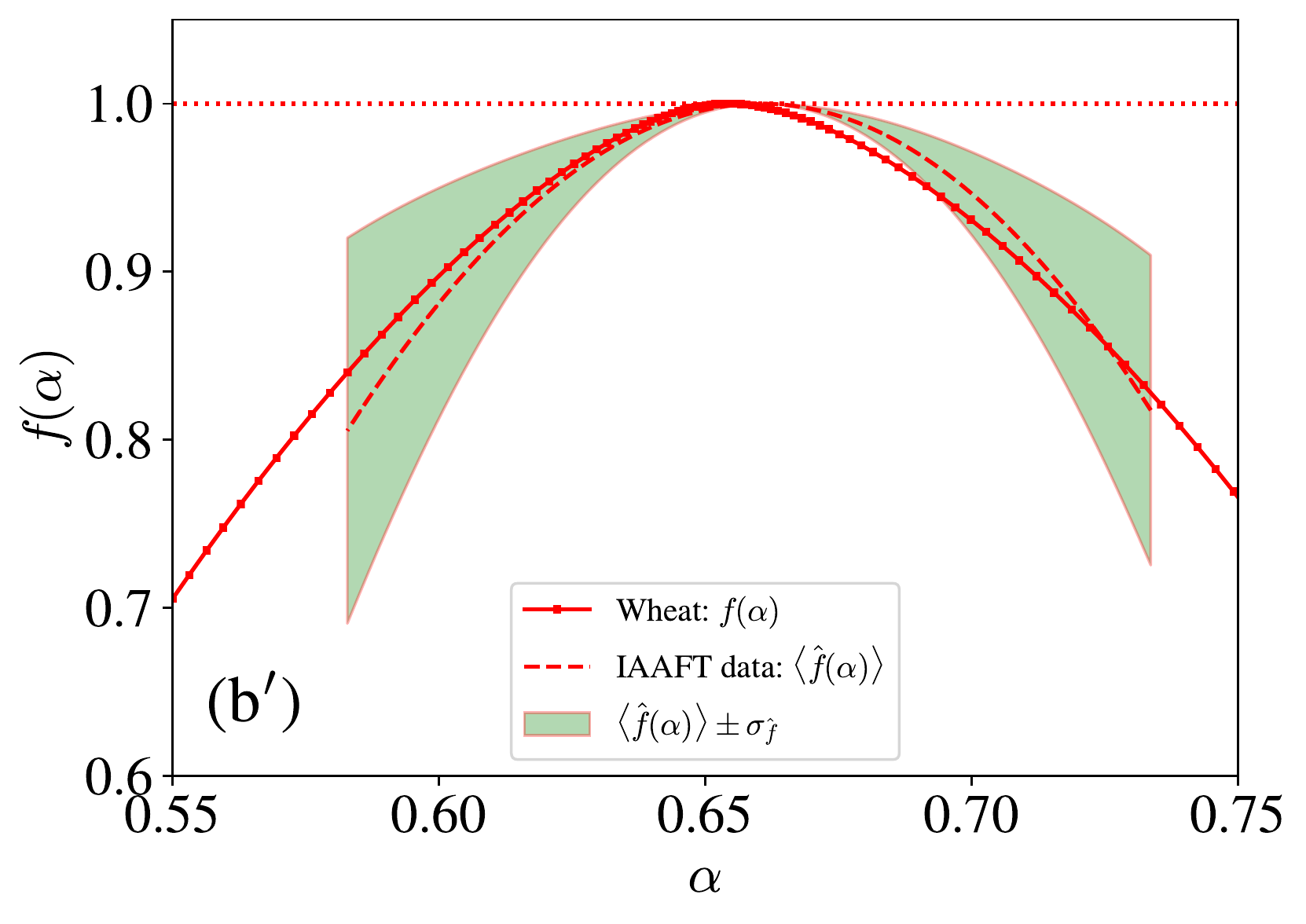}
    \includegraphics[width=0.325\linewidth]{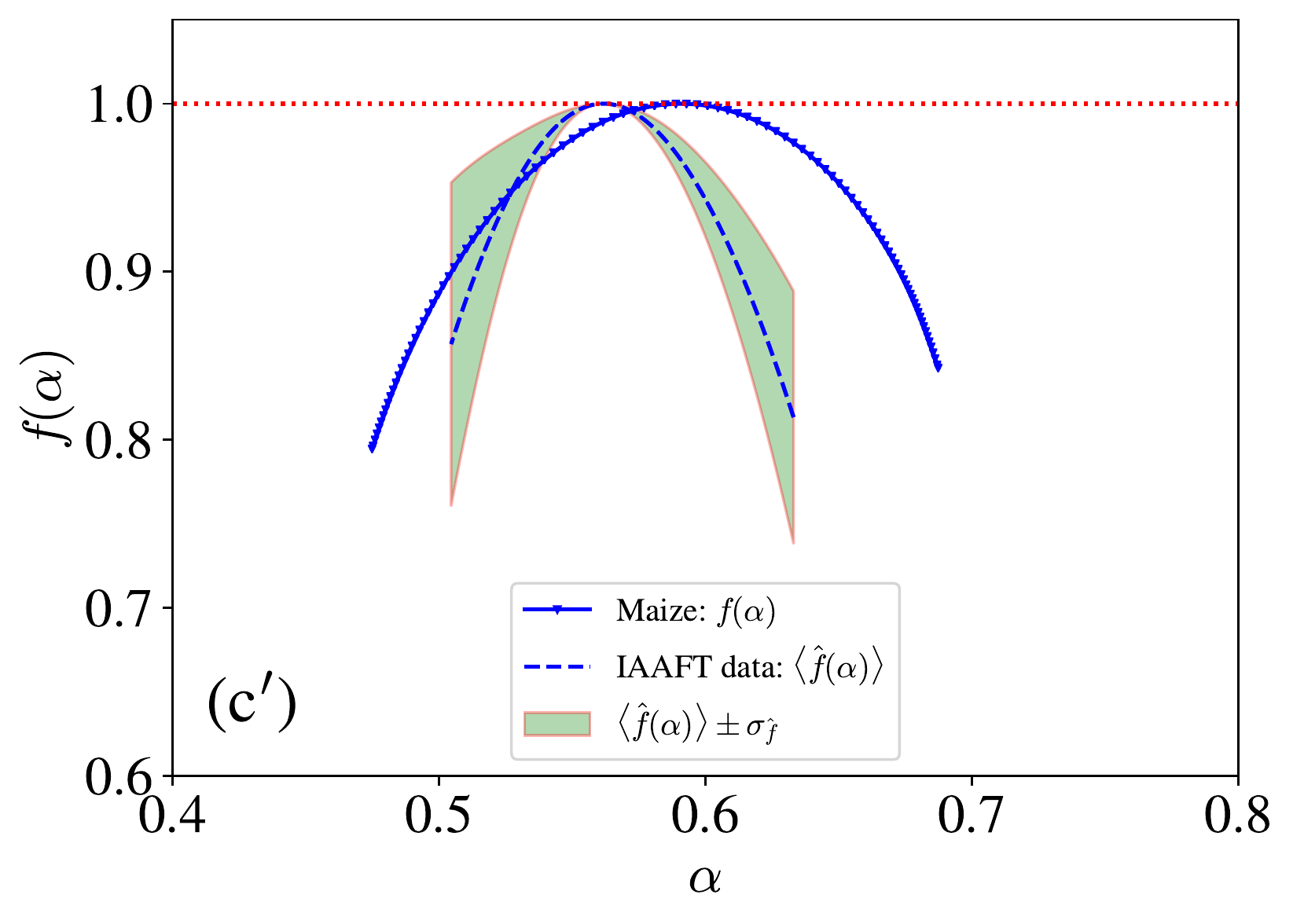}
    \includegraphics[width=0.325\linewidth]{Fig_GOI_Return_MFDFA1_f_alpha_Soyabeans.pdf}
    \includegraphics[width=0.325\linewidth]{Fig_GOI_Return_MFDFA1_f_alpha_Rice.pdf}
    \includegraphics[width=0.325\linewidth]{Fig_GOI_Return_MFDFA1_f_alpha_Barley.pdf}
    \includegraphics[width=0.325\linewidth]{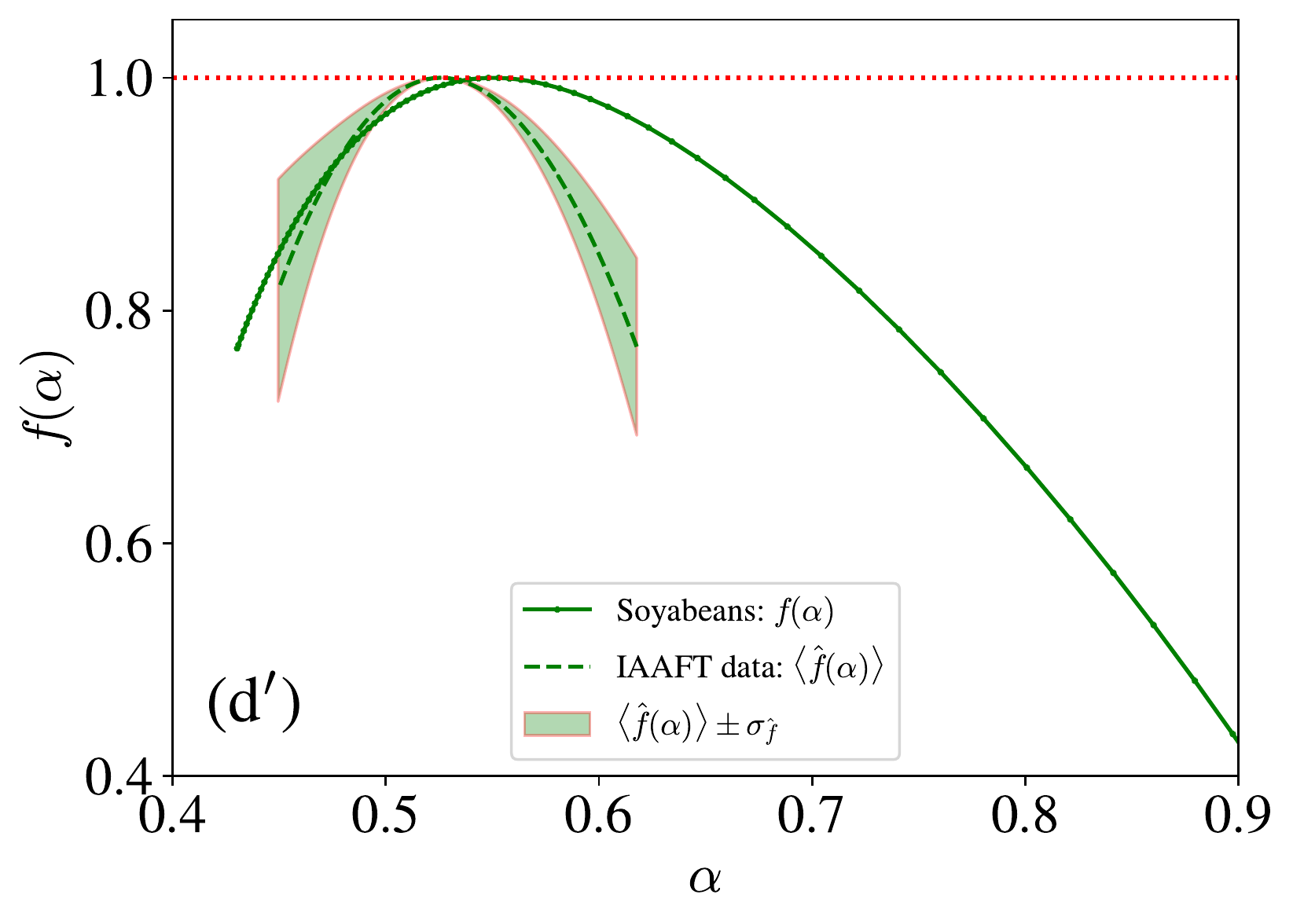}
    \includegraphics[width=0.325\linewidth]{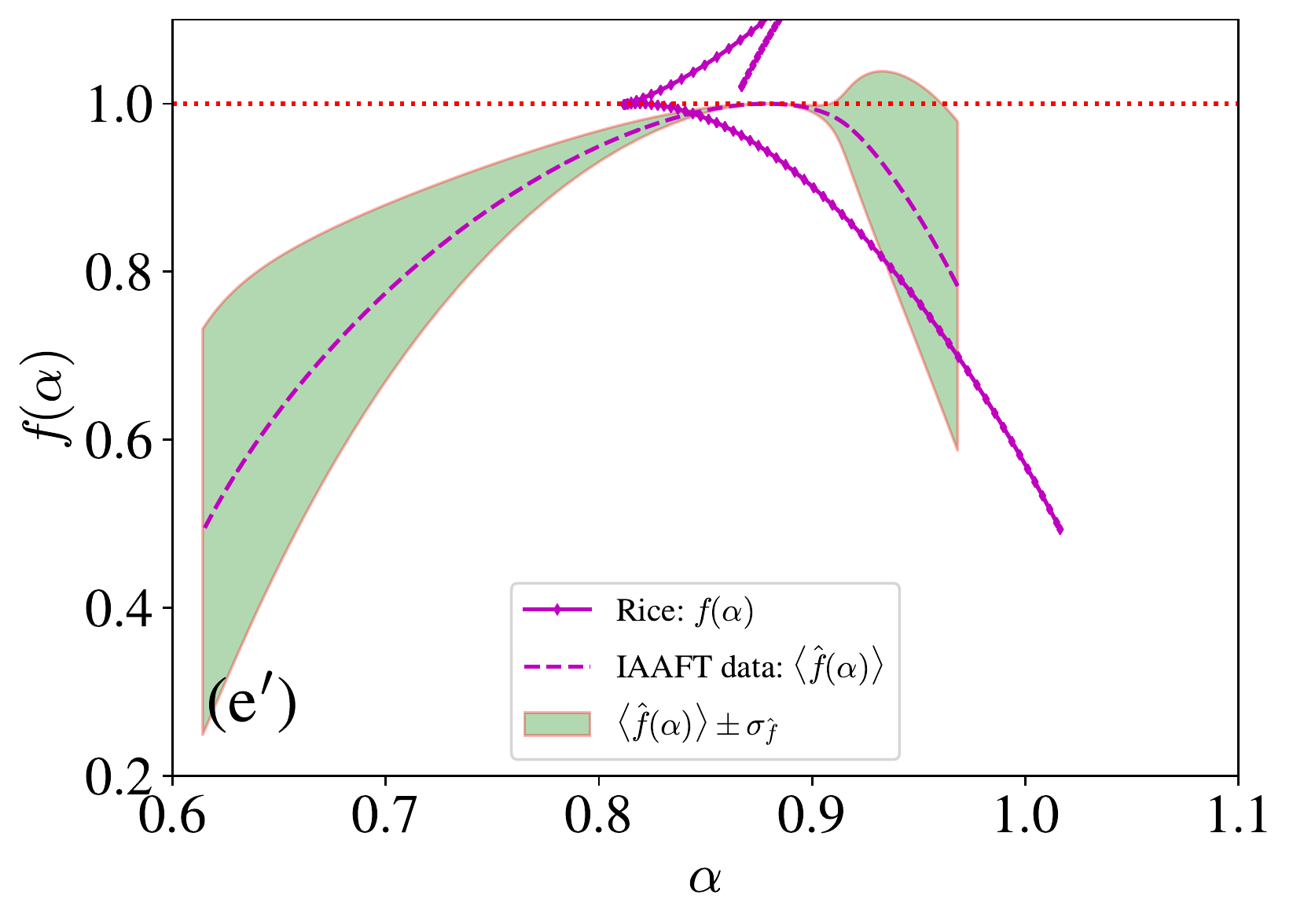}
    \includegraphics[width=0.325\linewidth]{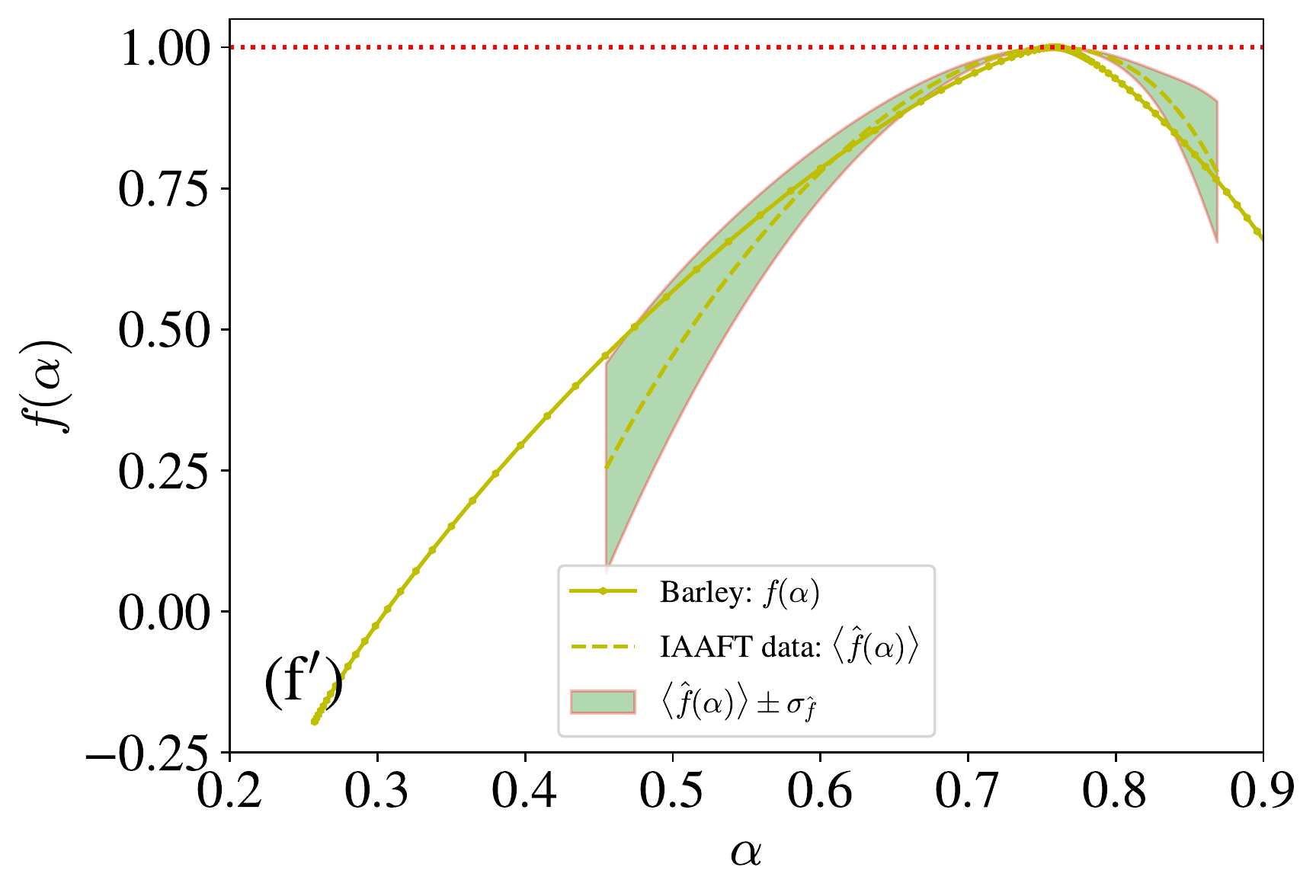}
    \caption{Singularity spectrum $f(\alpha)$ with respect to the singularity $\alpha$ for the GOI index [(a) for $\ell=1$ and (a$^\prime$) for $\ell=2$], the wheat sub-index [(b) for $\ell=1$ and (b$^\prime$) for $\ell=2$], the maize sub-index [(c) for $\ell=1$ and (c$^\prime$) for $\ell=2$], the soyabeans sub-index [(d) for $\ell=1$ and (d$^\prime$) for $\ell=2$], the rice sub-index [(e) for $\ell=1$ and (e$^\prime$) for $\ell=2$], and the barley sub-index [(f) for $\ell=1$ and (f$^\prime$) for $\ell=2$] released by the International Grains Council. For each GOI and sub-indices, we generate 1000 surrogate time series using the IAAFT algorithm and calculate the mean $\langle\hat{f}\rangle(\alpha)$ and standard deviation $\sigma_{\hat{f}}$.}
    \label{FigA:GOI:Return:MFDFA:f:alpah}
\end{figure}

\begin{figure}[!ht]
    \centering
    \includegraphics[width=0.325\linewidth]{Fig_GOI_Return_MFDFA1_dalpha_test_GOI.pdf}
    \includegraphics[width=0.325\linewidth]{Fig_GOI_Return_MFDFA1_dalpha_test_Wheat.pdf}
    \includegraphics[width=0.325\linewidth]{Fig_GOI_Return_MFDFA1_dalpha_test_Maize.pdf}
    \includegraphics[width=0.325\linewidth]{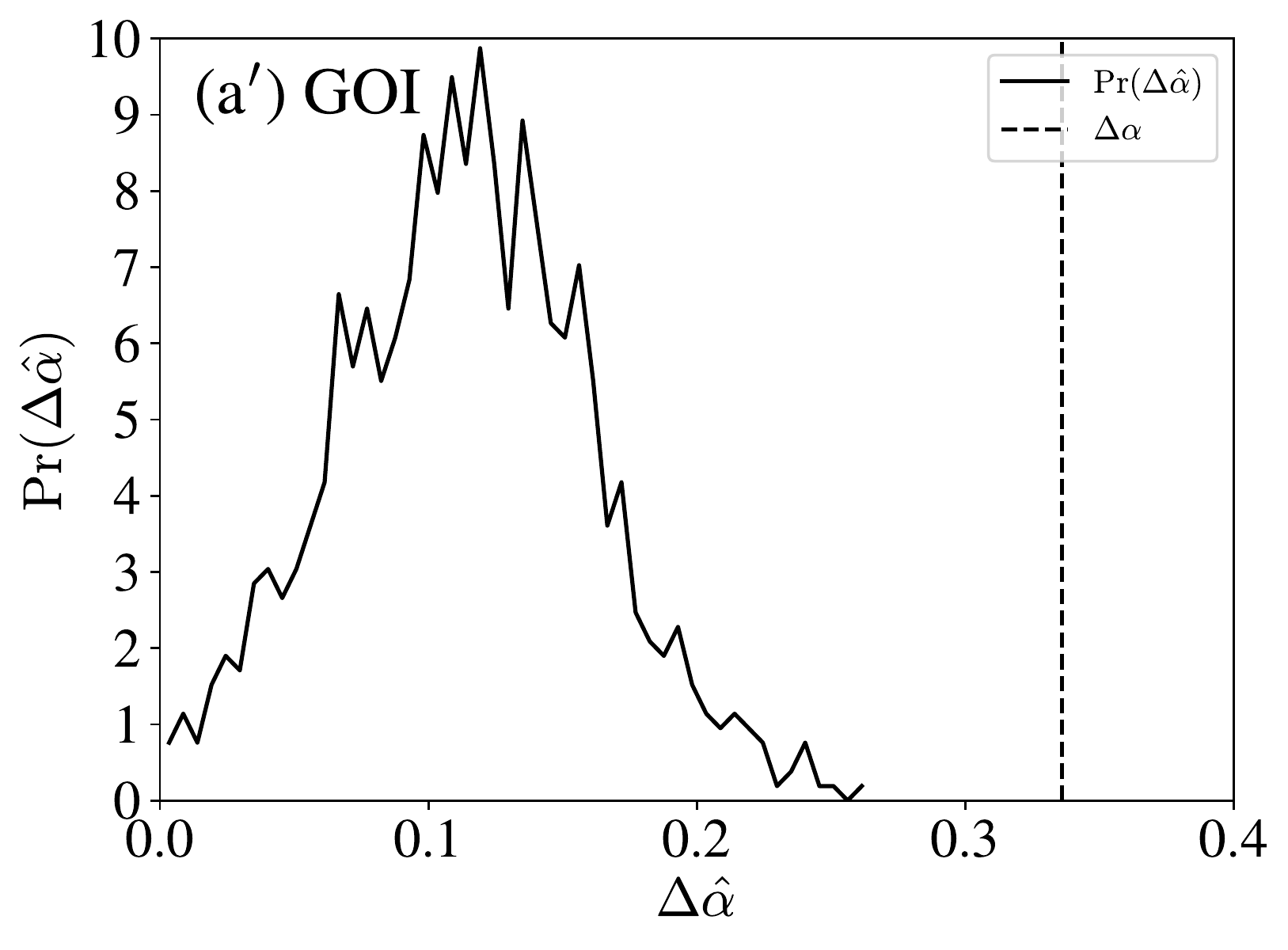}
    \includegraphics[width=0.325\linewidth]{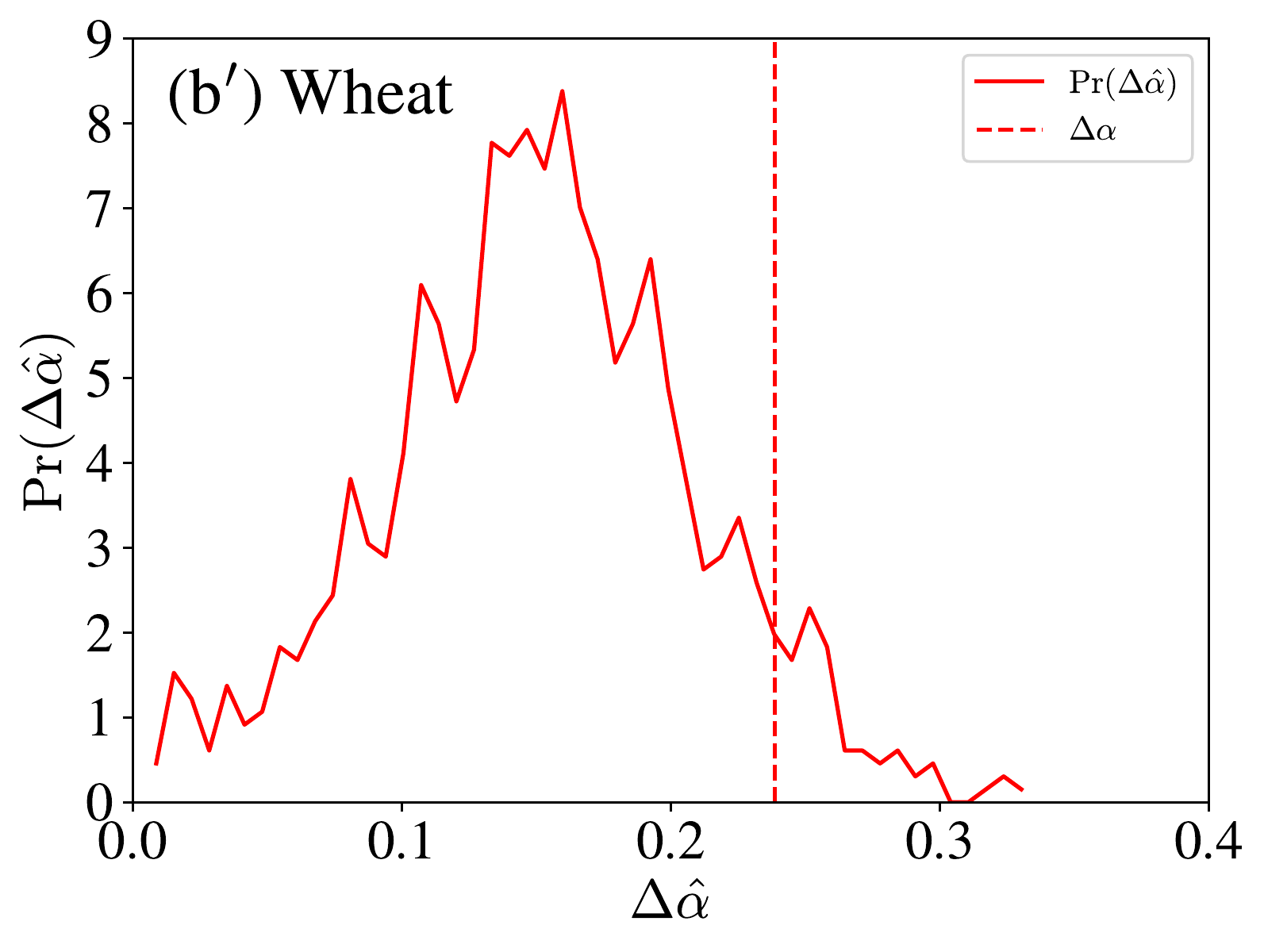}
    \includegraphics[width=0.325\linewidth]{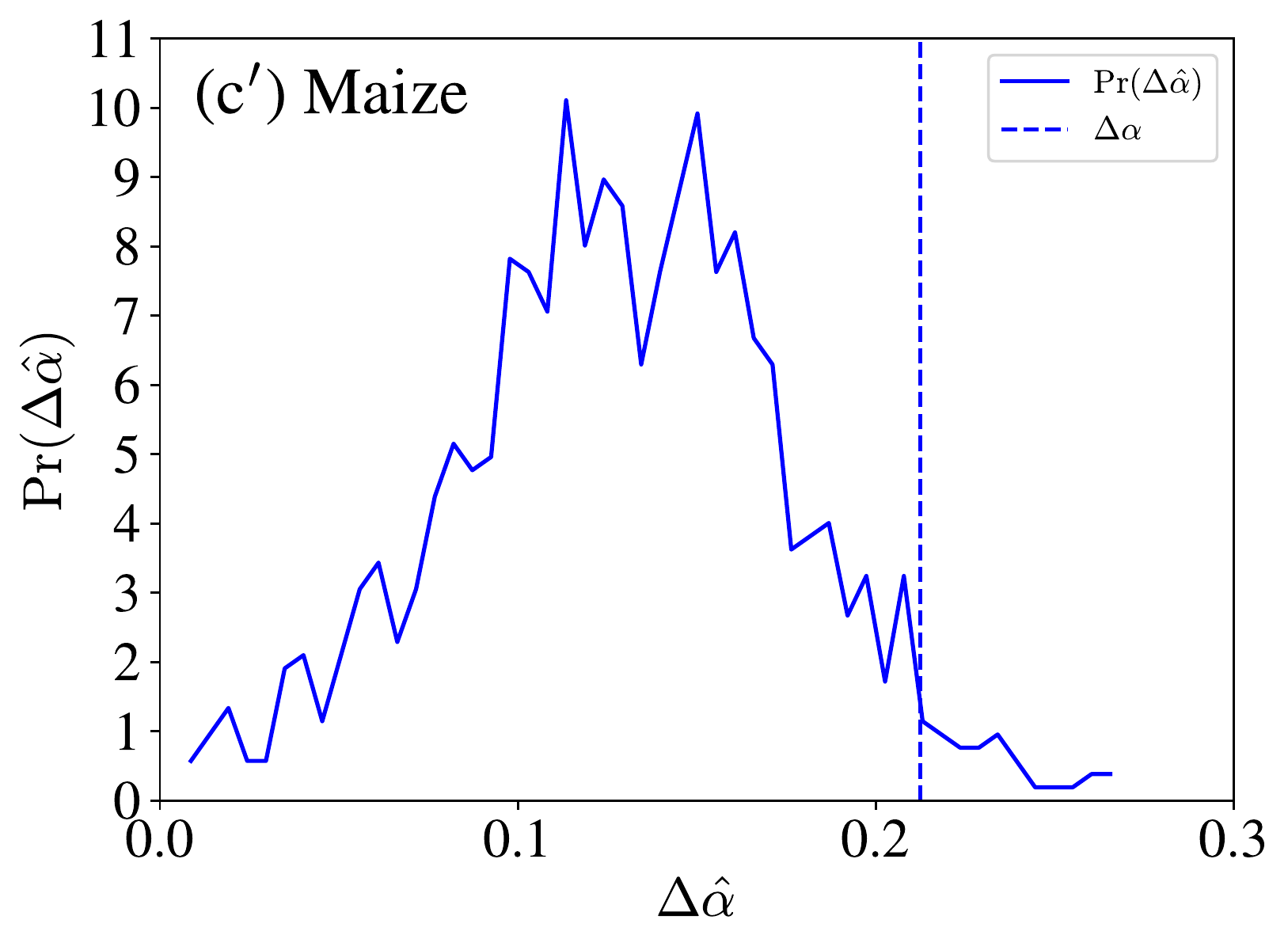}
    \includegraphics[width=0.325\linewidth]{Fig_GOI_Return_MFDFA1_dalpha_test_Soyabeans.pdf}
    \includegraphics[width=0.325\linewidth]{Fig_GOI_Return_MFDFA1_dalpha_test_Rice.pdf}
    \includegraphics[width=0.325\linewidth]{Fig_GOI_Return_MFDFA1_dalpha_test_Barley.pdf}
    \includegraphics[width=0.325\linewidth]{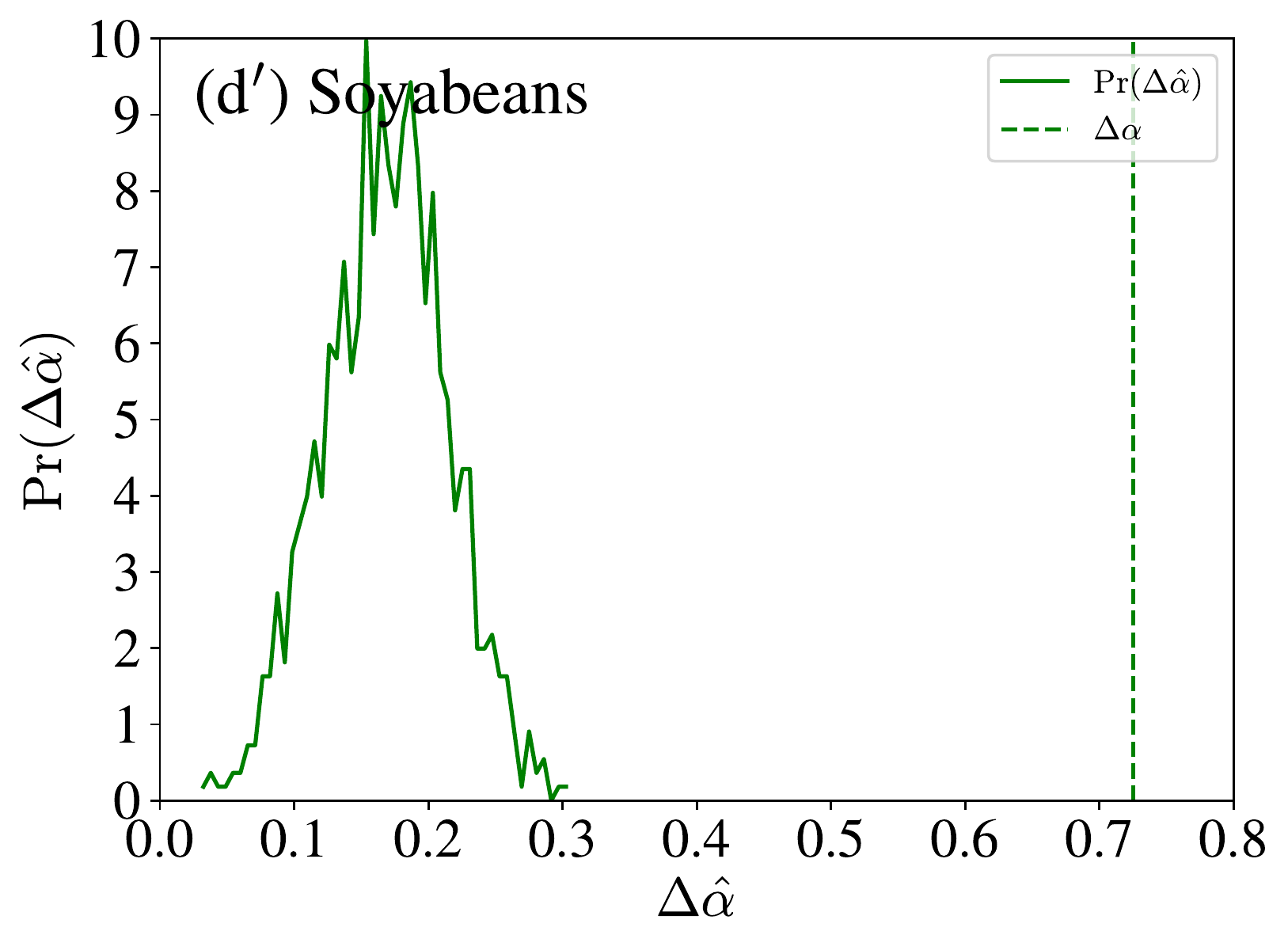}
    \includegraphics[width=0.325\linewidth]{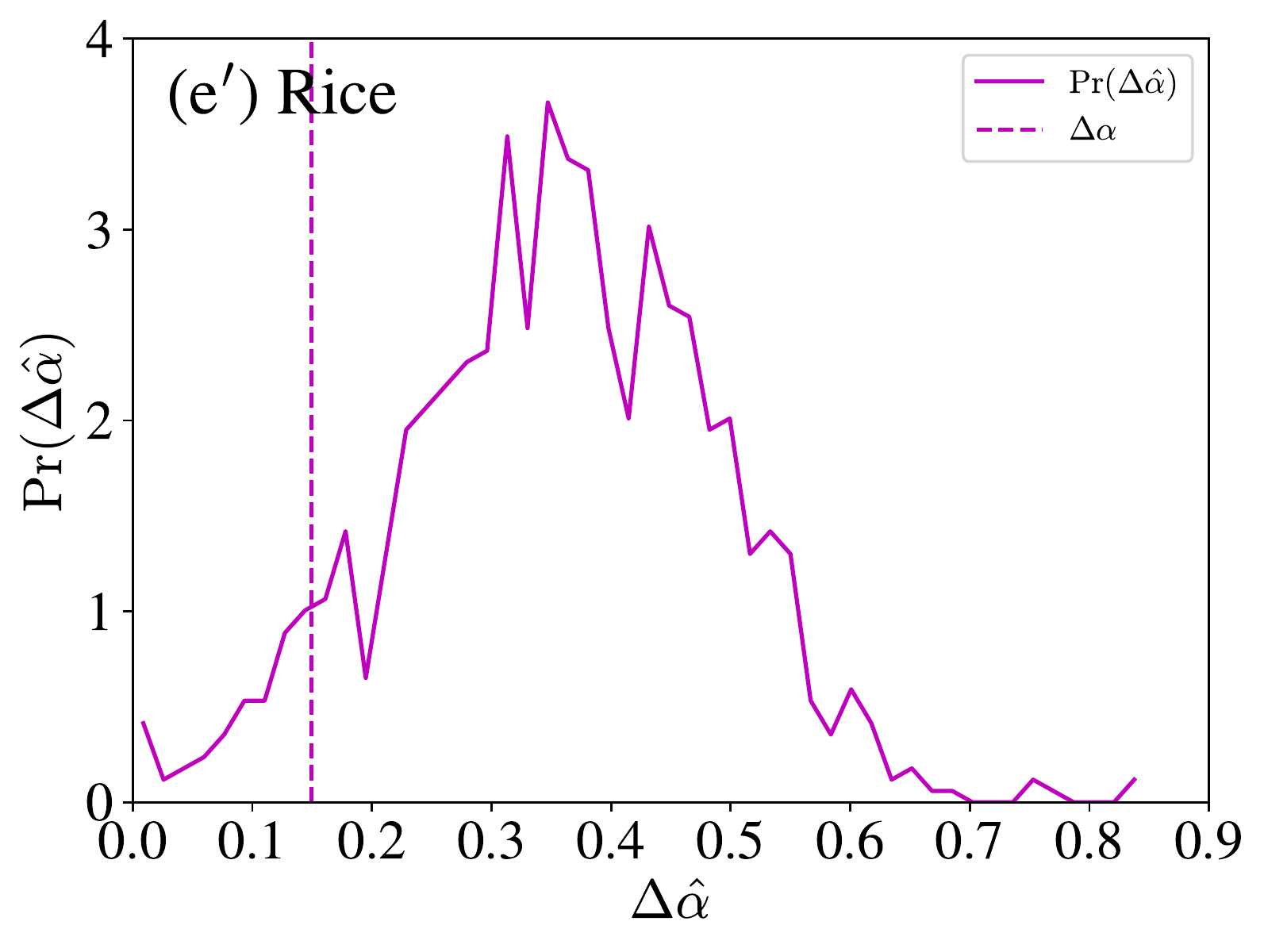}
    \includegraphics[width=0.325\linewidth]{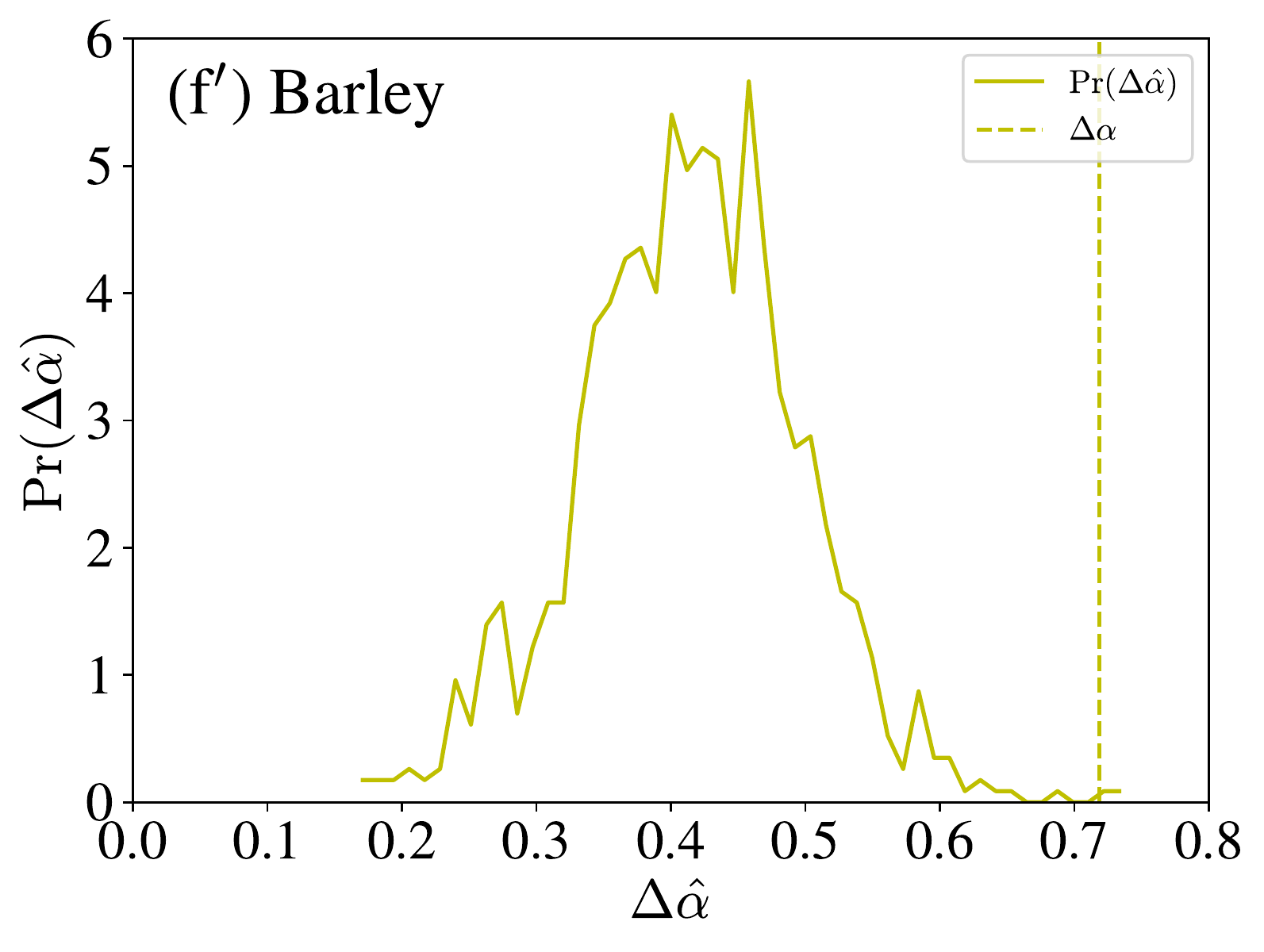}
    \caption{Empirical distribution of the singularity widths $\Delta\hat{\alpha}$ of the 1000 IAAFT surrogates for the GOI index [(a) for $\ell=1$ and (a$^\prime$) for $\ell=2$], the wheat sub-index [(b) for $\ell=1$ and (b$^\prime$) for $\ell=2$], the maize sub-index [(c) for $\ell=1$ and (c$^\prime$) for $\ell=2$], the soyabeans sub-index [(d) for $\ell=1$ and (d$^\prime$) for $\ell=2$], the rice sub-index [(e) for $\ell=1$ and (e$^\prime$) for $\ell=2$], and the barley sub-index [(f) for $\ell=1$ and (f$^\prime$) for $\ell=2$] released by the International Grains Council. The vertical dashed lines are the corresponding singularity widths $\Delta \alpha$ of the original time series.}
    \label{FigA:GOI:Return:MFDFA:dalpah:test}
\end{figure}

\begin{figure}[!ht]
    \centering
    \includegraphics[width=0.325\linewidth]{Fig_GOI_Return_MFDFA1_df_test_GOI.pdf}
    \includegraphics[width=0.325\linewidth]{Fig_GOI_Return_MFDFA1_df_test_Wheat.pdf}
    \includegraphics[width=0.325\linewidth]{Fig_GOI_Return_MFDFA1_df_test_Maize.pdf}   \includegraphics[width=0.325\linewidth]{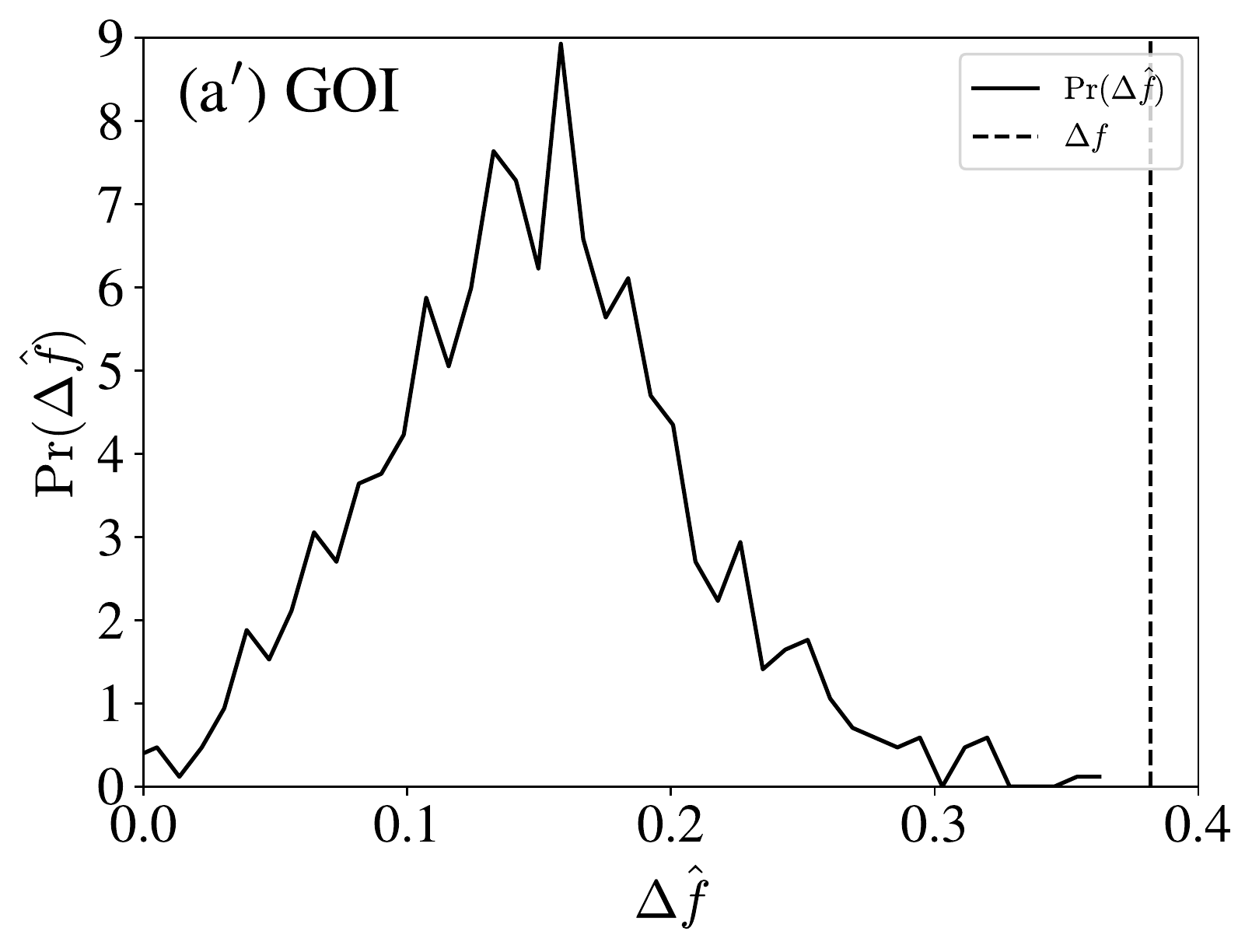}
    \includegraphics[width=0.325\linewidth]{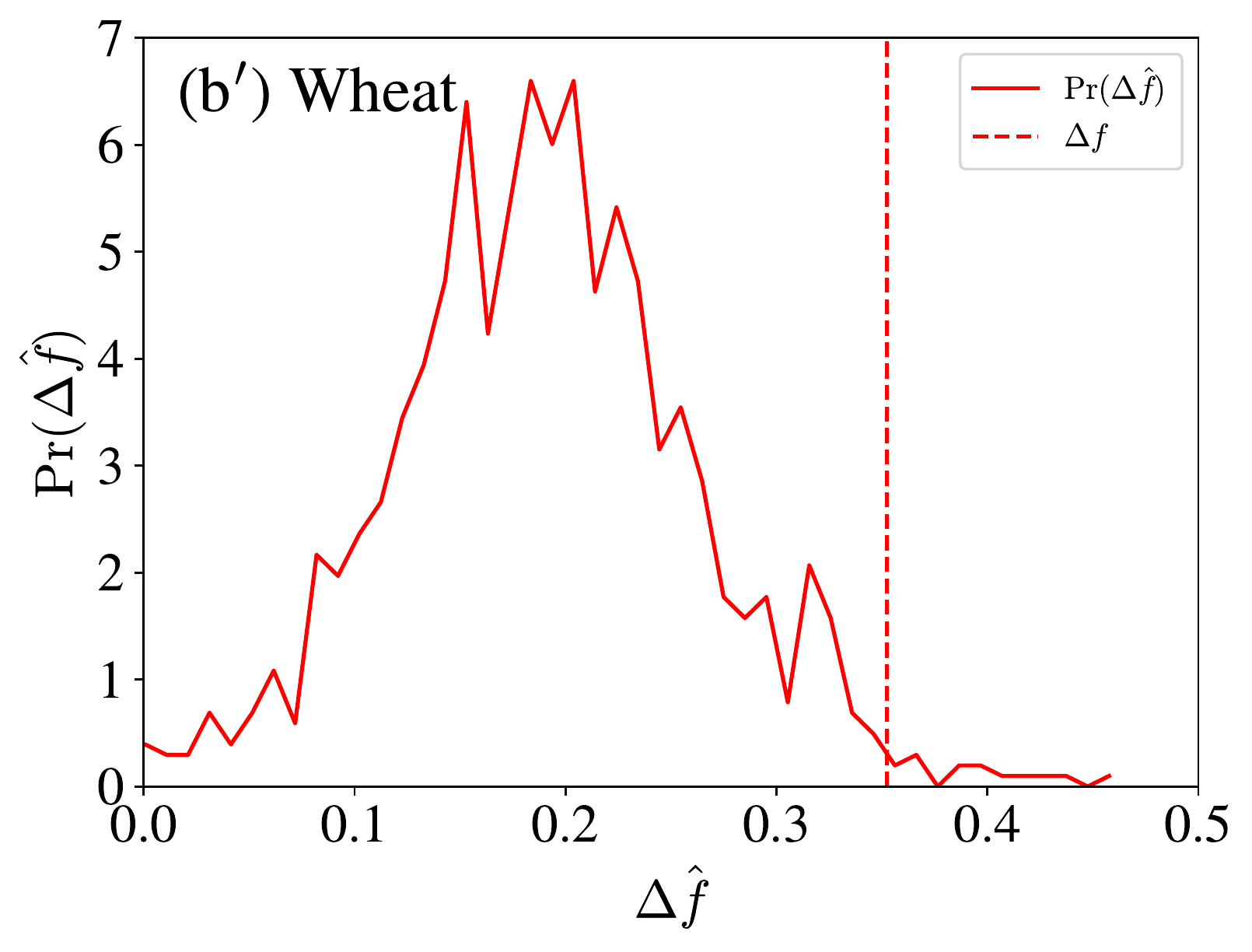}
    \includegraphics[width=0.325\linewidth]{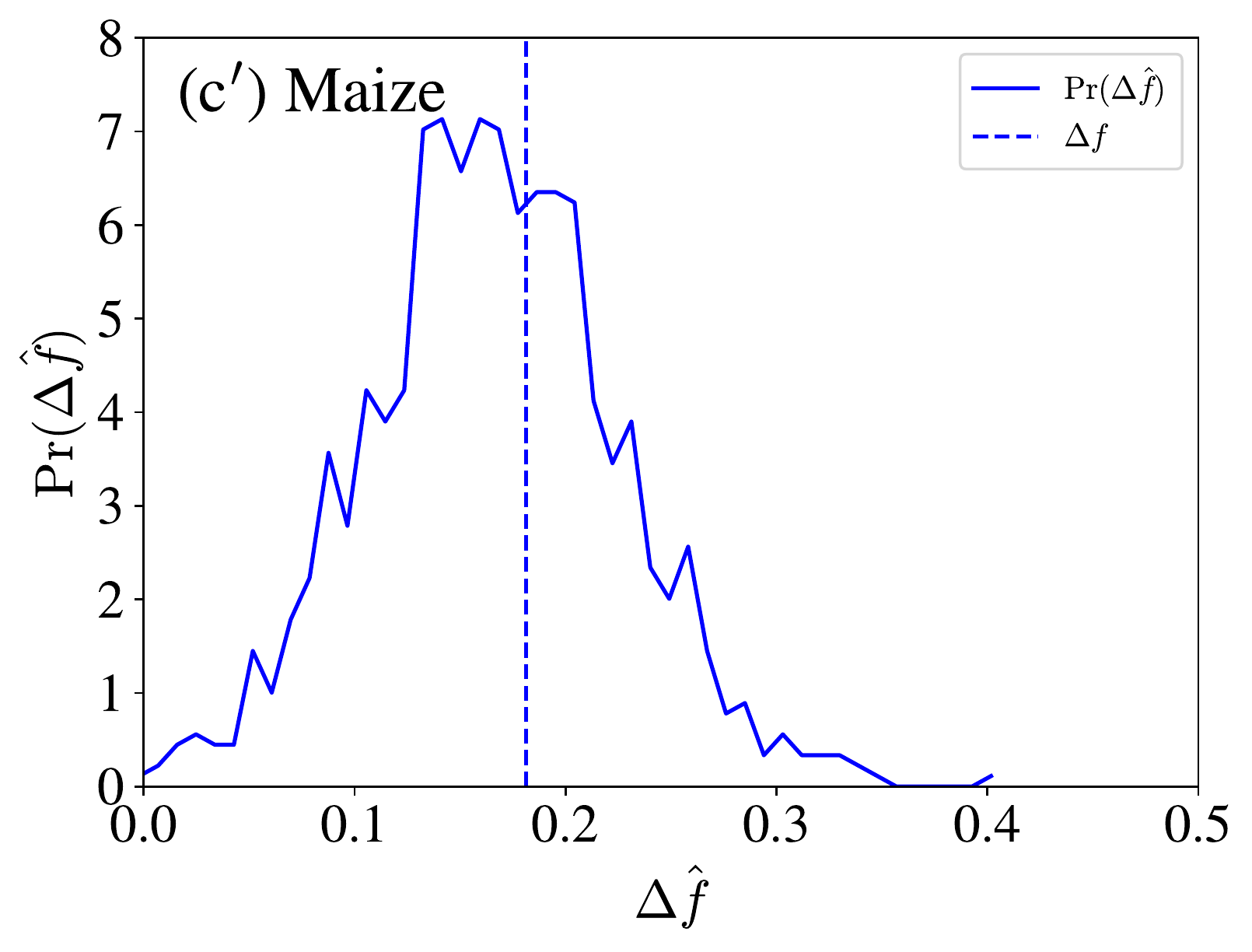}
    \includegraphics[width=0.325\linewidth]{Fig_GOI_Return_MFDFA1_df_test_Soyabeans.pdf}
    \includegraphics[width=0.325\linewidth]{Fig_GOI_Return_MFDFA1_df_test_Rice.pdf}
    \includegraphics[width=0.325\linewidth]{Fig_GOI_Return_MFDFA1_df_test_Barley.pdf}
    \includegraphics[width=0.325\linewidth]{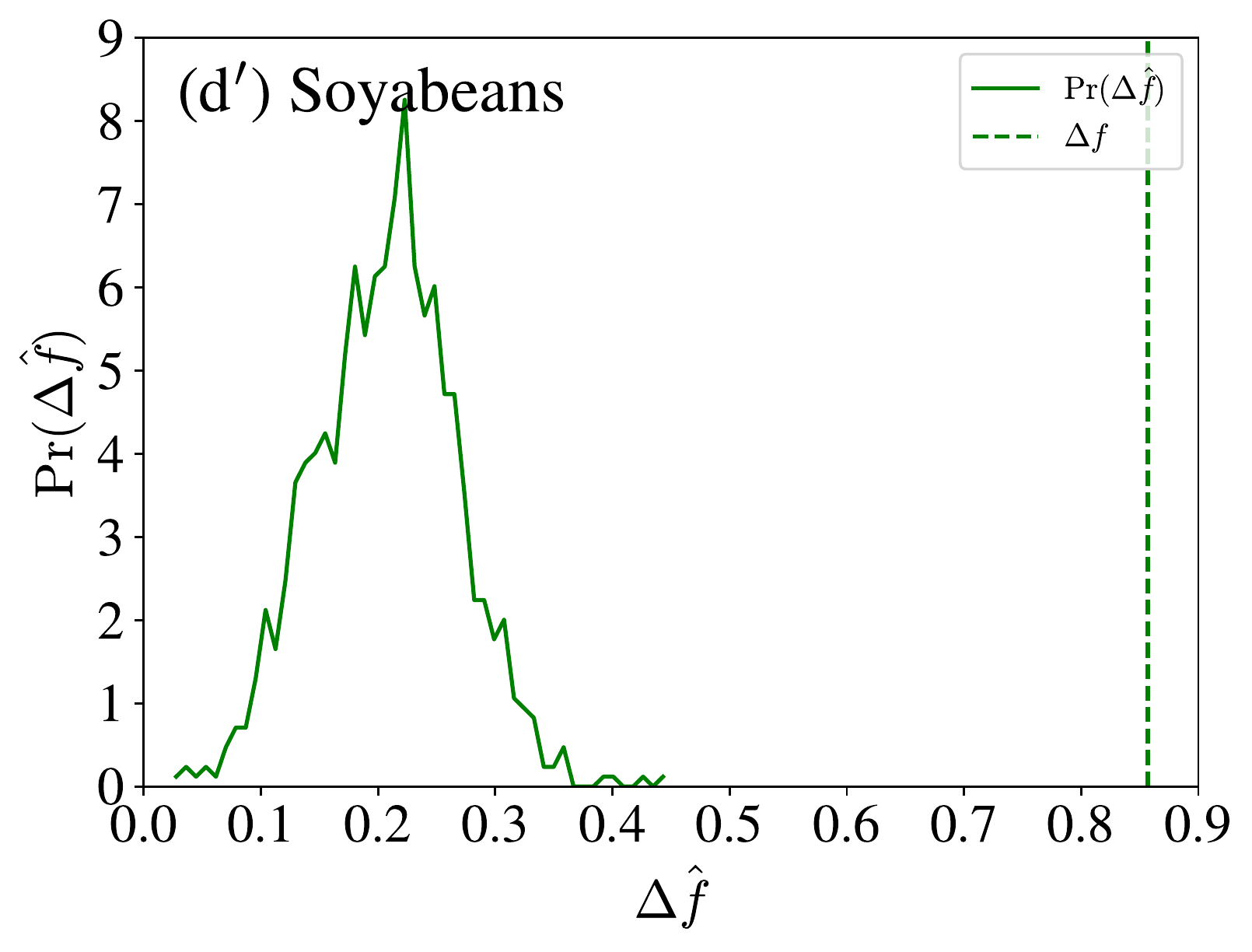}
    \includegraphics[width=0.325\linewidth]{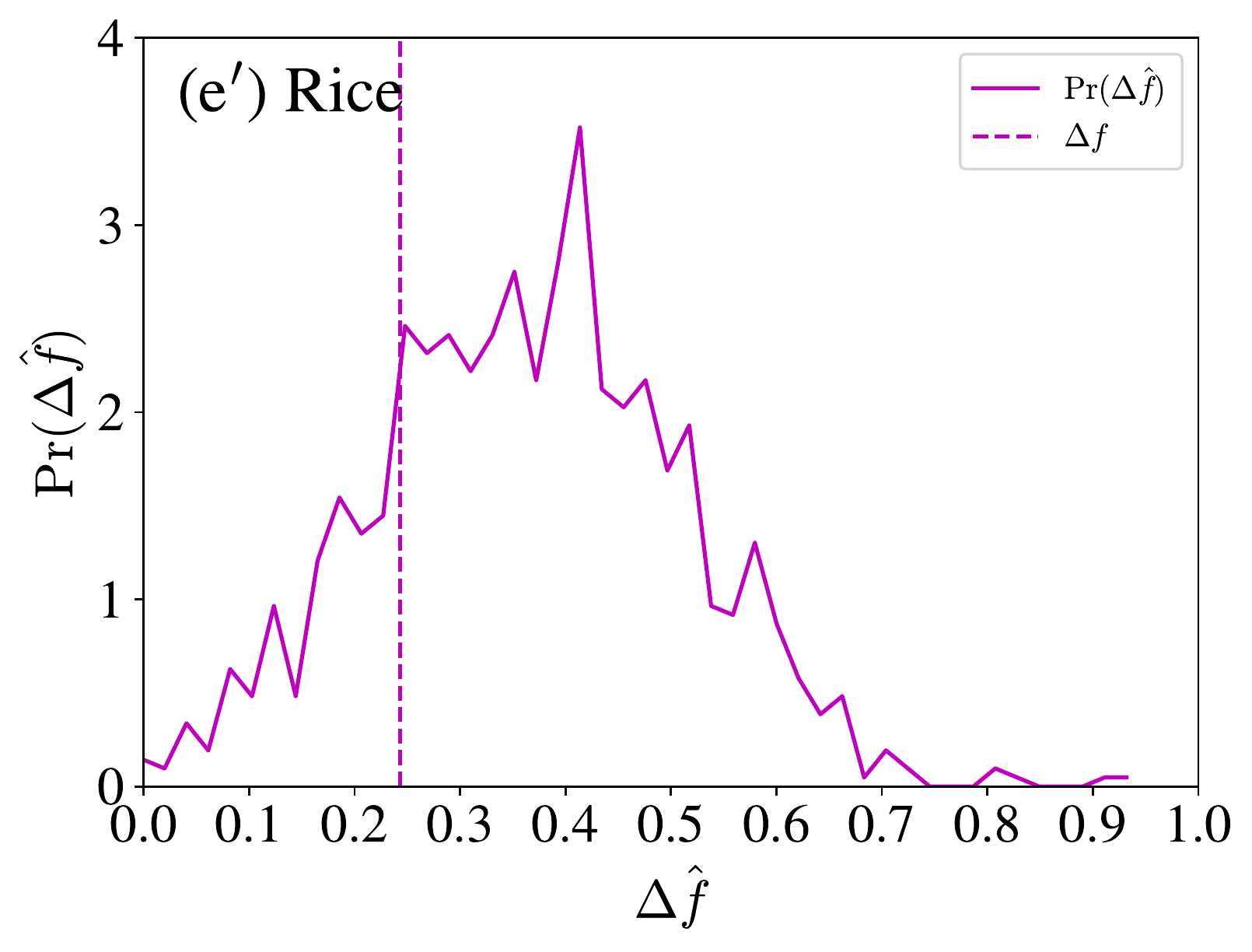}
    \includegraphics[width=0.325\linewidth]{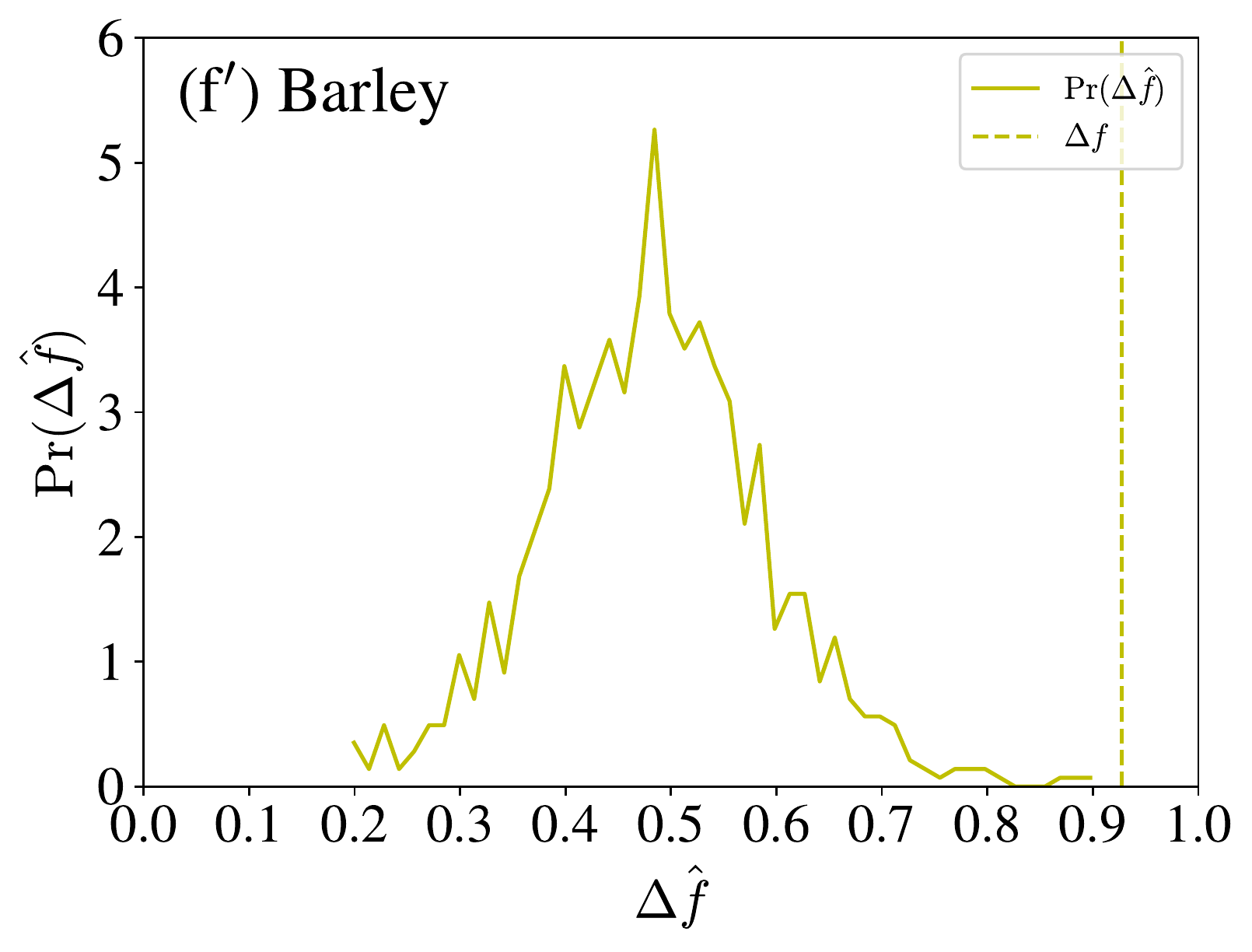}
    \caption{Empirical distribution of the spectrum differences $\Delta\hat{f}$ of the 1000 IAAFT surrogates for the GOI index [(a) for $\ell=1$ and (a$^\prime$) for $\ell=2$], the wheat sub-index [(b) for $\ell=1$ and (b$^\prime$) for $\ell=2$], the maize sub-index [(c) for $\ell=1$ and (c$^\prime$) for $\ell=2$], the soyabeans sub-index [(d) for $\ell=1$ and (d$^\prime$) for $\ell=2$], the rice sub-index [(e) for $\ell=1$ and (e$^\prime$) for $\ell=2$], and the barley sub-index [(f) for $\ell=1$ and (f$^\prime$) for $\ell=2$] released by the International Grains Council. The vertical dashed lines are the corresponding spectrum differences $\Delta{f}$ of the original time series.}
    \label{FigA:GOI:Return:MFDFA:df:test}
\end{figure}

\end{document}